\documentclass[twocolumn,pra,superscriptaddress]{revtex4}
\usepackage{dcolumn}
\usepackage{graphicx}
\usepackage{bbm}
\usepackage{amssymb,amsmath}
\usepackage{latexsym}
\usepackage{amsfonts}
\usepackage{epsfig}
\usepackage{psfrag}
\usepackage{graphicx}
\usepackage{hyperref}
\usepackage{color}
\usepackage{soul,xcolor}
\usepackage[normalem]{ulem}
\usepackage[titletoc]{appendix}

\newcommand{\blue}[1]{{\color{blue}#1}}

\newcommand{\be}{\begin{equation}}
\newcommand{\ee}{\end{equation}}
\newcommand{\beq}{\begin{eqnarray}}
\newcommand{\eeq}{\end{eqnarray}}

\newcommand{\hide}[1]{}

\newcommand{\ket}[1]{\left| #1 \right\rangle}
\newcommand{\bra}[1]{\left\langle #1 \right|}

\begin{document}

\title{Effective spin-spin interactions in bilayers of Rydberg atoms and polar molecules}

\author{Elena Kuznetsova}
\address{IQSE, Texas A$\&$M University, College Station, TX, 77840, USA}
\address{Rzhanov Institute of Semiconductor Physics, Novosibirsk, 630090, Russia}
\author{Seth T. Rittenhouse}
\address{Department of Physics, The United States Naval Academy, Annapolis, MD 21402, USA}
\author{I. I. Beterov}
\address{Rzhanov Institute of Semiconductor Physics, Novosibirsk, 630090, Russia}
\address{Novosibirsk State University, Novosibirsk, 630090, Russia}
\author{Marlan O. Scully}
\address{IQSE, Texas A$\&$M University, College Station, TX, 77840, USA}
\address{Department of Mechanical and Aerospace Engineering, Princeton University, Princeton, New Jersey 08544, USA}
\address{Department of Physics, Baylor University, Waco, Texas 76706, USA}
\author{Susanne F. Yelin}
\address{Department of Physics, University of Connecticut, 2152 Hillside Road, Storrs, CT 06269}
\address{Department of Physics, Harvard University, 17 Oxford Street, Cambridge, MA 02138, USA}
\address{ITAMP, Harvard-Smithsonian Center for Astrophysics, 60 Garden Street, Cambridge, MA 02138, USA}
\author{H. R. Sadeghpour}
\address{ITAMP, Harvard-Smithsonian Center for Astrophysics, 60 Garden Street, Cambridge, MA 02138, USA}

\date{\today}

\begin{abstract}
We show that indirect spin-spin interactions between effective spin-1/2 systems can be realized in two parallel 1D optical lattices loaded with polar molecules and/or Rydberg atoms. 
The effective spin can be encoded into low-energy rotational states of polar molecules or long-lived states of Rydberg atoms, tightly trapped in a deep optical lattice. 
The spin-spin interactions can be mediated by Rydberg atoms, placed in a parallel shallow optical lattice, interacting with the effective spins by charge-dipole 
(for polar molecules) or dipole-dipole (for Rydberg atoms) interaction. Indirect XX, Ising and XXZ interactions with interaction 
coefficients $J^{\bot}$ and $J^{zz}$ sign varying with interspin distance can be realized, in particular, the $J_{1}-J_{2}$ XXZ model with frustrated ferro-(antiferro-)magnetic
nearest (next-nearest) neighbor interactions.

\end{abstract}

\maketitle

\section{Introduction}

Polar molecules and Rydberg atoms interact via strong, anisotropic and long-range dipole-dipole intraspecies
and charge-dipole interspecies interactions. Both systems have long-lived internal states, which can encode qubits and effective spins, 
such as low-energy rotational states of the ground electronic and vibrational state of molecules and long-lived high-$n$ states of Rydberg atoms. 
The long-lived qubit/effective spin states and strong long-range interactions make for highly attractive quantum computation and quantum simulation platforms \cite{Pol-mol-review,Rydb-QC-review}. In periodic trap arrays, these systems offer the additional advantage of scalability to qubit numbers sufficient  for large-scale simulations \cite{Lukin-51-qubit,Monroe-53-qubits}.

Quantum magnetism is particularly amenable to simulations with ultracold 
atomic and molecular systems because various types of magnetism models can be modelled due to exquisite control over
atomic interactions.
In particular, polar molecules can efficiently simulate various quantum magnetism models \cite{Pol-mol-QSim-magnetism}, e.g.  effective XX spin-exchange has been realized in a 3D lattice of KRb molecules \cite{Spin-exchange-pol-mol-experim}. Rydberg atoms have also been proposed for quantum simulation of magnetism phenomena \cite{QSim-magnetism-Rydbergs}, starting 
with the seminal work on realization of an Ising model  with Rydberg crystals \cite{Rydb-crystals-theory}, 
recently demonstrated in \cite{Rydb-crystals-experim}, and extending to simulation of exotic frustrated magnetic states such as quantum spin-ice \cite{Zoller-spin-ice}.

One particularly interesting class of magnetic interactions are indirect, {\it i.e. mediated}, spin-spin interactions. Examples  include superexchange  \cite{Superexchange,Superexchange-QDots}, electron-spin mediated interaction between nuclear spins in molecules (J-coupling) \cite{NMR-J-coupling}, and Ruderman-Kittel-Kasuya-Yosida (RKKY) interaction between localized magnetic impurities in metals and semiconductors, mediated by
coupling to conduction electron spins \cite{RKKY,GMR,Heavy-RE-RKKY,Magn-semicond-RKKY}. The RKKY interaction is of special interest in that it has a sign periodically varying with the distance between the impurities,  
which can lead to frustration and random magnetization, producing non-trivial magnetic phases such as 
 spin glass \cite{Spin-glass-book}. Frustrated magnetic systems with sign-changing interactions, in particular, with competing ferromagnetic nearest and antiferromagnetic next nearest neighbor
 interactions, such as copper oxide spin chains \cite{LiCu2O2, NaCu2O2, PbCuSO4OH, LiCuSbO4,Multiferroic-cuprates},  have attracted active interest 
in recent years \cite{Frustrated-interactions,J1-J2-XXZ-model,J1-J2-XXZ-phase-diagram} due to unusual magnetic properties of the 
corresponding materials stemming from large degeneracies of their ground states induced by frustration.

In the present work we consider indirect interaction in a setup comprised of effective spin-1/2 systems, 
encoded 
in either rotational molecular or atomic Rydberg states, mediated by their respective interactions with auxiliary spin-1/2 systems, encoded in Rydberg atom states. 
The effective and mediator spins can be trapped in two 1D parallel optical lattices or trap arrays such that the effective spins are
tightly trapped in their sites, while the mediator spins are loosely trapped and because of spatial delocalization of their motional wavefunction can simultaneously 
interact with several effective spins.

Drawing an analogy with the RKKY interaction the tightly trapped polar 
molecules/Rydberg atoms act as 
 localized magnetic impurities, and the weakly trapped mediator Rydberg atoms play the role of conduction electrons. We show that this indirect interaction 
can change sign depending on interspin distance analogous to the one of RKKY. The resulting interaction extends beyond nearest neighbors,  e.g. nearest and 
next-nearest neighbors can interact with comparable strengths, allowing to realize the $J_{1}-J_{2}$ XXZ model \cite{J1-J2-XXZ-model}. By making the next-nearest neighbor interaction 
antiferromagnetic, frustrated interactions in the $J_{1}-J_{2}$ model similar to those in 1D copper oxide spin chains, can be realized.

Indirect magnetic interactions such as superexchange have been previously simulated with ultracold atoms in an optical lattice \cite{Superexchange-Trotzky}. Past 
proposals also include simulation of phonon-mediated electron interactions in a hybrid system of trapped ions and ground state atoms \cite{Bilayer-atoms-ions} 
and in a bilayer of Rydberg atoms \cite{Bilayer-Rydberg}. Spin-spin interactions with distance dependent tunable interaction 
strengths, giving rise to frustration, were simulated in a linear chain of three ions \cite{Frust-Ising-ions}, where the spin-spin interactions were mediated by phonons 
in the ion chain \cite{Spin-spin-ions-phonons}. A similar approach 
to realize phonon-mediated spin-spin interactions with distance-dependent interaction strength was considered in  Ref.\cite{Ising-pol-mol} 
for polar molecules arranged in a dipolar crystal. Spin-spin interactions can also be mediated via interaction with electromagnetic modes 
of a cavity \cite{QSim-cavity-modes}. 
Nuclear spins of 
ultracold atoms of two internal electronic states, tightly and weakly trapped in an optical lattice, were proposed to simulate the RKKY interaction \cite{Gorshkov-SU(N)-magnetism}. The atoms interact 
via short-range $s$-wave potential, making the corresponding interaction strengths much smaller compared to the long-range dipole-dipole or charge-dipole interactions considered here.

The systems envisioned in this work offer the possibility that a large realizable atomic and molecular parameter set can be exploited to simulate a range of many-body interactions. 
The paper is organized as follows. In Section II we describe the system and derive the effective Hamiltonian for indirect interaction 
between effective spins encoded in polar molecules or Rydberg atoms. In Section III two examples of simulation of indirect interaction are discussed: 
i) XX interaction by encoding spins into low-energy rotational states of polar molecules, interacting via mediator Rydberg atoms; ii) 
XXZ interaction by spin encoding into states of Rydberg atoms, mediated by a Rydberg atom in a different state. Finally, we conclude in 
Section IV.

\section{Model description}
 In this section, we will introduce the physical system (cf, Fig.~\ref{fig:setup-scheme}), derive its Hamiltonian, and simplify it into the effective 
Hamiltonian by tracing out and averaging over the degrees of freedom of the mediating Rydberg atoms. It will then be obvious that, like in the RKKY case, 
this system leads to interspin distance-dependent sign-changing interaction,   which constitutes the main result of this article.

\subsection{Effective interaction Hamiltonian}

\begin{figure}
\center{
\includegraphics[width=9.cm]{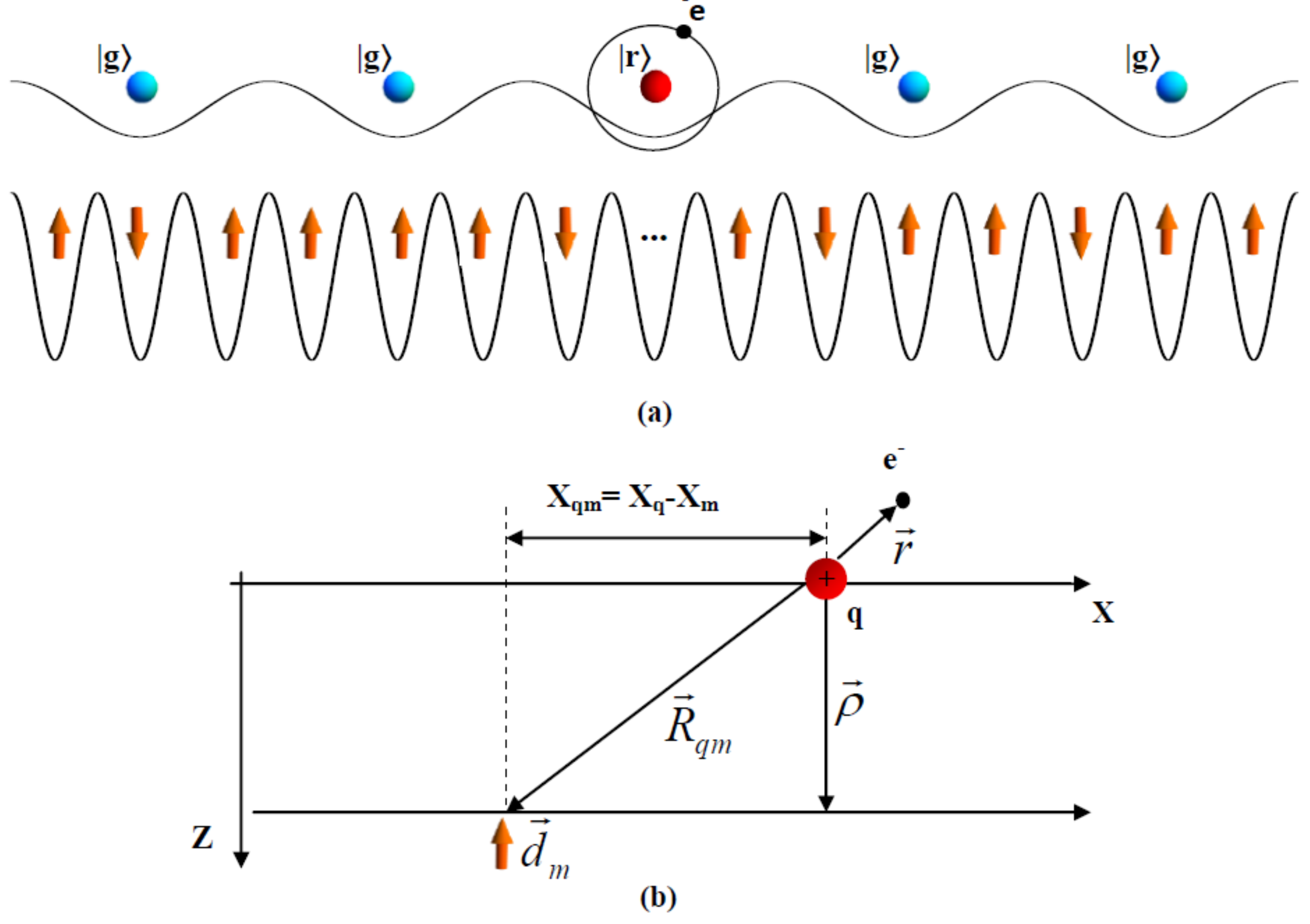}
\caption{\label{fig:setup-scheme} {\it Setup schematic.} (a) Illustration of the bilayer setup, in which spin-encoding polar molecules or Rydberg 
atoms are trapped in a deep optical lattice or trap array, and the mediating Rydberg atom(s) is placed in a shallow optical lattice such 
that its spatial wave function is delocalized over several sites, and it can simultaneously interact with several spins; 
(b) Geometry of the setup: 
an effective m$^{\rm th}$ spin with a dipole moment $\vec{d}_{m}$ interacts with a mediator atom via charge-dipole in case of polar molecule 
spin encoding or dipole-dipole interaction in the case of Rydberg atom spin encoding. The distance between the two parallel lattices is $|\vec{\rho}|$, 
$\vec{R}_{qm}$ is the vector connecting the ionic core of the q$^{\rm th}$ mediator atom to the spin, $X_{qm}=X_{q}-X_{m}$ is the 
distance between the mediator atom and the spin along the $X$ axis, $\vec{r}$ is the vector connecting the mediator Rydberg electron and the 
ionic core.  }
}
\end{figure}

We consider a setup with two parallel 1D optical lattices or trap arrays, one filled with polar molecules or Rydberg atoms representing effective spin-1/2 
systems, and another filled with auxiliary Rydberg atoms, mediating the interaction between the effective spins, as illustrated in Fig.\ref{fig:setup-scheme}a. 
The effective spins are assumed to be tightly trapped in their optical lattice such 
that tunneling between sites is strongly suppressed. They interact with Rydberg atoms trapped in a parallel shallow optical lattice, in which 
tunneling is significant. 
The mediator atoms in the second lattice can interact with the spin-encoding species via charge-dipole interaction \cite{Charge-dipole-interaction}
\begin{eqnarray}
\label{eq:ch-dip-inter}
V_{\rm cd}=\frac{e\vec{d}_{\rm spin}\cdot\vec{R}}{R^{3}}-\frac{e\vec{d}_{\rm spin}\cdot\left(\vec{R}-\vec{r}\right)}{\left|\vec{R}-\vec{r}\right|^{3}},
\end{eqnarray}
if spins are encoded in polar molecules and via dipole-dipole interaction
\begin{eqnarray}
\label{eq:dip-dip-inter}
V_{\rm dd}=\frac{\vec{d}_{\rm spin}\cdot\vec{d}_{\rm Ryd}}{R^{3}}-\frac{3\left(\vec{d}_{\rm spin}\cdot\vec{R}\right)\left(\vec{d}_{\rm Ryd}\cdot\vec{R}\right)}{R^{5}},
\end{eqnarray}
if spins are encoded in Rydberg atoms. Here $\vec{d}_{\rm spin}$ is the spin electric dipole moment, $\vec{d}_{\rm Ryd}$ is the electric dipole moment of the mediator 
Rydberg atom, $\vec{R}$ is the distance between the mediator Rydberg atom ionic core and the spin-encoding system, 
and $\vec{r}$ is the distance between the Rydberg electron and the ionic core. The mediator atoms are assumed to be initially prepared in 
the $ns$ Rydberg state, and can couple to neighboring $np_{j}$, $(n-1)p_{j}$ states due to the interaction with the spins. {\blue In the following we actually will need to consider only the coupling to $np_{j}$ states.}  

The effective spins are assumed to occupy the lowest energy band of the deep optical lattice in both $\ket{\downarrow}$, $\ket{\uparrow}$ 
spin states, such that the spin-mediator interaction does not excite the spins to higher-energy bands. The mediator atoms, on the  other hand, are assumed to be  
initially prepared in the lowest-energy band in their lattice in the $\ket{ns,m_{j}}$ Rydberg state, such that 
they can be excited to $\ket{np_{j'},m_{j}'}$ internal and higher-energy motional states by the interaction with the spins. The mediator states are denoted as 
$\ket{{\bf n},k_{\nu}}_{q}=\phi(X_{q},k_{\nu\;q})\ket{nl_{j},m_{j}}_{q}$, 
describing the motional $\phi(X_{q},k_{\nu\;q})$ and internal $\ket{nl_{j},m_{j}}_{q}$ states of the q$^{\rm th}$ mediator 
atom, and ${\bf n}=\{n,l,j,m_{j}\}$ is a short-hand notation for internal quantum numbers. Here we 
assume for simplicity that the 
trapping potentials and therefore the motional states for $\ket{ns,m_{j}}$ and $\ket{np_{j'},m_{j}'}$ internal states are the same, which is not a 
principal requirement, and will be only used to simplify numerical analysis in the next section. Rydberg atoms can be trapped in intensity minima in a ponderomotive \cite{Rydberg-ponderomotive-lattice} or a blue-detuned 
\cite{Rydberg-blue-detuned-lattice} optical lattice. The latter  also can be used to trap atoms in their ground state, which can be used if a superatom or a dressed 
Rydberg mediator state is used, as will be discussed later. The motional wavefunction of 
the q$^{\rm th}$ mediator atom is then given by 
the Bloch function of a $\nu^{\rm th}$ Bloch band, corresponding to the quasimomentum $k$:
\begin{equation}
\phi(X_{q},k_{\nu\;q})=u_{k}^{(\nu)}(X_{q})e^{ikX_{q}},
\end{equation}
where $X_{q}$ is the coordinate for the atom along the lattice, $u_{k}^{(\nu)}(X_{q}+L_{\rm at})=u_{k}^{(\nu)}(X_{q})$ is periodic with the period $L_{\rm at}$ 
of the mediator atom's lattice. We also assume periodic boundary conditions $N_{\rm latt\;at}kL_{\rm at}=2\pi n_w$, where 
$N_{\rm latt\;at}$ is the number of sites in the mediator lattice and $n_w$ is the periodicity integer.

The Hamiltonian without the spin-mediator interaction has the form:
\begin{eqnarray}
\lefteqn{\hat{H}_{0} = \sum_{i=1}^{N}E_{\rm spin}\ket{\uparrow}_{i}\bra{\uparrow}_{i}+}\nonumber \\
&&\qquad+\sum_{q=1}^{N_{a}}\sum_{\substack{{\bf n} \\ k,\nu}}{\cal E}_{\bf n}(k_{\nu})\ket{{\bf n},k_{\nu}}_{q}\bra{{\bf n},k_{\nu}}_{q}, \nonumber 
\end{eqnarray}
where $\textbf{n}=\{n,l,j,m_j\}$ is the state of the mediator atom. The summation is over $i=1,...,N$ effective spins and $q=1,..,N_{a}$ mediator atoms in the first and second optical lattices, respectively; over  
$nl_{j}=ns,np_{1/2,3/2}$; $m_{j}=\pm 1/2,\pm 3/2$ internal states of the mediator atoms, and their quasimomenta $k$ 
in the first Brillouin zone of $\nu=1,...,\infty$ Bloch bands. 
Here $E_{\rm spin}=E_{\uparrow}-E_{\downarrow}$ is the spin transition energy and the mediator atom energy ${\cal E}_{\bf n}(k_{\nu})$ includes both 
internal $E_{\bf n}$ and motional energy of 
the corresponding Bloch states.

The spin-mediator interaction Hamiltonian can be written in the combined basis of spin and mediator states:
\begin{align}
\hat{V}=\sum_{i=1}^{N}\sum_{q=1}^{N_{a}}\sum_{\substack{ {\bf n},{\bf n'} \\ \alpha,\beta=\uparrow,\downarrow \\ k,k' \\ \nu,\nu' }}
\ket{{\bf n},k_{\nu}}_{q}\ket{\alpha_{i}}\bra{\beta_{i}}\bra{{\bf n'},k'_{\nu'}}_{q} \times \nonumber \\
\times \bra{\alpha_{i}}\bra{{\bf n},k_{\nu}}_{q}\hat{V}\ket{{\bf n'},k'_{\nu'}}_{q}\ket{\beta_{i}}, 
\label{eq:Inter}
\end{align}
where $\hat{V}=\hat{V}_{\rm cd}$ for polar molecule spin encoding and $\hat{V}=\hat{V}_{\rm dd}$ for Rydberg atom spin encoding.

Next we show how the spin-mediator interaction Eq.(\ref{eq:Inter}) gives rise to indirect interaction between the effective spins.  
The interaction Hamiltonian in the basis of two-spin states $\ket{\alpha_{i}\beta_{m}}$  
 is:
\begin{align}
\hat{V}   =&\sum_{i,m=1}^{N}\sum_{q=1}^{N_{a}}\sum_{\substack{\alpha
,\beta,\gamma,\delta\\\mathbf{n},\mathbf{n}^{\prime},\mathbf{n}^{\prime\prime}}%
}\sum_{\substack{k,k^{\prime},k^{\prime\prime}\\\nu,\nu^{\prime},\nu
^{\prime\prime}}}\left[  \left\vert \mathbf{n},k_{\nu}\right\rangle
_{q}\left\vert {\alpha_{i}\beta_{m}}\right\rangle \right.\\ 
& \times \left.\left(  V_{\mathbf{n},k_{\nu},\alpha;\mathbf{n}^{\prime},k_{\nu^{\prime}}^{\prime},\gamma}%
^{iq}\delta_{\beta_{m},\delta_{m}}+\right.  \right.  \nonumber\\
& \left.  \left.  +V_{\mathbf{n},k_{\nu},\beta;\mathbf{n}^{\prime}%
,k_{\nu^{\prime}}^{\prime},\delta}^{mq}\delta_{\alpha_{i},\gamma_{i}%
}\right)  \left\langle {\gamma_{i}\delta_{m}}\right\vert \left\langle
\mathbf{n}{^{\prime},k}_{\nu^{\prime}}^{\prime}\right\vert _{q}+\mathrm{H.c.}%
\right]  +\nonumber\\
& +\left[  \left\vert \mathbf{n},k_{\nu}\right\rangle _{q}\left\vert
{\alpha_{i}\beta_{m}}\right\rangle \left(  V_{\mathbf{n},k_{\nu}%
,\alpha;\mathbf{n},k_{\nu^{\prime}}^{\prime},\gamma}^{iq}\delta_{\beta
_{m},\delta_{m}}+\right.  \right.  \nonumber\\
& \left.  \left.  +V_{\mathbf{n},k_{\nu},\beta;\mathbf{n},k_{\nu^{\prime}%
}^{\prime},\delta}^{mq}\delta_{\alpha_{i},\gamma_{i}}\right)  \left\langle
{\gamma_{i}\delta_{m}}\right\vert \left\langle \mathbf{n},k_{\nu^{\prime}%
}^{\prime}\right\vert _{q}+\mathrm{H.c.}\right]  +\nonumber\\
& +\left[  \left\vert \mathbf{n}{^{\prime}},k_{\nu^{\prime}}^{\prime
}\right\rangle _{q}\left\vert {\alpha_{i}\beta_{m}}\right\rangle \left(
V_{\mathbf{n}^{\prime},k_{\nu^{\prime}}^{\prime},\alpha;\mathbf{n}%
^{\prime\prime},k_{\nu^{\prime\prime}}^{\prime\prime},\gamma}^{iq}%
\delta_{\beta_{m},\delta_{m}}+\right.  \right.  \nonumber\\
& \left.  \left.  +V_{\mathbf{n}^{\prime},k_{\nu^{\prime}}^{\prime}%
,\beta;\mathbf{n}^{\prime\prime},k_{\nu^{\prime\prime}}^{\prime\prime}%
,\delta}^{mq}\delta_{\alpha_{i},\gamma_{i}}\right)  \left\langle
{\gamma_{i}\delta_{m}}\right\vert \left\langle \mathbf{n}{^{\prime\prime}%
,}k_{\nu^{\prime\prime}}^{\prime\prime}\right\vert _{q}+\mathrm{H.c.}\right]
,\label{eq:Inter-two-spin}%
\end{align}
where the sums over the index vectors are restricted to ${\bf n}=\{ns,j=1/2,m_j=\pm1/2\}$, 
${\bf n'}=\{np,j=3/2,1/2,m_j=\pm 3/2,\pm 1/2\}$,and  
 ${\bf n}''=\{np,j=3/2,1/2,m_j=\pm 3/2,\pm 1/2\}$ for a fixed radial quantum number $n$. The interaction matrix element between 
the m$^{\rm th}$ spin and the q$^{\rm th}$ mediator atom is 
$V_{{\bf n},k_{\nu},\xi;{\bf n'},k'_{\nu'},\eta}^{mq}=\bra{{\bf n},k_{\nu}}_{q}\bra{\xi_{m}}\hat{V}\ket{\eta_{m}}\ket{{\bf n'},k'_{\nu'}}_{q}$, 
which describes the process in which the $q^{\rm th}$ mediator atom is transferred from the $\ket{{\bf n},k_{\nu}}$ to the $\ket{{\bf n'},k'_{\nu'}}$ state, and 
the $m^{\rm th}$ spin goes from the $\ket{\xi}$ to the $\ket{\eta}$ state.
 In particular,  
the term in the first square bracket describes the interaction in which a mediator atom changes parity $\ket{ns.m_{j}} \leftrightarrow \ket{np_{j'},m_{j'}'}$, the terms in the  
second and third square brackets 
 describe  the mediator being transferred to electronic states of the same parity 
$\ket{ns,m_{j}}\leftrightarrow \ket{ns,m_{j}}$ and $\ket{np_{j},m_{j}}\leftrightarrow \ket{np_{j'},m_{j'}'}$, which is allowed by the 
charge-dipole interaction. The  parity  conserving interaction terms are present only in the polar molecule spin encoding setup and 
can be neglected for sufficiently large spin-mediator distances, 
allowing  for the 
charge-dipole interaction  to be approximated by the dipole-dipole one, because at these distances the corresponding interaction matrix elements differ by a very small amount, as is discussed in Appendix D, end of part A.

The interaction Hamiltonian in the limit of weak interaction $|\hat{V}|\ll E_{\rm spin},|E_{np_{j'}}-E_{ns}|,|E_{np_{j'}}-E_{ns}\pm E_{\rm spin}|$ induces 
energy shifts and couplings among many-body spin states $\ket{\alpha_{1}\alpha_{2}...\alpha_{N}}$, corresponding to the same mediator state 
$\ket{{\bf n},k_{\nu}}_{q}$, which have a form of interaction between the effective spins. 
This can be shown using the Schrieffer-Wolff transformation
\begin{eqnarray}
e^{\hat{S}}\hat{H}e^{-\hat{S}}=\hat{H}+\left[\hat{S},\hat{H}\right]+\frac{\left[\hat{S},\left[\hat{S},\hat{H}\right]\right]}{2}+O\left(\hat{S}^{3}\right),
\end{eqnarray}
in which the terms of the first order in $\hat{V}$ are eliminated by setting $\left[\hat{S},\hat{H}_{0}\right]=-\hat{V}$, where the corresponding generator $\hat{S}$ is 
given in Appendix A. The transformed Hamiltonian has the form:
\begin{eqnarray}
\label{eq:Schrieffer-Wolff}
e^{\hat{S}}\hat{H}e^{-\hat{S}}=\hat{H}_{0}+\frac{\left[\hat{S},\hat{V}\right]}{2}+O\left(|\hat{V}|^{3}\right),
\end{eqnarray} 
in which the effective interaction $\hat{V}_{\rm eff}=\left[\hat{S},\hat{V}\right]/2$ terms are now of the second order in $\hat{V}$.

Assuming for concreteness that the mediator atoms are initially prepared in a single or a superposition of $\ket{{\bf ns},k_{\nu}}$ states,  
 we are interested in the projection of the effective interaction on these states: 
\begin{align}
\hat{V}_{\rm eff}^{ns}=\hat{P}_{ns}\hat{V}_{\rm eff}\hat{P}_{ns}=
\sum_{q=1}^{N_{a}}\sum_{\substack{ k,\nu \\ m_{j}=\pm 1/2 }}\ket{{\bf ns},k_{\nu}}_{q}\bra{{\bf ns},k_{\nu}}_{q}\times \nonumber \\
\times \left(\sum_{i,m=1}^{N}\sum_{\alpha,\beta,\gamma,\delta=\uparrow,\downarrow}K_{\alpha_{i} \beta_{m},\gamma_{i} \delta_{m}}^{q,k_{\nu}}\ket{\alpha_{i} \beta_{m}}\bra{\gamma_{i} \delta_{m}}\right),  
\label{eq:Veff}
\end{align}
where  
\begin{eqnarray}
\label{eq:ns-projector} 
\hat{P}_{ns}&=&\sum_{q=1}^{N_a}\sum_{\substack{k,\nu \\ m_{j}=\pm 1/2}}\ket{{\bf ns},k_{\nu}}_q\bra{{\bf ns},k_{\nu}}_{q}= \\
&=&\sum_{q=1}^{N_a}\sum_{\substack{ k, \nu \\ m_{j}=\pm 1/2}}\ket{ns_{1/2},m_{j},k_{\nu}}_{q}\bra{ns_{1/2},m_{j},k_{\nu}}_{q} \nonumber
\end{eqnarray} 
is the projection operator on the $ns$ state and the $K_{\alpha \beta,\gamma \delta}^{q,k_{\nu}}$ coefficients are given in Appendix B.

Replacing $\ket{\alpha_{i}\beta_{m}}\bra{\gamma_{i}\delta_{m}}$ by $\hat{S}_{i}^{\pm,z}\hat{S}_{m}^{\pm,z}$ spin-1/2 operators 
as shown in Appendix B, we can rewrite Eq.(\ref{eq:Veff}) in the following way:
\begin{eqnarray*}
\lefteqn{\hat{V}_{\rm eff}^{ns}=\sum_{q=1}^{N_{a}}\sum_{\substack{k,  \nu \\ m_{j}=\pm 1/2}}\ket{{\bf ns},k_{\nu}}_{q}\bra{{\bf ns},k_{\nu}}_{q} \times } \\
&&\times \sum_{i,m=1}^{N}\Bigg[J_{im}^{zz\;q,k_{\nu}}\hat{S}_{i}^{z}\hat{S}_{m}^{z}+J_{im}^{+-\;q,k_{\nu}}\hat{S}_{i}^{+}\hat{S}_{m}^{-}+ \\
&& \qquad\qquad +\Big(J_{im}^{+-\;q,k_{\nu}}\big)^{*}\hat{S}_{i}^{-}\hat{S}_{m}^{+} 
+J_{im}^{++\;q,k_{\nu}}\hat{S}_{i}^{+}\hat{S}_{m}^{+}+ \\
&&\qquad \qquad + \big(J_{im}^{++\;q,k_{\nu}}\big)^{*}\hat{S}_{i}^{-}\hat{S}_{m}^{-}
+J_{im}^{z+\;q,k_{\nu}}\hat{S}_{i}^{z}\hat{S}_{m}^{+}+ \\
&&\qquad\qquad +\big(J_{im}^{z+\;q,k_{\nu}}\big)^{*}\hat{S}_{i}^{z}\hat{S}_{m}^{-}
+b_{im}^{z\;q,k_{\nu}}n_{m}\hat{S}_{i}^{z}+\\
&&\qquad\qquad + b_{im}^{+\;q,k_{\nu}}n_{m}\hat{S}_{i}^{+}+ 
\big(b_{im}^{+\;q,k_{\nu}}\big)^{*}n_{m}\hat{S}_{i}^{-}+b_{0\;im}^{q,k_{\nu}}n_{i}n_{m}\Bigg], 
\end{eqnarray*}  
where $n_{i}$ is the number of spins at site $i$, and the interaction coefficients $J_{im}^{+- \;q,k_{\nu}}$, $J_{im}^{zz \; q,k_{\nu}}$, $J_{im}^{++\;q,k_{\nu}}$, $J_{im}^{z+\;q,k_{\nu}}$ and the 
coefficients $b_{i}^{z\;q,k_{\nu}}$, $b_{i}^{+\;q,k_{\nu}}$ and $b_{0}^{q,k_{\nu}}$ are given in Appendix C.

If the effective interaction is weak such that $|J_{im}|\ll E_{\rm spin},|E_{np_{j'}}-E_{ns}|,|E_{np_{j'}}-E_{ns} \pm E_{\rm spin}|$,   
 the non-resonant terms $J_{im}^{++\;q,k_{\nu}}\hat{S}_{i}^{+}\hat{S}_{m}^{+}$, $\left(J_{im}^{++\;q,k_{\nu}}\right)^{*}\hat{S}_{i}^{-}\hat{S}_{m}^{-}$, 
$J_{im}^{z+\;q,k_{\nu}}\hat{S}_{i}^{z}\hat{S}_{m}^{+}$, $\left(J_{im}^{z+\;q,k_{\nu}}\right)^{*}\hat{S}_{i}^{z}\hat{S}_{m}^{-}$, $b_{i}^{z\;q,k_{\nu}}\hat{S}_{i}^{+}$ and  
$\left(b_{i}^{z\;q,k_{\nu}}\right)^{*}\hat{S}_{i}^{-}$, coupling collective spin states with energies differing by the spin transition energy or twice this energy, 
can be neglected. As a result, 
we are left with the effective Hamiltonian: 
\begin{align}
\hat{V}_{\rm eff}^{ns}=\sum_{i,m=1}^{N}\sum_{q=1}^{N_{a}}\sum_{\substack{ k,\nu \\ m_{j}=\pm 1/2} }\ket{{\bf ns},k_{\nu}}_{q}\bra{{\bf ns},k_{\nu}}_{q}\times \nonumber \\
\times \left(J_{im}^{zz\;q,k_{\nu}}\hat{S}_{i}^{z}\hat{S}_{m}^{z}+
\frac{J_{im}^{\bot \;q,k_{\nu}}}{2}\left(\hat{S}_{i}^{+}\hat{S}_{m}^{-}+\hat{S}_{i}^{-}\hat{S}_{m}^{+}\right)+ \right. \nonumber \\
\left. +b_{im}^{z\;q,k_{\nu}}n_{m}\hat{S}_{i}^{z}+b^{q,k_{\nu}}_{0\;im}n_{i}n_{m}\right),
\label{eq:V-eff-ns}
\end{align}
where we assumed $\left(J_{im}^{+-\;q,k_{\nu}}\right)^{*}=J_{im}^{+-\;q,k_{\nu}}=J_{im}^{\bot \;q,k_{\nu}}/2$. 

\subsection{Averaging over initial mediator states}

The Hamiltonian, acting only on effective spins can be obtained by taking the expectation value of Eq.(\ref{eq:V-eff-ns}) with respect to an 
unperturbed initial state of the mediator atoms. As a first example we consider the mediator atoms prepared in a Rydberg $\ket{{\bf ns},k_{0\;\nu_{0}}}$ 
superatom state:
\begin{align}
\ket{\Psi}_{\rm supat}=&\sum_{q=1}^{N_{a}}\sum_{\substack{ k',k_{0} \\ \nu',\nu_{0}}}\prod_{q' \ne q}\frac{c_{k'_{\nu'},k_{0\;\nu_{0}}}}{\sqrt{N_a}}\phi_{g_{q'}}\left(X_{q'},k'_{\nu'\;q'}\right) \nonumber \\
&\times \Phi_{ns_{q}}\left(X_{q},k_{0\;\nu_{0} \; q}\right)\ket{g_{1},...(ns_{1/2},m_{j})_{q},...,g_{N_{a}}},
\label{eq:Psi-supat}
\end{align}
where $\phi_{g_{q'}}\left(X_{q'},k'_{\nu'\;q'}\right)$ is the spatial wave function of a q'$^{\rm th}$ atom in the ground state; 
$\Phi_{ns_{q}}\left(X_{q},k_{0\; \nu_{0} \;q}\right)$ is the spatial wave function of the q$^{\rm th}$ atom in the $\ket{ns}$ state; in the general case the 
atoms in the ground and Rydberg states are assumed to be prepared in a wave packet of Bloch states with quasimomenta $k'_{\nu'}$ and $k_{0\;\nu_{0}}$, respectively, 
weighted by the coefficients $c_{k'_{\nu'},k_{0\;\nu_{0}}}$. In this case 
the spin Hamiltonian takes the form:
\begin{eqnarray}
\label{eq:V-eff-supat}
\hat{V}_{\rm eff\;spin}^{ns}&=&\bra{\Psi_{\rm supat}}\hat{V}_{\rm eff}^{ns}\ket{\Psi_{\rm supat}}= \\
&=&\sum_{i,m=1}^{N}
\Bigg(J_{im}^{zz}\hat{S}_{i}^{z}\hat{S}_{m}^{z}+\frac{J_{im}^{\bot}}{2}\big(\hat{S}_{i}^{+}\hat{S}_{m}^{-}+\hat{S}_{i}^{-}\hat{S}_{m}^{+}\big)+  \nonumber\\
&&\qquad\qquad +b_{im}^{z}n_{m}\hat{S}_{i}^{z}+b_{0\;im}n_{i}n_{m}\Bigg), \nonumber
\end{eqnarray}
with the averaged interaction coefficients:
\begin{align}
J_{im}^{zz(\bot)}=\frac{1}{N_{a}}\sum_{q=1}^{N_{a}}\sum_{\substack{ k,k',k_{0},k'_{0} \\ \nu,\nu',\nu_{0},\nu'_{0}}}c_{k'_{\nu'},k'_{0\;\nu_{0}'}}(c_{k'_{\nu'},k_{0\; \nu_{0}}})^{*}\times \nonumber \\
\times J_{im}^{zz(\bot)\;q,k_{\nu}}\int dX_{q}\Phi_{ns_{q}}^{*}(X_{q},k_{0\;\nu_{0}\;q})\phi(X_{q},k_{\nu\;q})\times \nonumber \\
\times \int dX_{q}\Phi_{ns_{q}}(X_{q},k'_{0\;\nu'_{0}\;q})\phi^{*}(X_{q},k_{\nu\;q}),
\label{eq:averaged-J-coeff-supat}
\end{align}
where $\phi(X_{q},k_{\nu\;q})$ is the spatial part of the mediator atom wave function in the $\hat{P}_{ns}$ projector Eq.(\ref{eq:ns-projector}). 
The averaged effective magnetic field $b_{im}^{z}$ and $b_{0\;im}$ satisfy the same relation.

Next, we use the assumption that initially the mediator atoms are prepared in a superposition of Bloch states and write explicitly Bloch functions as the spatial parts of the mediator atom wavefunction 
$\Phi_{ns_{q}}(X_{q},k_{0\; \nu_{0}\;q})=u_{k_{0}}^{(\nu_{0})}e^{ik_{0}X_{q}}$. 
 In this case $\int \Phi_{ns_{q}}^{*}(X_{q},k_{0\;\nu_{0}\;q})u_{k}^{(\nu)}e^{ikX_{q}}=\delta_{k,k_{0}}\delta_{\nu,\nu_{0}}$, giving the 
averaged interaction coefficients (same for $b_{im}^{z}$, $b_{0\;im}$):
\begin{align}
J_{im}^{zz(\bot)}=\frac{1}{N_{a}}\sum_{q=1}^{N_{a}}\sum_{\substack {k',k_{0} \\ \nu',\nu_{0}}}|c_{k'_{\nu'},k_{0\;\nu_{0}}}|^{2}J_{im}^{zz(\bot)\;q,k_{0\;\nu_{0}}}= \nonumber \\
=\frac{1}{N_{a}}\sum_{q=1}^{N_{a}}\sum_{\substack {k_{0},\nu_{0}}}|c_{k_{0\;\nu_{0}}}|^{2}J_{im}^{zz(\bot)\;q,k_{0\;\nu_{0}}},
\label{eq:averaged-J-supat}
\end{align}
where $|c_{k_{0\;\nu_{0}}}|^{2}=\sum_{k',\nu'}|c_{k'_{\nu'},k_{0\;\nu_{0}}}|^{2}$.
In particular, for mediator atoms initially prepared in a stationary 
BEC $k_{0}=0$, $\nu_{0}=1$ the averaged interaction coefficients are $J_{im}^{zz(\bot)}=\frac{1}{N_{a}}\sum_{q=1}^{N_{a}}J_{im}^{zz(\bot)\;q,k_{0}=0_{\nu_{0}=1}}$ \cite{BEC-init-state}. In a more general case the 
initial superatom state is a superposition of Bloch states with quasimomenta $k_{0}$ and Bloch bands $\nu_{0}$ determined by the distribution $|c_{k_{0\;\nu_{0}}}|^{2}$.  
 Assuming for simplicity that the only dependence of the $J_{im}^{zz(\bot)\;q,k_{0\;\nu_{0}}}$ coefficients 
on the initial quasimomentum $k_{0}$ is given by a prefactor $J_{im}^{zz(\bot)\;q,k_{0\;\nu_{0}}}\sim e^{-ik_{0}(X_{i}-X_{m})}$, results in the  
 averaged interaction coefficients, following from 
Eq.(\ref{eq:averaged-J-supat}) 
coefficients 
\begin{align}
J_{im}^{zz(\bot)} \sim \frac{1}{N_{a}}\sum_{q=1}^{N_{a}}J_{im}^{zz(\bot)\;q}\sum_{k_{0},\nu_{0}} \nonumber \\
 |c_{k_{0\;\nu_{0}}}|^{2}e^{-ik_{0}(X_{i}-X_{m})}. 
\label{eq:averaged-J-supat-simplified}
\end{align}
Assuming e.g. a Gaussian distribution $|c_{k_{0\;\nu_{0}}}|^{2} \sim e^{-k_{0}^2/\kappa_{0}^{2}}$ will give  
$\sum_{k_{0},\nu_{0}}|c_{k_{0\;\nu_{0}}}|^{2}e^{-ik_{0}(X_{i}-X_{m})} \sim e^{-(X_{i}-X_{m})^{2}\kappa_{0}^{2}/4}$ for a narrow 
wave packet with $\kappa_{0} \ll \pi/L_{\rm at}$, resulting in an additional factor of decay of interaction coefficients with an interspin distance, controlled by the 
 wave packet distribution width $\kappa_{0}$.

Another possible initial mediator state is the Rydberg dressed state
\begin{align}
\ket{\Psi}_{\rm dress}&=&\prod_{q=1}^{N_{a}}\sum_{\substack{ k',k_{0}  \\ \nu',\nu_{0}}} c_{k'_{\nu'},k_{0\;\nu_{0}}}\Bigg(c_{g}\phi_{g_{q}}(X_{q},k'_{\nu'\;q})\ket{g}_{q}+  \nonumber \\
&&\qquad\qquad +c_{ns}\Phi_{ns_{q}}(X_{q},k_{0\;\nu_{0}\;q})\ket{{\bf ns}}_{q}\Bigg),   
\label{eq:Psi-dressed}
\end{align}
created when all mediator atoms interact with a dressing laser field of Rabi frequency $\Omega$ and detuning $\Delta$ from the Rydberg state, 
and $c_{g}=\sqrt{\sqrt{\Delta^{2}/4+\Omega^{2}}+\Delta/2}/[\sqrt{2}\left(\Delta^{2}/4+\Omega^{2}\right)^{1/4}]$, $c_{ns}=\sqrt{\sqrt{\Delta^{2}/4+\Omega^{2}}-\Delta/2}/[\sqrt{2}\left(\Delta^{2}/4+\Omega^{2}\right)^{1/4}]$.   
Here $\phi_{g_{q}}(X_{q},k_{\nu\;q})$ is the spatial wave function of a q$^{\rm th}$ atom in the ground state; $\Phi_{ns_{q}}(X_{q},k_{0 \;\nu_{0} \; q})$ 
is the spatial wave function of the q$^{\rm th}$ atom in the Rydberg $ns$ state. 

In 
this case the spin Hamiltonian takes the same form as Eq.(\ref{eq:V-eff-supat})
with the averaged interaction coefficients (the same for $b_{im}^{z}$, $b_{0\;im}$):
\begin{eqnarray}
\lefteqn{J_{im}^{zz(\bot)}=\left|c_{ns}\right|^{2}\sum_{q=1}^{N_{a}}\sum_{\substack{ k,k',k_{0},k'_{0} \\ \nu,\nu',\nu_{0},\nu'_{0}}}c_{k'_{\nu'},k'_{0\;\nu_{0}'}}(c_{k'_{\nu'},k_{0\; \nu_{0}}})^{*}\times }\nonumber \\
&& \times J_{im}^{zz(\bot)\;q,k_{\nu}}\int dX_{q}\Phi_{ns_{q}}^{*}(X_{q},k_{0\;\nu_{0}\;q})\phi(X_{q},k_{\nu\;q})\times \nonumber \\
&&\times \int dX_{q}\Phi_{ns_{q}}(X_{q},k'_{0\;\nu'_{0}\;q})\phi^{*}(X_{q},k_{\nu\;q}),
\label{eq:averaged-J-coeff-dress}
\end{eqnarray}
where we again consider a case when initially the ground and Rydberg state atoms are prepared 
in a superposition of Bloch states determined by the weights $|c_{k'_{\nu'},k_{0\;\nu_{0}}}|^{2}$. Plugging the Bloch functions for the Rydberg and ground 
motional wavefunctions, their averages will be 
given by the following expression:
\begin{align}
J_{im}^{zz(\bot)}=|c_{ns}|^{2}\sum_{q=1}^{N_{a}}\sum_{\substack{ k',k_{0} \\ \nu',\nu_{0}}}|c_{k'_{\nu'},k_{0\;\nu_{0}}}|^{2}J_{im}^{zz(\bot)\;q,k_{0\;\nu_{0}}} \nonumber \\
=|c_{ns}|^{2}\sum_{q=1}^{N_{a}}\sum_{\substack{k_{0},\nu_{0}}}|c_{k_{0\;\nu_{0}}}|^{2}J_{im}^{zz(\bot)\;q,k_{0\;\nu_{0}}}.
\label{eq:averaged-J-coeff-dress-reduced}
\end{align}
In particular, for the case of an initial BEC $k_{0}=0$, $\nu_{0}=1$ we have 
\begin{align}
J_{im}^{zz(\bot)}=|c_{ns}|^{2}\sum_{q=1}^{N_{a}}J_{im}^{zz(\bot)\;q,k_{0}=0_{\nu_{0}=1}}. \nonumber
\end{align}
The expressions (\ref{eq:averaged-J-coeff-supat}), (\ref{eq:averaged-J-coeff-dress}) are also valid in the case of a single mediator atom corresponding to $N_{a}=1$.

The total effective Hamiltonian will thus take the form: 
\begin{align}
\hat{H}_{\rm eff}=\hat{H}_{\rm 0\;spin}+\hat{V}_{\rm eff\;spin}^{ns}=\nonumber \\
=\sum_{i,m=1}^{N}
\left(J_{im}^{zz}\hat{S}_{i}^{z}\hat{S}_{m}^{z}+\frac{J_{im}^{\bot}}{2}\left(\hat{S}_{i}^{+}\hat{S}_{m}^{-}+\hat{S}_{i}^{-}\hat{S}_{m}^{+}\right)\right.+ \nonumber \\
\left.+\left(E_{\rm spin}\delta_{im}+b_{im}^{z}n_{m}\right)\hat{S}_{i}^{z}+b_{0\;im}n_{i}n_{m}\right). 
\label{eq:total-H-eff}
\end{align}
The effective Hamiltonian (\ref{eq:total-H-eff}) with averaged interaction coefficients from Eqs. (\ref{eq:averaged-J-coeff-supat}), and (\ref{eq:averaged-J-coeff-dress})
is the main result of our work. In the following we assume unit occupancy $n_{i}=1$ of the spin sites, which allows for the introduction of the effective magnetic field 
$b_{i}^{z}=\sum_{m \ne i}b_{im}^{z}$ at site $i$, and neglect a constant term $\sum_{i,m}b_{0\;im}n_{i}n_{m}$. In the case $b_{i}^{z}$ do not depend on $i$, 
the Hamiltonian couples collective spin states 
with the same $z$ component of the total spin $\hat{S}^{z}=\sum_{i=1}^{N}\hat{S}_{i}^{z}/N$, and describes the XXZ model of magnetism in the presence 
of a longitudinal magnetic field. 
As will be shown below, due to mediator atoms being spatially delocalized in their lattice, both the 
magnitude and the sign of the interaction coefficients $J_{im}^{zz}$, $J_{im}^{\bot}$ can depend on the distance 
between the spins. 

\subsection{Sign varying interactions}

Below we show that the interaction coefficients $J_{im}^{\bot}$, $J_{im}^{zz}$ can change sign depending on the distance between the spins. 
 From Eq.(\ref{eq:Inter-two-spin}) one can see that the coefficients depend 
on the interaction matrix elements, given by the expression:
\begin{eqnarray} 
\lefteqn{V_{{\bf ns},k_{0\;\nu_{0}},\alpha;{\bf np'},k_{\nu},\alpha'}^{mq}=\int dX_{q}\Phi_{ns_{q}}^{*}(X_{q},k_{0\;\nu_{0}\;q})\times } \nonumber \\  
&&\times \bra{{\bf ns}}\bra{\alpha_{m}}\hat{V}(\vec{R}_{qm})\ket{\alpha'_{m}}\ket{{\bf np'}}\phi(X_{q},k_{\nu\;q}).
\label{eq:inter-matr-elements}
\end{eqnarray}
Let us approximate the Bloch functions by plane waves as $\phi\left(X_{q},k_{\nu\;q}\right)=e^{ikX_{q}}/\sqrt{N_{\rm latt\; at}L_{\rm at}}$, 
$\Phi_{ns_{q}}\left(X_{q},k_{0\;\nu_{0}}\right)=e^{ik_{0}X_{q}}/\sqrt{N_{\rm latt\; at}L_{\rm at}}$. The matrix elements then have the form: 
\begin{eqnarray}
\lefteqn{V_{{\bf ns},k_{0\;\nu_{0}},\alpha;{\bf np'},k_{\nu},\alpha'}^{mq}=\frac{1}{N_{\rm latt \; at}L_{\rm at}}e^{i(k-k_{0})X_{m}}\times } \\
&&\times \int dX_{qm}\bra{{\bf ns}}\bra{\alpha_{m}}\hat{V}(\vec{R}_{qm})\ket{\alpha'_{m}}\ket{{\bf np'}}e^{i(k-k_{0})X_{qm}}, \nonumber 
\end{eqnarray}
where we introduced the $x$ coordinate $X_{m}$ of the $m^{\rm th}$ spin and the separation between the $q^{\rm th}$ mediator atom and the $m^{\rm th}$ spin along 
the $x$ axis $X_{qm}=X_{q}-X_{m}$ (see Fig.\ref{fig:setup-scheme}b). The integral in the above expression no longer depends on $X_{m}$ for sufficiently long spin and mediator atom arrays. 
As a result, the matrix element can be written as
\begin{align}
V_{{\bf ns},k_{0\;\nu_{0}},\alpha;{\bf np'},k_{\nu},\alpha'}^{mq}=c_{{\bf ns},k_{0\;\nu_{0}},\alpha;{\bf np'},k_{\nu},\alpha'}^{mq}e^{i(k-k_{0})X_{m}}.
\label{eq:int-matr-elements}
\end{align}
The interaction coefficients will be proportional to a sum over quasimomenta $k$ of the first Brillouin zone of the product of the matrix elements 
corresponding to the interaction of the q$^{\rm th}$ atom with i$^{\rm th}$ and m$^{\rm th}$ spins:
\begin{align}
J_{im}^{\bot(zz) \; q}\sim \sum_{k,\nu}
\frac{V_{{\bf ns},k_{0\;\nu_{0}},\alpha ; {\bf np'},k_{\nu},\alpha'}^{iq}\left(V_{{\bf ns},k_{0\;\nu_{0}},\beta ; {\bf np'},k_{\nu},\beta'}^{mq}\right)^{*}}{{\cal E}_{{\bf np'}}-{\cal E}_{{\bf ns}}+E_{\alpha'}-E_{\alpha}} \nonumber \\
\sim \sum_{\nu}\frac{c_{{\bf ns},k_{0\;\nu_{0}},\alpha ; {\bf np'},k_{\nu},\alpha'}^{iq}\left(c_{{\bf ns},k_{0\;\nu_{0}},\beta ; {\bf np'},k_{\nu},\beta'}^{mq}\right)^{*}}{{\cal E}_{{\bf np'}}-{\cal E}_{{\bf ns}}+E_{\alpha'}-E_{\alpha}}\times \nonumber \\
\times \sum_{k}e^{i(k-k_{0})(X_{i}-X_{m})}, \nonumber 
\end{align}
where it is assumed for simplicity that the $c^{qm}$ terms and the total energies of the mediator states weakly depend on the quasimomentum $k$. For high $\nu$ Bloch 
bands the motional energy will eventually become comparable to the internal energies, but for these bands the overlap integral between the 
interaction potential and Bloch functions in Eq.(\ref{eq:inter-matr-elements}) will already be negligible. The summation over the quasimomenta of the 
first Brillouin zone will give the factor
\begin{align}
\sum_{k}e^{ik(X_{i}-X_{m})}=\frac{\sin\left[\frac{\pi\left(X_{i}-X_{m}\right)\left(1+ N_{\rm latt \; at}\right)}{N_{\rm latt \; at} L_{\rm at}}\right]}{\sin [\frac{\pi\left(X_{i}-X_{m}\right)}{N_{\rm latt \; at}L_{\rm at}}]}=(-1)^{p}, \nonumber 
\end{align}   
for a spin lattice having the same period as the mediator lattice with $X_{i}-X_{m}=pL_{\rm at}$. In addition to varying the sign 
the interaction coefficients also fall off with the distance between the spins. The $c^{qm}$ factors have approximate dependence 
$\sim 1/R_{qm}^{3}$, 
resulting in $J_{im}^{\bot(zz)\;q}\sim 1/(R_{qi}^{3}R_{qm}^{3}) \sim 1/|X_{i}-X_{m}|^{6}$ for distant spins. 

This simplified derivation qualitatively shows that the interaction coefficients can change 
sign with an interspin distance, analogous to the RKKY effect. In the next section two examples of spin encoding in polar molecules and Rydberg atoms 
will be considered and the corresponding interaction coefficients $J_{im}^{\bot(zz)}$ will be numerically calculated.

\section{Modelling XX, Ising, XXZ interactions}

In this section effective spin-spin interactions that can be realized in the bilayer system are discussed by analyzing the interaction 
coefficients $J_{im}^{\bot}$ and $J_{im}^{zz}$, given in Appendix C. Examples of i) XX interaction using 
spin encoding in polar molecule states and ii) XXZ interaction using Rydberg atom spin encoding 
are considered.

\subsection{XX interaction with LiCs effective spins and Rb 
Rydberg mediator atoms}

\begin{figure*}
\center{
\includegraphics[width=17.cm]{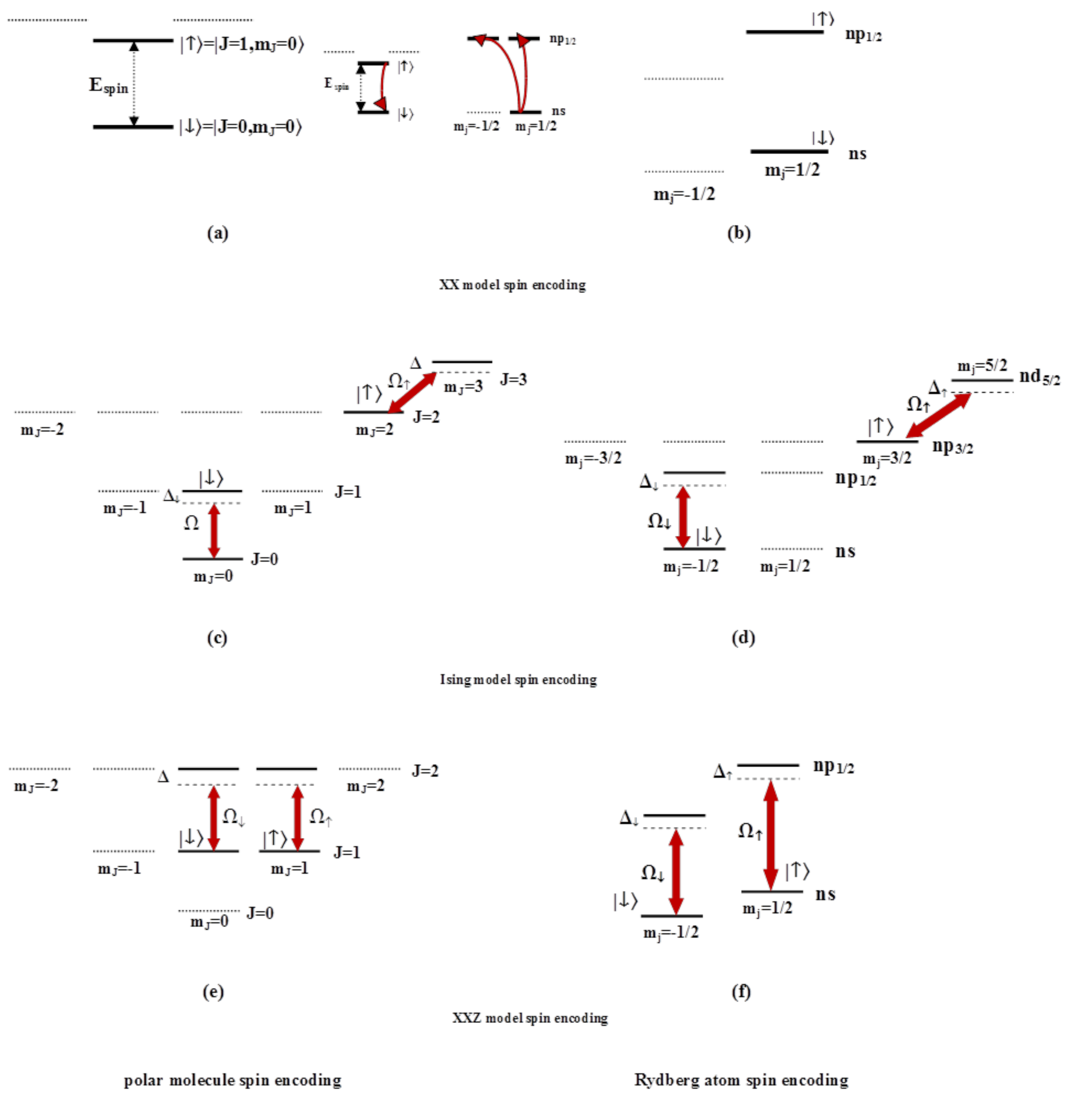}
\caption{\label{fig:level-scheme} {\it Schemes of different types of spin encoding into low-energy rotational states of polar molecules (left side) or 
long-lived states of Rydberg atoms (right side), allowing to model indirect interactions in a 1D effective spin chain}: (a) Schematic illustrating spin encoding into low-energy rotational states of a polar molecule and spin flipping between 
near resonant molecular rotational and mediator Rydberg atomic states, allowing to model XX type of interaction; (b) example of spin encoding into long-lived states of a Rydberg atom; (c) polar 
molecule rotational states not connected by a dipole-allowed transition can be used as spin states to realize effective Ising-type spin-spin interaction. 
The state dipole moments can be induced by near-resonantly coupling the spin states to opposite parity states by MW fields; 
(d) the same as in (c) for spins encoded in Rydberg states; (e) spin encoding in near degenerate rotational sublevels of a polar molecule 
such that $|E_{\rm spin}| \ll |{\cal E}_{np_{j'}}-{\cal E}_{ns}|$ is satisfied allows to realize an effective XXZ interaction;
 (f) the same as (e) for Rydberg atom spin encoding. The Rydberg encoded spin states can be initially additionally split 
by an external magnetic field. }
}
\end{figure*}

The XX interaction $\sum_{i,m=1}^{N}J_{im}^{\bot}\left(\hat{S}_{i}^{+}\hat{S}_{m}^{-}+\hat{S}_{i}^{-}\hat{S}_{m}^{+}\right)/2+\sum_{i=1}^{N}(E_{\rm spin}+b_{i}^{z})\hat{S}_{i}^{z}$ 
can be realized if the spin states have zero dipole moments $\bra{\uparrow}\vec{d}_{\rm spin}\ket{\uparrow}=\bra{\downarrow}\vec{d}_{\rm spin}\ket{\downarrow}=0$, 
non-zero spin transition dipole moment $\bra{\uparrow}\vec{d}_{\rm spin}\ket{\downarrow} \ne 0$, and spin 
$\ket{\uparrow}\leftrightarrow \ket{\downarrow}$ and mediator $\ket{ns}\leftrightarrow \ket{np_{j'}}$ transitions are close in energy such 
that their energy difference $\Delta E=E_{np_{j'}}-E_{ns}\pm E_{\rm spin}$ is $|\Delta E|\ll E_{\rm spin},|E_{np_{j'}}-E_{ns}|$ (but $|\Delta E|\gg |\hat{V}|$) 
(see Fig.\ref{fig:level-scheme}a,b). In 
this case from Eqs.(\ref{eq:Jzz})-(\ref{eq:biz}) one can see that $|J_{im}^{\bot\;q,k_{\nu}}|,|b_{i}^{z\;q,k_{\nu}}|\ne 0$, $|J_{im}^{zz\;q,k_{\nu}}|=0$, 
i.e. only the spin flipping terms $J_{im}^{\bot}\hat{S}_{i}^{\pm}\hat{S}_{m}^{\mp}$ will be present in the effective interaction.

In the polar molecules setup the spin-exchange interaction
between a polar molecule and a mediator Rydberg atom can be realized if a rotational molecular transition is close in energy to a Rydberg transition 
(atom-molecule Forster resonance). The Forster resonances between rotational states of 
a polar molecule and atomic Rydberg states have been studied recently in \cite{Pol-mol-Rydb-Forster-experim}, where 
the resonant exchange between a NH$_{3}$ molecule and a He atom has been experimentally observed. 
For effective spins, encoded in Rydberg states, atom-atom Forster resonances can be 
 used to realize the spin-mediator interaction. In this case e.g. different atomic species can be used to 
encode the spin and mediate the 
interaction, such as Rb and Cs. For this atomic pair there are several interspecies Forster resonances available such as 
$\ket{{\rm Rb} 59s_{1/2},{\rm Cs} 57s_{1/2}}\leftrightarrow \ket{{\rm Rb} 58p_{1/2},{\rm Cs} 57p_{1/2}}$ with an energy defect $\Delta E=-16.6$ MHz, 
$\ket{{\rm Rb} 81s_{1/2},{\rm Cs} 78s_{1/2}}\leftrightarrow \ket{{\rm Rb} 80p_{1/2},{\rm Cs} 78p_{1/2}}$ with $\Delta E=6.31$ MHz, 
$\ket{{\rm Rb} 82s_{1/2},{\rm Cs} 79s_{1/2}}\leftrightarrow \ket{{\rm Rb} 81p_{1/2},{\rm Cs} 79p_{1/2}}$ with $\Delta E=-6.41$ MHz, 
$\ket{{\rm Rb} 84s_{1/2},{\rm Cs} 89s_{1/2}}\leftrightarrow \ket{{\rm Rb} 84p_{1/2},{\rm Cs} 88p_{1/2}}$ with $\Delta E=-2.43$ MHz \cite{Ilya-Saffman-Forster}. In fact, recently 
signs of indirect interaction between two Rydberg atoms, mediated by a third one, have been observed in \cite{Three-Rydb-indirect}. An advantage of using 
different species for spin encoding and mediating the interaction is that they can be spectrally addressed using laser fields of different frequencies, allowing 
to separately initialize, control and read out their states. 
Finally, a spin 
can be encoded in two ground state sublevels of neutral atoms, which can be coupled to two Rydberg states such as $\ket{ns}$ and $\ket{np_{j}}$ to 
form Rydberg dressed states. In this case two dressed atoms encoding spins can indirectly interact via spin-exchange with a mediator Rydberg atom.

As an example of how indirect XX interaction can be modelled we consider a system of spin-encoding polar molecules and mediator Rydberg atoms in a superatom state 
Eq.(\ref{eq:Psi-supat}). 
In the case interactions between a Rydberg atom and ground state atoms are present in the superatom state \cite{Rydberg-ground-interactions}, a 
single mediator atom can be used.  
Each molecule is assumed 
to be in the ground electronic and vibrational state; two low-energy rotational states 
are used to encode the spin states, e.g. $\ket{\uparrow}=\ket{J=1,m_{J}=0}$, 
$\ket{\downarrow}=\ket{J=0,m_{J}=0}$. 
We consider the case 
shown in Fig.\ref{fig:level-scheme}a,  
where the spin $\ket{\uparrow} \leftrightarrow \ket{\downarrow}$ and mediator $\ket{ns} \leftrightarrow \ket{np_{j'},m_{j'}'}$ transitions are 
close in energy such that the states $\ket{ns,m_{j}}\ket{\uparrow}$ and $\ket{np_{j'},m_{j'}'}\ket{\downarrow}$ are almost degenerate and 
the spin-mediator system can 
coherently oscillate between these states.  In this case 
$|\Delta E=E_{np_{j'}}-E_{ns}-E_{\rm spin}| \ll E_{\rm spin},|E_{np_{j'}}-E_{ns}|$  and the atomic and molecular 
basis sets can be limited to only the $\ket{ns,m_{j}}$, $\ket{np_{j'},m_{j'}'}$ and $\ket{\downarrow}$, $\ket{\uparrow}$ states. 
 The coefficients $J_{im}^{\bot \;q,k_{0\;\nu_{0}}}$ and $b_{i}^{z \;q,k_{0\;\nu_{0}}}$ will have the form (see Eqs.(\ref{eq:Jpm}),(\ref{eq:biz})):
\begin{align}
J_{im}^{\bot\;q,k_{0\;\nu_{0}}}\approx & -\sum_{\substack{ {\bf np'} \\ k,\nu }} \frac{2V_{{\bf ns},k_{0\;\nu_{0}},\uparrow;{\bf np'},k_{\nu},\downarrow}^{iq}\left(V_{{\bf ns},k_{0\;\nu_{0}},\uparrow;{\bf np'},k_{\nu},\downarrow}^{mq}\right)^{*}}{{\cal E}_{{\bf np'}}(k_{\nu})-{\cal E}_{{\bf ns}}(k_{0\;\nu_{0}})-E_{\rm spin}}, \nonumber \\
b_{i}^{z\;q,k_{0\;\nu_{0}}}=& -\sum_{\substack{ m \ne i \\ {\bf np'} \\ k,\nu } }\frac{2\left|V_{{\bf ns},k_{0\;\nu_{0}},\uparrow;{\bf np'},k_{\nu},\downarrow}^{iq}\right|^{2}}{{\cal E}_{{\bf np'}}(k_{\nu})-{\cal E}_{{\bf ns}}(k_{0\;\nu_{0}})-E_{\rm spin}}
\label{eq:inter-coeff-XY}
\end{align} 
where ${\bf np'}=\{np,j',m_{j}'\}$ and summation is over $j'$, $m_{j}'$ quantum numbers; and $J_{im}^{zz \;q,k_{0\;\nu_{0}}}=0$ since spin states are assumed to have zero dipole moments.

As a concrete example we consider 
a 1D array of spin-encoding LiCs molecules and mediator Rb atoms in a superatom $65s_{1/2}$ Rydberg state (or a single mediator atom in the Rydberg state), 
placed in two parallel 1D optical lattices (see Fig.\ref{fig:setup-scheme}a). 
LiCs has the permanent dipole moment $d=5.39$ D \cite{LiCs-dip-mom-rot-const} and the rotational constant $B=0.218$ cm$^{-1}=6.535$ GHz 
\cite{LiCs-dip-mom-rot-const}, and is actively studied towards formation of ground state ultracold molecules \cite{LiCs-ground-state}.
The spin transition with the frequency $E_{\rm spin}/\hbar=2B\approx 13.071$ GHz is nearly resonant with the transition $65p_{1/2}-65s_{1/2}$ of Rb atom, having the 
frequency $(E_{65p_{1/2}}-E_{65s_{1/2}})/\hbar\approx 13.082$ GHz,  
where $E_{nlj}=-\frac{1}{2(n-\mu_{lj})^{2}}$ (in a.u.) and 
$\mu_{s}=3.1311804$, $\mu_{p_{1/2}}=2.6548849$ \cite{Rb-quantum-defects}. 
This gives the frequency defect of the mediator and spin transitions $(\Delta E=E_{65p_{1/2}}-E_{65s_{1/2}}-E_{\rm spin})/\hbar\approx 11$ MHz. 
At the same time the frequency defect between
the $65p_{3/2} - 65s_{1/2}$ and the spin transition is $\approx 370$ MHz, and spin flips involving this transition can be neglected.  
Other examples of near resonant molecular rotational $J=1 \leftrightarrow J=0$ and Rydberg transitions are listed in Table {\ref{table:atom-mol-Forster-resonances}}.

\begin{table*}
\caption{Examples of Forster resonances between $J=1 \leftrightarrow J=0$ rotational transitions of alkali dimer polar molecules and Rydberg transitions of alkali atoms}
\begin{tabular}{c c c c}
\hline
{\rm Species} & $B$,\;cm$^{-1}$ & $nl'_{j'}-nl_{j}$ & $(\Delta E=E_{nl'_{j'}}-E_{nl_{j}}-E_{\rm spin})/\hbar$,\; {\rm MHz} \\
              &  ($E_{\rm spin}/\hbar=2B$,\;{\rm GHz}) &     &                                                       \\  [0.5ex]
\hline
{\rm LiCs+Na} & $0.218$ & $64p_{3/2(1/2)}-64s_{1/2}$ & $-26.4\;(-42.7)$ \\ 
               & $(13.071)$       &                            &        \\ 
{\rm LiRb+Rb} &  $0.254$  & $62p_{1/2}-62s_{1/2}$ & $-52$     \\
              &  ($15.229$) &                     &           \\    
{\rm LiNa+Rb} &  $0.425$ &  $53p_{3/2}-53s_{1/2}$  &  $111$   \\
              &  ($25.482$) &                      &          \\  
{\rm LiK+Rb}  &  $0.293$  & $59p_{1/2}-59s_{1/2}$  & $176$    \\
              &  ($17.568$) &                      &          \\ [1ex]
\end{tabular}
\label{table:atom-mol-Forster-resonances}
\end{table*}

\begin{figure*}
\center{
\includegraphics[width=15.cm]{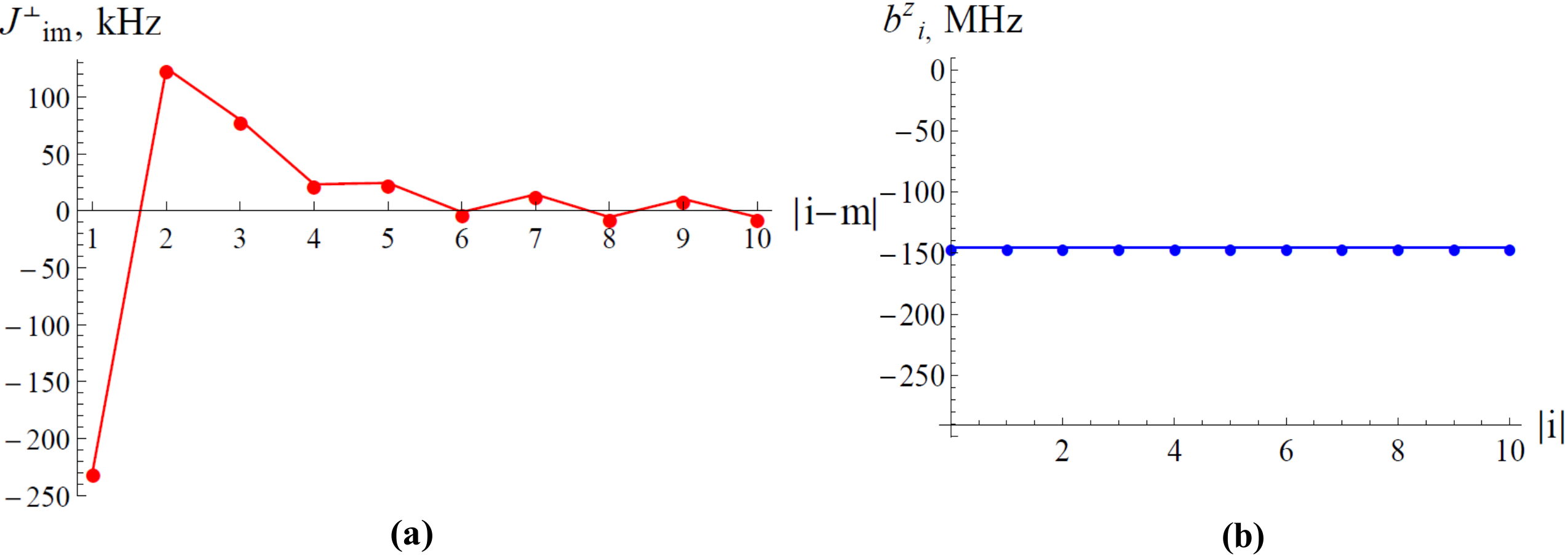}
\caption{\label{fig:J-coupl-LiCs} {\it Spin encoding in low-energy rotational states of LiCs molecules near resonantly interacting with a Rydberg transition of Rb 
allows 
to model XX interaction}. Numerically calculated from Eqs.(\ref{eq:inter-coeff-XY}): (a) Interaction coefficient $J_{im}^{\bot}$ and (b) effective magnetic field $b_{i}^{z}$ 
 for spins encoded in rotational states of LiCs $\ket{\downarrow}=\ket{J=0,m_{J}=0}$, $\ket{\uparrow}=\ket{J=1,m_{J}=0}$, near resonantly 
interacting with Rb mediator atoms at the transition $\ket{65p_{1/2},m_{j}=\pm 1/2}-\ket{65s_{1/2},m_{j}=1/2}$. The 
mediator atoms are assumed to be initially prepared in the superatom state Eq.(\ref{eq:Psi-supat}) in the $\ket{65s_{1/2},m_{j}=1/2}$ internal and in a BEC motional state with $k_{0}=0$, $\nu_{0}=1$. In the 
coefficients Eqs.(\ref{eq:inter-coeff-XY}) summation over quasimomenta in the 
first Brillouin zone for $\nu=1,...,5$ lowest Bloch bands of the mediator lattice with the depth $V_{0}=-E_{\rm rec}$ was performed. Setup dimensions: $\rho=500$ nm, 
 spin and mediator lattice periods $L_{\rm spin}=L_{\rm at}=500$ nm, $N=N_{\rm at}=100$. The calculations were done for a spin in the center of the array $m=0$.}
}
\end{figure*}

\begin{figure}
\center{
\includegraphics[width=9.cm]{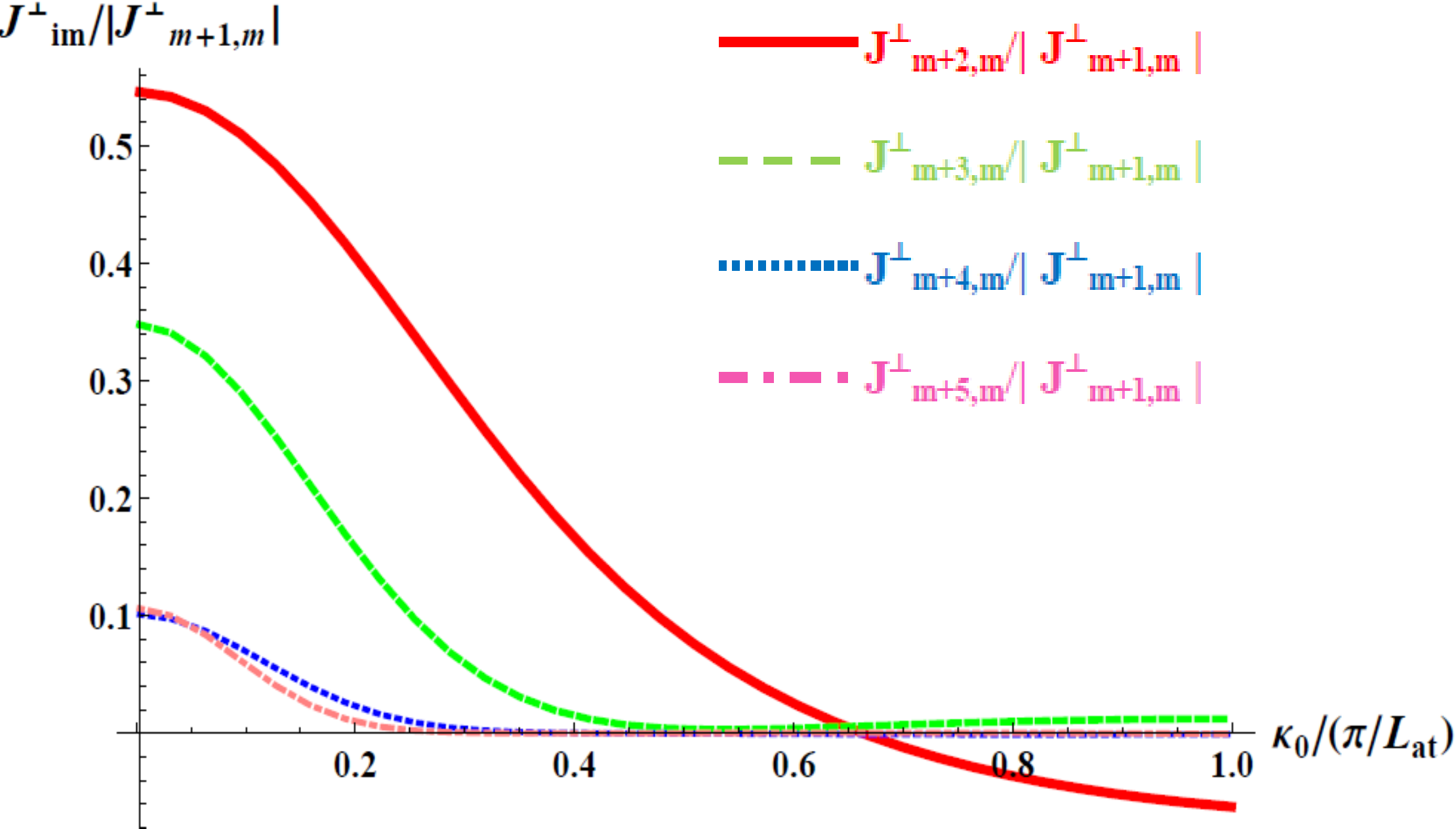}
\caption{\label{fig:JperpimJperp1-XX} {\it In the XX model with interaction coefficients Eqs.(\ref{eq:averaged-J-supat}),(\ref{eq:inter-coeff-XY}) 
the ratios of the strengths of next-nearest and more distant to the nearest neighbor interaction can be controlled by  the width of the distribution of 
the quasimomenta of the initial superposition of Bloch states of mediator atoms}. Assuming the Gaussian initial distribution of the quasimomenta in 
the lowest Bloch band $|c_{k_{0\;\nu_{0}=1}}|^{2} \sim e^{-k_{0}^{2}/\kappa_{0}^{2}}$, the ratios of interaction coefficients $J^{\bot}_{im}/|J^{\bot}_{m+1,m}|$ 
can be controlled by the distribution width $\kappa_{0}$: $J^{\bot}_{m+2,m}/|J^{\bot}_{m+1,m}|$ (red solid curve), $J^{\bot}_{m+3,m}/|J^{\bot}_{m+1,m}|$ 
(green dashed curve), $J^{\bot}_{m+4,m}/|J^{\bot}_{m+1,m}|$ (blue dotted curve), $J^{\bot}_{m+5,m}/|J^{\bot}_{m+1,m}|$ (pink dashed-dotted curve). 
The interaction coefficients shown in Fig.\ref{fig:J-coupl-LiCs} correspond to $\kappa_{0}=0$.}
}
\end{figure} 

Details of the calculations of the matrix elements of the spin-mediator charge-dipole interaction and the coefficients Eqs.(\ref{eq:inter-coeff-XY}) are given in Appendix D. 
Fig.\ref{fig:J-coupl-LiCs} shows the numerically calculated 
$J_{im}^{\bot}=\sum_{q=1}^{N_{a}}J_{im}^{\bot \;q,k_{0}=0_{\nu_{0}=1}}/N_{a}$  
and $b_{i}^{z}=\sum_{q=1}^{N_{a}}b_{i}^{z \;q,k_{0}=0_{\nu_{0}=1}}/N_{a}$ coefficients for an $m^{\rm th}$ spin at the center ($m=0$) of the array interacting with an $i^{\rm th}$ spin 
depending on their spatial separation assuming that initially the mediator atoms are prepared in a BEC state $k_{0}=0$, $\nu_{0}=1$ and in the $\ket{65s_{1/2},m_{j}=1/2}$ 
internal state. The following parameters were used in the calculations: the spin-mediator arrays distance $\rho=500$ nm; 
the spin and mediator lattices periods $L_{\rm spin}=L_{\rm at}=500$ nm; number of spins and mediator atoms $N=N_{a}=100$; 
the mediator atoms lattice depth $V_{0}=-E_{\rm rec}$, where $E_{\rm rec}=\hbar^{2}\left(\pi/L_{\rm at}\right)^{2}/2M_{\rm at}$ is the recoil energy of mediator 
atoms; and $\nu=1,...,5$ Bloch bands 
of the mediator lattice were taken into account. One can see that i) the $J_{im}^{\bot}$ changes sign with an interspin distance. The nearest neighbor interaction 
is ferro- and the next nearest neighbor one is antiferromagnetic; ii) the interaction extends beyond nearest neighbors and falls off at about 
$|i-m|\sim 5$; iii)  the interaction strengths $\sim$ hundreds kHz can be realized;  iv) 
$b_{i}^{z}$ do not depend on $i$ and only change the initial spin transition frequency. Two inmmediate consequencies follow: first, the XX interaction conserves the z component of the total spin 
$\hat{S}^{z}=\sum_{i=1}^{N}\hat{S}_{i}^{z}/N$, so for homogeneous $b_{i}^{z}$ the term $\sum_{i=1}^{N}(E_{\rm spin}+b_{i}^{z})\hat{S}_{i}^{z}$ gives a constant energy offset which can be neglected. 
Second, the effective spin-spin interaction strength $\sim$ hundreds kHz is an order of magnitude larger than the strength of the direct dipole-dipole 
interaction between LiCs molecules $V_{dd}\sim d_{\rm spin\;z}^{2}/L_{\rm spin}^{3} \sim 12$ kHz, and is limited by the requirement that the spin-encoding molecules are not excited to higher-energy 
Bloch bands by the effective interaction, i.e. it should be smaller than the spin lattice trapping frequency. This requirement could be 
lifted and even larger interaction strengths could be achieved if 
molecular motion could be cooled, but motional cooling of molecules in an optical lattice currently presents a challenge. 

From Fig.\ref{fig:J-coupl-LiCs}a one can see that if the mediator atoms are initially prepared in a BEC state the interaction is significant between spins separated by up to five lattice sites.  
The contribution of 
the next-nearest and more distant neighbors can be controlled by preparing the initial motional state of the mediator atoms as a superposition of Bloch states Eqs.(\ref{eq:Psi-supat}),
(\ref{eq:averaged-J-supat}). Assuming a Gaussian distribution of quasimomenta of initial Bloch states in the lowest Bloch band with $|c_{k_{0\;\nu_{0}=1}}|^{2} \sim e^{-k_{0}^{2}/\kappa_{0}^{2}}$, 
the ratio of the next-nearest and more distant to the nearest neighbor interaction coefficients can be controlled by the quasimomenta distribution width $\kappa_{0}$, 
as shown in Fig.\ref{fig:JperpimJperp1-XX}. In particular, for $\kappa_{0}/(\pi/L_{\rm at})\approx 0.65$ the $J^{\bot}_{im} \ll J^{\bot}_{m+1,m}$ for 
$2 \le|i-m|\le 5$, resulting in the exactly solvable XX model with only nearest neighbor interactions \cite{XX-model-exact}. In the range $0.3 \lesssim \kappa_{0}/(\pi/L_{\rm at}) \lesssim 0.65$ 
the ratio $J^{\bot}_{m+3,m}/J^{\bot}_{m+2,m} \lesssim 0.2$ 
and the interaction can be approximated as the $J_{1}-J_{2}$ XX model 
$\sum_{l=1,2}J_{l}\left(\hat{S}^{x}_{i}\hat{S}^{x}_{i+l}+\hat{S}^{y}_{i}\hat{S}^{y}_{i+l}\right)$. Its phase diagram for ferromagnetic nearest neighbour 
$J_{1}=J^{\bot}_{m+1,m}<0$ and antiferromagnetic next-nearest neighbor $J_{2}=J^{\bot}_{m+2,m}>0$ interactions was analyzed in \cite{J1-J2-XXZ-phase-diagram}, 
where it was found that in the range $J_{1}/J_{2} \lesssim -3.1$ (corresponding to $J_{2}/|J_{1}|\le 0.32$) realized in our case, the system is in a spin liquid phase.  

The mediator atoms have to stay in the $ns$ state long enough for the indirect interaction to take place, i.e. the effective interaction strength should be 
larger than the mediator Rydberg state decay rate. For Rb $65s_{1/2}$ state the lifetimes, including contributions from spontaneous emission and interaction with 
black-body radiation, are $\tau_{65s_{1/2}}=325.03$ $\mu$s at $T=4.2$ K and $\tau_{65s_{1/2}}=126.32$ $\mu$s at $T=300$ K \cite{Rb-lifetimes}. The $J_{im}^{\bot}\sim 100$ kHz corresponds 
to interactions times $\sim 1$ $\mu$s, which are two orders of magnitude shorter than the mediator Rydberg state decay times, making the interaction observable.

Finally, we note that in order to selectively address the spin transition the degeneracy between the $\ket{\uparrow}=\ket{J=1,m_{J}=0}$ state and 
the $\ket{J=1,m_{J}=\pm 1}$ rotational states should be lifted such that their energy difference exceeds the effective interaction strength. The degeneracy 
can be lifted by a DC electric field, but in this case the field required to induce energy shifts $(d_{\rm spin}E)^{2}/E_{\rm spin} \sim$ hundreds kHz is 
of the order of $E\sim 10^{5}$ V/cm. The fields of such strength will induce shifts of the mediator Rydberg states $\sim 100$ GHz and their ionization. 
Another way to lift the degeneracy of the rotational states is by MW fields, coupling the $\ket{J=1,m_{J}=0,\pm 1}$ to $\ket{J=2,m_{J}=0,\pm 1,\pm 2}$ states. For example,
for a $\sigma^{+}$ polarized MW field the ratio of the dipole moments for the transitions $\ket{1,-1}\leftrightarrow \ket{2,0}$, $\ket{1,0} \leftrightarrow \ket{2,1}$ and 
$\ket{1,1} \leftrightarrow \ket{2,2}$ is $|d_{1,-1;2,0}|/|d_{1,0;2,1}|/|d_{1,1;2,2}|=(1/\sqrt{6})/(\sqrt{2})/(2)$ and the states will be shifted differently by the 
MW field. For a MW
field with detuning ${\tilde \Delta} \sim 100$ MHz and Rabi frequency ${\tilde \Omega}\sim 10$ MHz shifts $|\tilde \Omega|^{2}/{\tilde \Delta}\sim 1$ MHz will 
be induced, exceeding the effective interaction strength by an order of magnitude. There is a subtle point, however, that the coupling of $\ket{\uparrow}$ 
to the $\ket{J=2,m_{J}=1}$ state will induce the state dipole moment 
$\bra{\uparrow}\vec{d}_{\rm spin}\ket{\uparrow}\sim d_{\rm spin}{\tilde \Omega}/{\tilde \Delta}$. In turn, the non-zero dipole moment of the $\ket{\uparrow}$ 
states will give rise to non-zero matrix elements of the charge-dipole interaction $V^{mq}_{{\bf ns},k_{0\;\nu_{0}},\uparrow;{\bf ns},k_{\nu},\uparrow}$, which in 
turn give rise to terms $\sim V^{iq}_{{\bf ns},k_{0\;\nu_{0}},\uparrow;{\bf ns},k_{\nu},\uparrow}\left(V^{mq}_{{\bf ns},k_{0\;\nu_{0}},\uparrow;{\bf ns},k_{\nu},\uparrow}\right)^{*}/({\cal E}_{{\bf ns}}(k_{\nu})-{\cal E}_{{\bf ns}}(k_{0\;\nu_{0}}))$ and 
$\sim \left|V^{iq}_{{\bf ns},k_{0\;\nu_{0}},\uparrow;{\bf ns},k_{\nu},\uparrow}\right|^{2}/({\cal E}_{{\bf ns}}(k_{\nu})-{\cal E}_{{\bf ns}}(k_{0\;\nu_{0}}))$ in $J_{im}^{zz \;q,k_{0\;\nu_{0}}}$ and $b_{i}^{z\;q,k_{0\;\nu_{0}}}$, 
respectively, where ${\cal E}_{{\bf ns}}(k_{\nu})-{\cal E}_{{\bf ns}}(k_{0}=0_{\nu_{0}=1}) \sim \hbar^{2}(2\pi)^{2}/2m_{\rm Rb}(N_{\rm at\;latt}L_{\rm at})^{2}=4E_{\rm rec}/N_{\rm at\;latt}^{2} \approx 0.87$ Hz 
for a Rb mediator atom.  It shows that 1) the $V^{iq}_{{\bf ns},k_{0\;\nu_{0}},\uparrow;{\bf ns},k_{\nu},\uparrow}$ matrix elements should be much smaller than the latter energy 
difference for the Schrieffer-Wolff expansion to be valid and 2) the resulting $J_{im}^{zz\;q,k_{0\;\nu_{0}}}$ and $b_{i}^{z\;q,k_{0\;\nu_{0}}}$ terms should 
be much smaller than $J_{im}^{\bot \;q,k_{0\;\nu_{0}}}$ 
in order to be neglected. Both of these requirements are indeed satisfied due to small values of the $V^{iq}_{{\bf ns},k_{0\;\nu_{0}},\uparrow;{\bf ns},k_{\nu},\uparrow}$ matrix 
elements, as discussed in Appendix D, end of part A.

\subsection{Ising interaction}

Ising interaction can be realized in the bilayer system if the spin states $\ket{\uparrow}$, $\ket{\downarrow}$ are not coupled by a dipole-allowed transition, 
i.e. $\bra{\uparrow}\vec{d}_{\rm spin}\ket{\downarrow}=0$, but at the same time have non-zero dipole moments 
$\bra{\uparrow}\vec{d}_{\rm spin}\ket{\uparrow} \ne 0$, $\bra{\downarrow}\vec{d}_{\rm spin}\ket{\downarrow} \ne 0$ 
(see Fig.\ref{fig:level-scheme}c,d). 
One can see from Eqs.(\ref{eq:Jzz})-(\ref{eq:biz}) that in this case $J_{im}^{\bot \;q,k_{\nu}}=0$ and $J_{im}^{zz \;q,k_{\nu}} \ne 0$, $b_{i}^{z\;q,k_{\nu}}\ne 0$, which describes the Ising interaction in the presence of 
a longitudinal magnetic field. Fig.\ref{fig:level-scheme}c shows an example of spin encoding in the polar molecule case, 
where $\ket{\downarrow}=\ket{J=1,m_{J}=0}$, $\ket{\uparrow}=\ket{J=2,m_{J}=2}$ spin states are 
not coupled by a dipole-allowed transition. The dipole moments in the 
states $\ket{\uparrow}$, $\ket{\downarrow}$ can be induced by MW fields, e.g. by coupling
the $\ket{\downarrow}=\ket{J=1,m_{J}=0}$ to $\ket{J=0,m_{J}=0}$ and $\ket{\uparrow}=\ket{J=2,m_{J}=2}$  
to $\ket{J=3,m_{J}=3}$ state by near-resonant MW fields (Fig.2c). 
An additional advantage of the 
MW dressing is that it shifts the energies of the dressed states with respect to near energy states, allowing for the 
$\ket{\uparrow} \leftrightarrow \ket{\downarrow}$ transition to be addressed spectroscopically. In the Rydberg atom encoding case the MW dressing can also induce non-zero spin state dipole moments, 
which can be done in the way, shown in 
Fig.\ref{fig:level-scheme}d.

\subsection{XXZ interaction with Rb (${\tilde n}=50$) effective spins and Rb ($n=100$) mediator atoms}

\begin{figure}
\center{
\includegraphics[width=9.5cm]{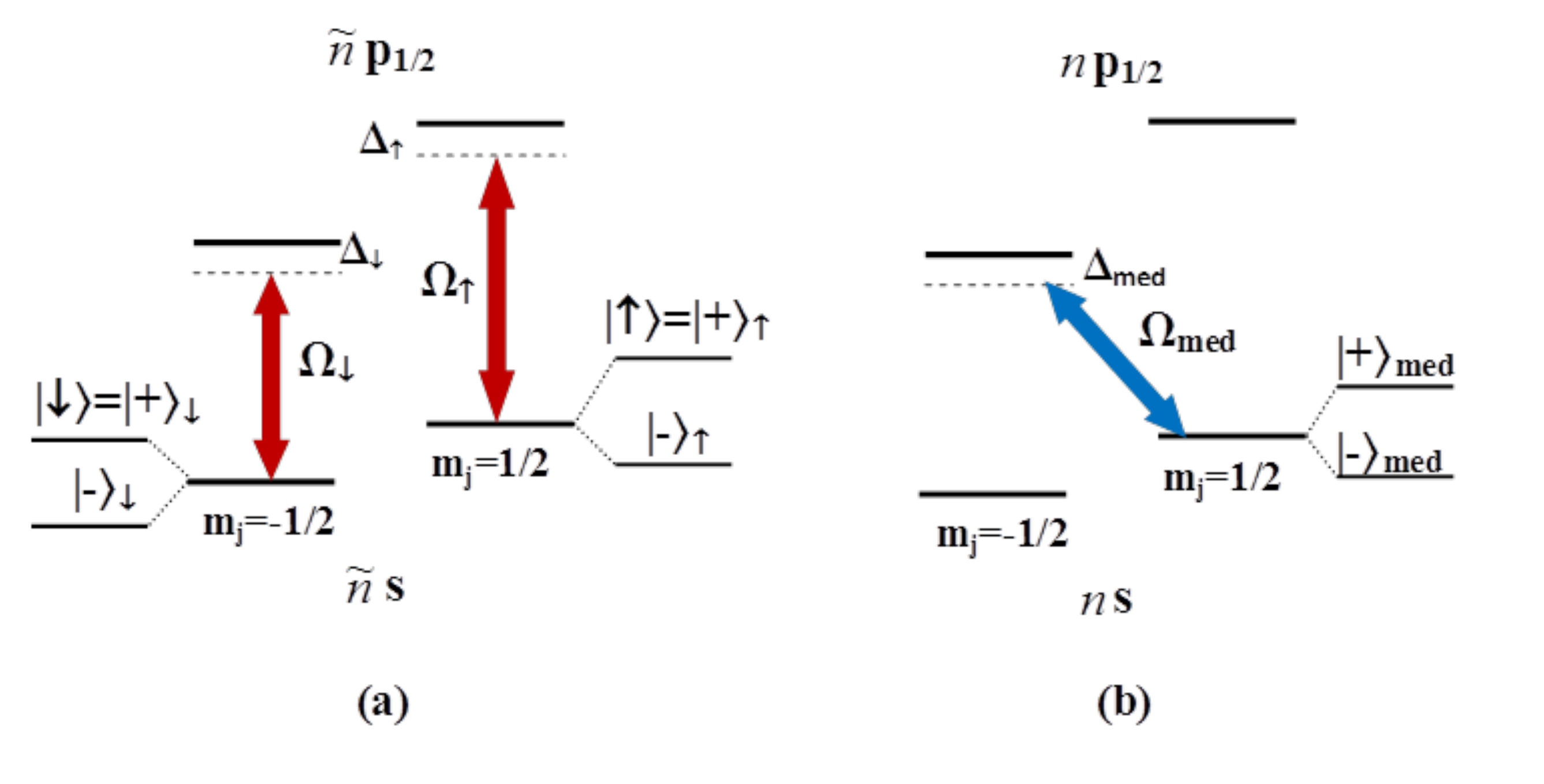}
\caption{\label{fig:level-scheme-XXZ} {\it Spin encoding in MW dressed states of a Rydberg atom interacting with a MW dressed mediator Rydberg atom allows 
to realize the XXZ interaction.} (a) Spin-1/2 system can be encoded in long-lived $\ket{{\tilde n}s_{1/2},m_{j}=\pm 1/2}$ 
Rydberg states dressed by MW fields near resonantly coupling them to $\ket{{\tilde n}p_{1/2},m_{j}=\pm 1/2}$ states. The sublevels 
of the ${\tilde n}p_{1/2}$ and ${\tilde n}s_{1/2}$ states can be additionally split by a DC magnetic field to make the spin transition frequency 
different from frequencies between other dressed states; (b) The $J_{im}^{\bot}$ and $J_{im}^{zz}$ interaction coefficients can be made
$\sim$ tens kHz by using mediator virtual transitions between the initial $\ket{ns_{1/2},m_{j}=-1/2}$ and dressed by a MW field states $\ket{\pm}_{\rm med}$, 
which can be separated by $\lesssim$ hundred MHz. }
}
\end{figure}

There is a growing number of theoretical proposals and experimental demonstrations on simulation of many-body interacting systems using 
Rydberg atoms \cite{QSim-magnetism-Rydbergs}. 
Below we discuss how the XXZ interaction can be realized in the case of effective spins encoded into long-lived states of 
Rydberg atoms, which interact indirectly via mediator Rydberg atoms.
The XXZ interaction, in which both $J_{im}^{\bot}$ and $J_{im}^{zz}$ are non-zero, can be realized if both transitional $\bra{\uparrow}\vec{d}_{\rm spin}\ket{\downarrow}$ and state 
$\bra{\uparrow}\vec{d}_{\rm spin}\ket{\uparrow}$ and/or $\bra{\downarrow}\vec{d}_{\rm spin}\ket{\downarrow}$ dipole moments are non-zero and of 
comparable strength and the 
spin transition frequency is smaller or comparable to the mediator transition frequencies $E_{\rm spin} \lesssim |{\cal E}_{np_{j'}}-{\cal E}_{ns}|$. 
It can be seen then from Eqs.(\ref{eq:Jzz})-(\ref{eq:biz}) that in this case $J_{im}^{\bot \;q,k_{\nu}} \sim J_{im}^{zz\;q,k_{\nu}}$. These requirements 
can be met by choosing the 
spin states to be nearly degenerate such as e.g. 
$\ket{\uparrow}=\ket{J=1,m_{J}=1}$, $\ket{\downarrow}=\ket{J=1,m_{J}=0}$ in the polar molecule case and $\ket{\uparrow}=\ket{ns_{1/2},m_{j}=1/2}$, 
$\ket{\downarrow}=\ket{ns_{1/2},m_{j}=-1/2}$ 
in the Rydberg atom case, where the latter can be additionally split by a DC magnetic field (see Fig.\ref{fig:level-scheme}e,f). The spin states can 
acquire dipole moments if dressed 
with MW fields, nearly resonantly coupling them to states of opposite parity, e.g. in the way, shown in Fig.\ref{fig:level-scheme-XXZ}a:
\begin{eqnarray}
\ket{\pm}_{\uparrow}=a^{\pm}_{\uparrow}\ket{{\tilde n}p_{1/2},m_{j}=\frac{1}{2}}+b^{\pm}_{\uparrow}\ket{{\tilde n}s_{1/2},m_{j}=\frac{1}{2}}, \nonumber \\
\ket{\pm}_{\downarrow}=a^{\pm}_{\downarrow}\ket{{\tilde n}p_{1/2},m_{j}=-\frac{1}{2}}+b^{\pm}_{\downarrow}\ket{{\tilde n}s_{1/2},m_{j}=-\frac{1}{2}}, \nonumber
\end{eqnarray}
with 
\begin{eqnarray}
a^{\pm}_{\uparrow(\downarrow)}=\frac{\sqrt{\sqrt{\left(\frac{\Delta_{\uparrow(\downarrow)}}{2}\right)^{2}+\Omega_{\uparrow(\downarrow)}^{2}}\pm \frac{\Delta_{\uparrow(\downarrow)}}{2}}}{\sqrt{2}\left(\left(\frac{\Delta_{\uparrow(\downarrow)}}{2}\right)^{2}+\Omega_{\uparrow(\downarrow)}^{2}\right)^{1/4}}, \nonumber \\
b^{\pm}_{\uparrow(\downarrow)}=\pm \frac{\sqrt{\sqrt{\left(\frac{\Delta_{\uparrow(\downarrow)}}{2}\right)^{2}+\Omega_{\uparrow(\downarrow)}^{2}}\mp \frac{\Delta_{\uparrow(\downarrow)}}{2}}}{\sqrt{2}\left(\left(\frac{\Delta_{\uparrow(\downarrow)}}{2}\right)^{2}+\Omega_{\uparrow(\downarrow)}^{2}\right)^{1/4}}. \nonumber 
\end{eqnarray}
where $\Omega_{\uparrow(\downarrow)}$ and $\Delta_{\uparrow(\downarrow)}$ are the dressing fields Rabi frequencies and detunings. 
The $\ket{+}_{\uparrow}$ and $\ket{+}_{\downarrow}$ or $\ket{-}_{\uparrow}$ and $\ket{-}_{\downarrow}$ dressed states can be chosen as spin states.
This encoding makes both the spin transition and spin states dipole moments to be non-zero and, by tuning the $a^{\pm}_{\uparrow,\downarrow}$, 
$b^{\pm}_{\uparrow,\downarrow}$ coefficients, of comparable size: $\bra{\uparrow}\vec{d}_{\rm spin}\ket{\uparrow}=-2a^{\pm}_{\uparrow}b^{\pm}_{\uparrow}\vec{e}_{z}d_{{\tilde n}p,{\tilde n}s}/3$, 
$\bra{\downarrow}\vec{d}_{\rm spin}\ket{\downarrow}=2a^{\pm}_{\downarrow}b^{\pm}_{\downarrow}d_{{\tilde n}p,{\tilde n}s}\vec{e}_{z}/3$ and 
$\bra{\uparrow}\vec{d}_{\rm spin}\ket{\downarrow}=-(a^{\pm}_{\uparrow}b^{\pm}_{\downarrow}+b^{\pm}_{\uparrow}a^{\pm}_{\downarrow})(\vec{e}_{x}-i\vec{e}_{y})d_{{\tilde n}p,{\tilde n}s}/3$. 
Here $d_{{\tilde n}p,{\tilde n}s}=e\int_{0}^{\infty}R_{{\tilde n}p}(r)R_{{\tilde n}s}(r)r^{3}dr$ is the radial part of the dipole moment between the ${\tilde n}p$ 
and ${\tilde n}s$ states.
We also assume that the ${\tilde n}s_{1/2},m_{j}=\pm 1/2$ states are split by a DC magnetic field such 
that their splitting is much larger 
than the energy differences between the $\ket{\pm}_{\uparrow}$, $\ket{\pm}_{\downarrow}$ dressed states. 
Additionally, the effective interaction 
strength should exceed the decoherence rate of the system, given mainly by the lifetimes of the Rydberg states. 
The interaction can be made $\sim$ tens kHz and faster than the decay, 
if the mediator atom transition frequency 
$(|{\cal E}_{n'p_{j'}}-{\cal E}_{ns}| \sim E_{\rm spin})/\hbar\lesssim$ hundreds MHz. These requirements can be met by using
dressed mediator states instead of bare ones in the way shown in Fig.\ref{fig:level-scheme-XXZ}b. One can initially prepare the mediator atom in the $\ket{{\tilde n}s_{1/2},m_{j}=-1/2}$ state and 
then the closest in energy virtual states, to which it
can be transferred from the initial state by the dipole-dipole interaction with a spin will be
\begin{eqnarray}
\ket{+}_{\rm med}=c_{+}\ket{ns_{1/2},m_{j}=\frac{1}{2}}+d_{+}\ket{np_{1/2},m_{j}=-\frac{1}{2}}, \nonumber \\
\ket{-}_{\rm med}=c_{-}\ket{ns_{1/2},m_{j}=\frac{1}{2}}+d_{-}\ket{np_{1/2},m_{j}=-\frac{1}{2}}. \nonumber
\end{eqnarray}
Other possible virtually excited mediator states are separated by $\sim 3.5$ GHz and can be therefore neglected.

\begin{figure}
\center{
\includegraphics[width=9.cm]{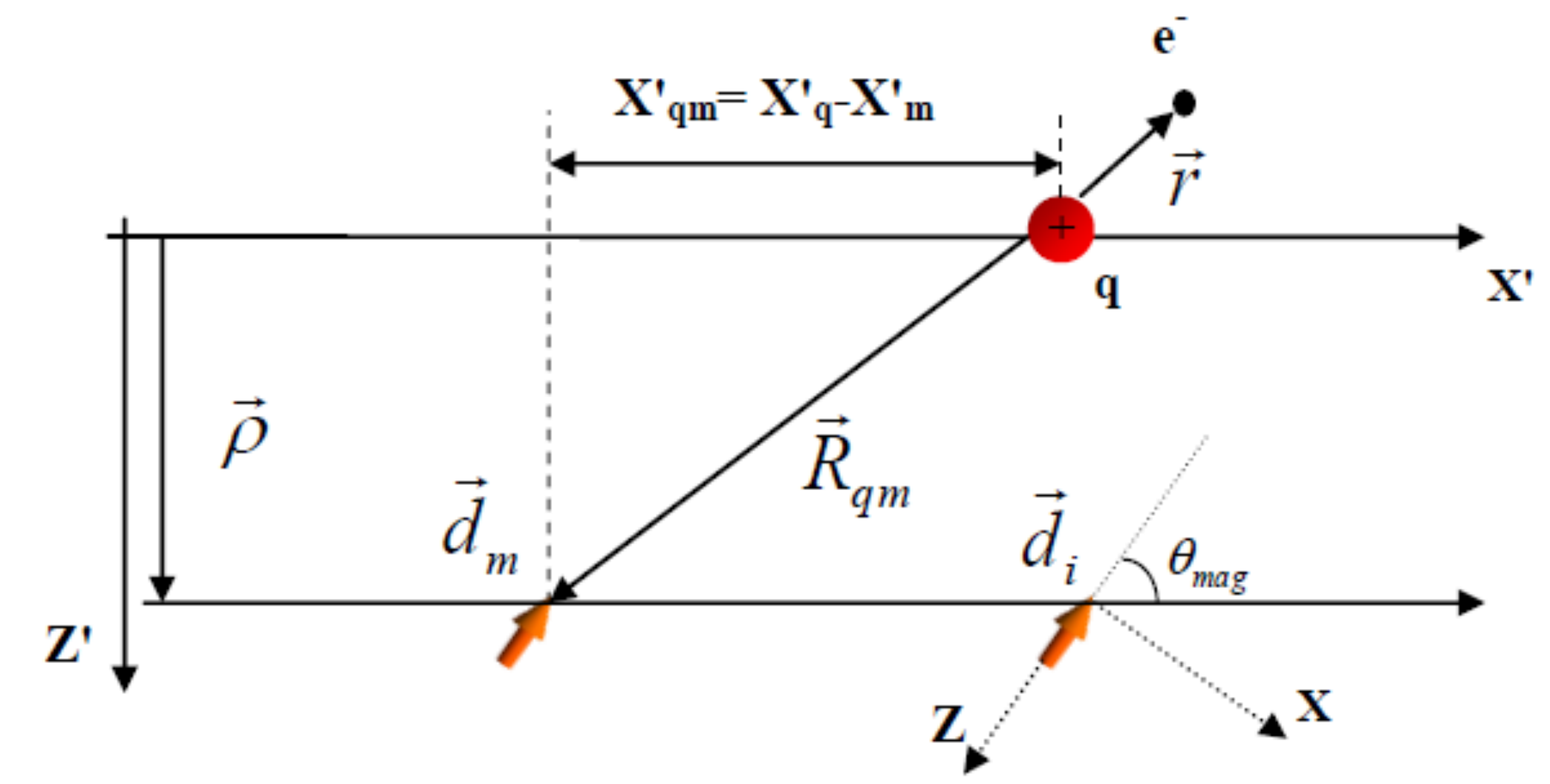}
\caption{\label{fig:setup-tilted} {\it Tilting the effective spins with respect to interspin distance vectors allows to cancel direct dipole-dipole 
interaction in the case Rydberg atom spin encoding.} In the case of spin encoding into long-lived Rydberg states the direct dipole-dipole interaction between 
the effective spins can be cancelled by tilting the spin quantiation axis $Z$ with respect to the line, connecting the spins. For the 
angle $\theta=\arccos(1/\sqrt{3})$ between the $-Z$ and $X'$ axes the resonant part of the direct dipole-dipole interaction is zero. }
}
\end{figure}

The $J_{im}^{\bot(zz)\;q,k_{0\;\nu_{0}}}$ coefficients and the effective magnetic field $b_{i}^{z\;q,k_{0\;\nu_{0}}}$ in this case will be given by the following expressions:
\begin{widetext}
\begin{align}
J_{im}^{\bot \; q,k_{0\;\nu_{0}}}= & \sum_{\substack{{\rm med=\pm} \\ k,\nu}}-2\left(\frac{V_{{\bf ns},k_{0\;\nu_{0}},\uparrow;{\rm med},k_{\nu},\downarrow}^{iq}\left(V_{{\bf ns},k_{0\;\nu_{0}},\uparrow;{\rm med},k_{\nu},\downarrow}^{mq}\right)^{*}}{{\cal E}_{\rm med}(k_{\nu})-{\cal E}_{{\bf ns}}(k_{0\;\nu_{0}})-E_{\rm spin}}
+\frac{V_{{\bf ns},k_{0\;\nu_{0}},\downarrow;{\rm med},k_{\nu},\uparrow}^{mq}\left(V_{{\bf ns},k_{0\;\nu_{0}},\downarrow;{\rm med},k_{\nu},\uparrow}^{iq}\right)^{*}}{{\cal E}_{\rm med}(k_{\nu})-{\cal E}_{{\bf ns}}(k_{0\;\nu_{0}})+E_{\rm spin}}\right), \nonumber \\
J_{im}^{zz \; q,k_{0\;\nu_{0}}}= & \sum_{\substack{{\rm med=\pm} \\ k,\nu}} - \frac{\left(V_{{\bf ns},k_{0\;\nu_{0}},\uparrow;{\rm med},k_{\nu},\uparrow}^{iq}-V_{{\bf ns},k_{0\;\nu_{0}},\downarrow;{\rm med},k_{\nu},\downarrow}^{iq}\right)\left(\left(V_{{\bf ns},k_{0\;\nu_{0}},\uparrow;{\rm med},k_{\nu},\uparrow}^{mq}\right)^{*}-\left(V_{{\bf ns},k_{0\;\nu_{0}},\downarrow;{\rm med},k_{\nu},\downarrow}^{mq}\right)^{*}\right)}{{\cal E}_{\rm med}(k_{\nu})-{\cal E}_{{\bf ns}}(k_{0\;\nu_{0}})}
+{\rm c.c.}, \nonumber \\
b_{i}^{z\;q,k_{0\;\nu_{0}}}= & \sum_{\substack{ m\ne i \\ {\rm med=\pm} \\ k,\nu}}\frac{2\left(\left|V_{{\bf ns},k_{0\;\nu_{0}},\downarrow;{\rm med},k_{\nu},\downarrow}^{iq}\right|^{2}-\left|V_{{\bf ns},k_{0\;\nu_{0}},\uparrow;{\rm med},k_{\nu},\uparrow}^{iq}\right|^{2}\right)}{{\cal E}_{\rm med}(k_{\nu})-{\cal E}_{{\bf ns}}(k_{0\;\nu_{0}})} \nonumber \\
&-\frac{2\left|V_{{\bf ns},k_{0\;\nu_{0}},\uparrow;{\rm med},k_{\nu},\downarrow}^{iq}\right|^{2}}{{\cal E}_{\rm med}(k_{\nu})-{\cal E}_{{\bf ns}}(k_{0\;\nu_{0}})-E_{\rm spin}}
+\frac{2\left|V_{{\bf ns},k_{0\;\nu_{0}},\downarrow;{\rm med},k_{\nu},\uparrow}^{iq}\right|^{2}}{{\cal E}_{\rm med}(k_{\nu})-{\cal E}_{{\bf ns}}(k_{0\;\nu_{0}})+E_{\rm spin}}+ \nonumber \\
& +\left[\frac{\left(V_{{\bf ns},k_{0\;\nu_{0}},\uparrow;{\rm med},k_{\nu},\uparrow}^{mq}\left(V_{{\bf ns},k_{0\;\nu_{0}},\downarrow;{\rm med},k_{\nu},\downarrow}^{iq}\right)^{*}-
V_{{\bf ns},k_{0\;\nu_{0}},\downarrow;{\rm med},k_{\nu},\downarrow}^{mq}\left(V_{{\bf ns},k_{0\;\nu_{0}},\uparrow;{\rm med},k_{\nu},\uparrow}^{iq}\right)^{*}\right)}{{\cal E}_{\rm med}(k_{\nu})-{\cal E}_{{\bf ns}}(k_{0\;\nu_{0}})}+{\rm c.c.}\right]. 
\label{eq:J-coupl-XXZ}
\end{align}
\end{widetext}
For spins encoded in Rydberg states an additional complication arises from the fact that the direct dipole-dipole interaction will be larger than the effective one, even for 
 not high ${\tilde n}$, and should be cancelled. The direct interaction between the i$^{\rm th}$ and m$^{\rm th}$ spins placed in the x-z plane has the form:
\begin{align}
\hat{V}_{\rm dd}=\frac{\hat{\vec{d}}_{i}\hat{\vec{d}}_{m}-3\left(\hat{\vec{d}}_{i}\vec{R}_{im}\right)\left(\hat{\vec{d}}_{m}\vec{R}_{im}\right)/R_{im}^{2}}{R_{im}^{3}}= \nonumber \\
=\frac{1}{2R_{im}^{3}}\left(\left(1-3\cos^{2}\theta\right)\left(\hat{d}_{i+}\hat{d}_{m-}+\hat{d}_{i-}\hat{d}_{m+}+2\hat{d}_{iz}\hat{d}_{mz}\right)+ \right. \nonumber \\
\left. +\frac{3}{\sqrt{2}}\sin \theta \cos \theta\left(\hat{d}_{i+}\hat{d}_{mz}-\hat{d}_{i-}\hat{d}_{mz}+\hat{d}_{iz}\hat{d}_{m+}-\hat{d}_{iz}\hat{d}_{m-}\right) \right. \nonumber \\
\left. -\frac{3}{2}\sin^{2}\theta \left(\hat{d}_{i+}\hat{d}_{m+}+\hat{d}_{i-}\hat{d}_{m-}\right)\right), \nonumber
\end{align}
where $\theta$ is the angle between the vector $\vec{R}_{im}$ connecting the two dipoles and their quantization axis $Z$, 
$\hat{d}_{i\pm}=\mp (\hat{d}_{i\;x} \mp i\hat{d}_{i\;y})/\sqrt{2}$. This expression shows that the resonant interaction terms 
$\hat{d}_{i+}\hat{d}_{m-}$, $\hat{d}_{i-}\hat{d}_{m+}$, $\hat{d}_{iz}\hat{d}_{mz}$, connecting states 
with the same energies, can be cancelled by tilting the dipoles such that $\cos^{2}\theta=1/3$, which corresponds to $\theta \approx 54.73^{o}$.  
The dipoles quantization axis can be set by applying a magnetic field along the $Z$ axis, as shown in Fig.\ref{fig:setup-tilted}.

As a concrete example we consider a 1D bi-layer setup shown in Fig.\ref{fig:setup-scheme}, in which the effective spins are encoded 
in the dressed $\ket{\uparrow}=\ket{+}_{\uparrow}$, $\ket{\downarrow}=\ket{+}_{\downarrow}$ states of ${\tilde n}=50$ of Rb, and a Rb superatom mediator state 
(or a single mediator Rb atom if interaction between Rydberg and ground state atoms is to be avoided \cite{Rydberg-ground-interactions}) 
initially prepared in the state Eq.(\ref{eq:Psi-supat}) with the Rydberg atoms in the $\ket{ns_{1/2},m_{j}=-1/2}$ state of $n=100$. 
Assuming also that initially the mediator atoms were prepared in a BEC state with $k_{0}=0$, $\nu_{0}=1$, the 
$J_{im}^{\bot,zz}=\sum_{q=1}^{N_{a}}J_{im}^{\bot,zz\;q,k_{0}=0_{\nu_{0}=1}}/N_{a}$ and $b_{i}^{z}=\sum_{q=1}^{N_{a}}b_{i}^{z\;q,k_{0}=0_{\nu_{0}=1}}/N_{a}$ coefficients 
were calculated from Eqs.(\ref{eq:J-coupl-XXZ}) using the $V_{{\bf ns},k_{0\;\nu_{0}},\alpha;{\rm med},k_{\nu},\beta}^{mq}$ 
matrix elements given in Appendix D, part B. The $b_{i}^{z}$ coefficients can be minimized and the $J^{zz}_{im}$ coefficients 
can be maximized at the same time by setting $a^{+}_{\uparrow,\downarrow}=b^{+}_{\uparrow,\downarrow}=c_{\pm}=d_{\pm}=1/\sqrt{2}$, which can be realized by using 
dressing microwave fields resonant to the corresponding 
transitions. In this case 
$V_{{\bf ns},k_{0\;\nu_{0}},\uparrow;\pm,k_{\nu},\uparrow}^{mq}=-V_{{\bf ns},k_{0\;\nu_{0}},\downarrow;\pm,k_{\nu},\downarrow}^{mq}$, and only the second and third terms in 
$b_{i}^{z\;q,k_{0\;\nu_{0}}}$ are non-zero.

\begin{figure*}
\center{
\includegraphics[width=18.cm]{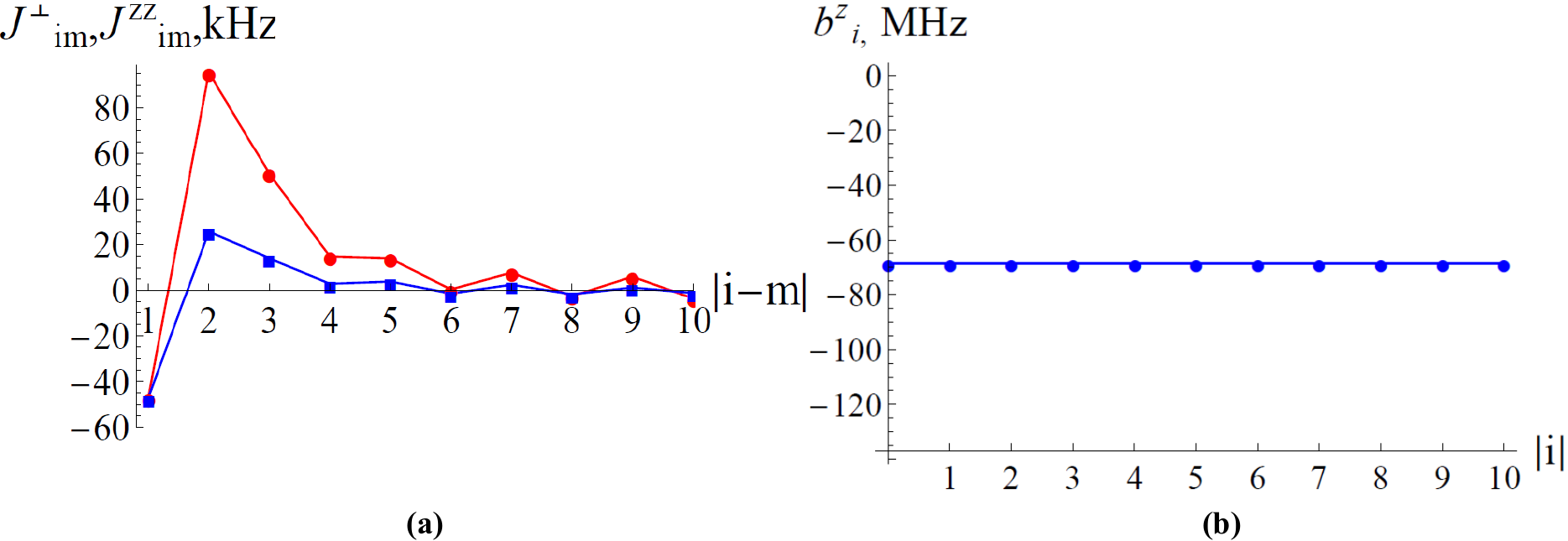}
\caption{\label{fig:J-coupl-Rydb} {\it Interaction coefficients in the case of spin encoding into long-lived Rydberg states, allowing to realize XXZ interaction}: 
(a) $J_{im}^{\bot}$ (circles, red curve), $J_{im}^{zz}$ (squares, blue curve) and 
(b) effective magnetic field $b_{i}^{z}$ for spins encoded in 
$\ket{\uparrow}=\ket{+}_{\uparrow}=(\ket{{\tilde n}p_{1/2},m_{j}=1/2}+\ket{{\tilde n}s_{1/2},m_{j}=1/2})/\sqrt{2}$, 
$\ket{\downarrow}=\ket{+}_{\downarrow}=(\ket{{\tilde n}p_{1/2},m_{j}=-1/2}+\ket{{\tilde n}s_{1/2},m_{j}=-1/2})/\sqrt{2}$ states of Rb with 
${\tilde n}=50$. The mediator atoms are initially prepared in a superatom state (\ref{eq:Psi-supat}) in the $\ket{100s_{1/2},m_{j}=-1/2}$ internal state and 
a BEC motional state with $k_{0}=0$, $\nu_{0}=1$. 
 In calculations of  $J_{im}^{\bot,zz}$, $b_{i}^{z}$ from Eqs.(\ref{eq:J-coupl-XXZ})  quasimomenta in the 
first Brillouin zone were summed for $\nu=1,...,5$ lowest Bloch bands of the mediator atom lattice. Setup dimensions: inter-layer distance $\rho=7$ $\mu$m, 
spin and mediator lattice periods $L_{\rm spin}=L_{\rm at}=7$ $\mu$m.  
The calculations were done for a spin in the center of the array $m=0$. Other parameters were as follows: 
$E_{ns_{1/2},m_{j}=1/2}-E_{ns_{1/2},m_{j}=-1/2}=E_{{\tilde n}s_{1/2},m_{j}=1/2}-E_{{\tilde n}s_{1/2},m_{j}=-1/2}=148.5$ MHz, 
$\Omega_{\uparrow}=2.5$ MHz, $\Omega_{\uparrow}=1$ MHz, $\Omega_{\rm med}=2.655$ MHz,  $\Delta_{\uparrow}=\Delta_{\downarrow}=\Delta_{\rm med}=0$, 
resulting in $E_{\rm +\;med}-E_{ns_{1/2},m_{j}=-1/2}=151.155$ MHz, $E_{\rm -\;med}-E_{ns_{1/2},m_{j}=-1/2}=145.845$ MHz. 
}
}
\end{figure*}

The numerically calculated  $J_{im}^{\bot(zz)}$ and  $b_{i}^{z}$ coefficients are shown in Fig.\ref{fig:J-coupl-Rydb}. One can see that similar to the case 
of XX interaction considered in the previous subsection, the XXZ interaction 
i) changes sign with the interspin distance. The nearest neighbor interaction is ferro- and the next nearest neighbor and more distant ones are antiferromagnetic;
ii) extends beyond nearest neighbors and falls off at the distances $|i-m|\sim 5$; 
(iii) interaction strengths $|J_{im}^{\bot}|,|J_{im}^{zz}|\sim$ tens kHz can be realized; iv) $b_{i}^{z}$ do not depend on spin position $i$. 
In the general case the interaction is of XXZ type, but tuning the spin and mediator transition frequencies by adjusting the corresponding MW dressing fields 
Rabi frequencies it can be made of symmetric Heisenberg type with $J^{\bot}_{im}=J^{zz}_{im}$ for chosen $im$ spin neighbors. In particular, in Fig.\ref{fig:J-coupl-Rydb} 
the nearest neighbor interaction with $|i-m|=1$ is of the Heisenberg type, and the next-nearest-neighbor and more distant interaction terms are of XXZ type. 
This case can be realized with the following parameters: spin and mediator lattice periods $L_{\rm spin}=L_{\rm at}=7$ $\mu$m, $\rho=7$ $\mu$m, 
$E_{ns_{1/2},m_{j}=1/2}-E_{ns_{1/2},m_{j}=-1/2}=E_{{\tilde n}s_{1/2},m_{j}=1/2}-E_{{\tilde n}s_{1/2},m_{j}=-1/2}=148.5$ MHz, 
$E_{\rm spin}=E_{\uparrow}-E_{\downarrow}=150$ MHz, the energy splittings of the mediator states $E_{\rm +\;med}-E_{ns_{1/2},m_{j}=-1/2}=151.155$ MHz, 
$E_{\rm -\;med}-E_{ns_{1/2},m_{j}=-1/2}=145.845$ MHz, the spin and mediator states dressing fields Rabi frequencies $\Omega_{\uparrow}=2.5$ MHz, 
$\Omega_{\downarrow}=1$ MHz, $\Omega_{\rm med}=2.655$ MHz and detunings $\Delta_{\uparrow}=\Delta_{\downarrow}=\Delta_{\rm med}=0$.

\begin{figure}
\center{
\includegraphics[width=9.cm]{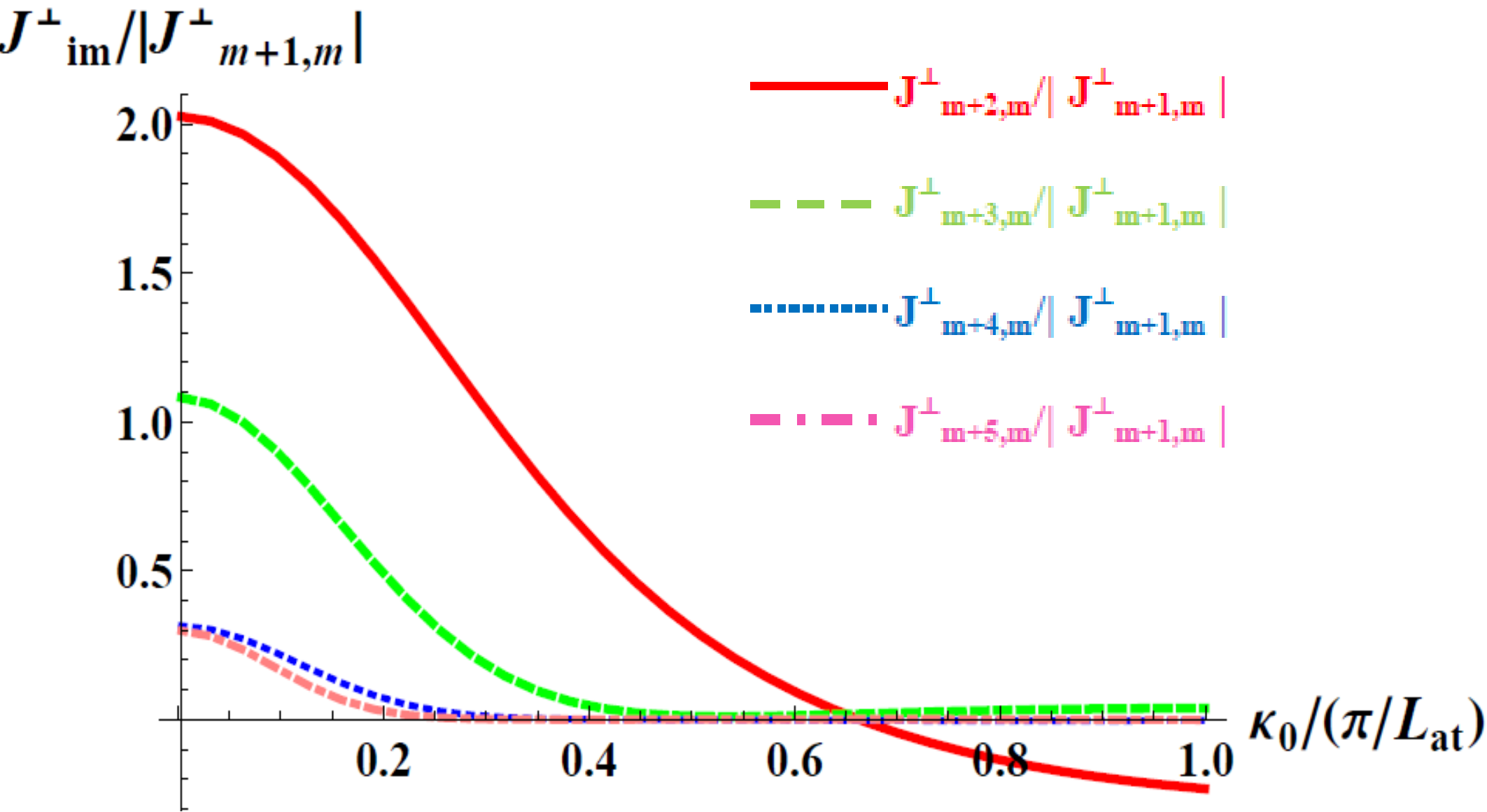}
\caption{\label{fig:JperpimJperp1-XXZ} {\it In the XXZ model with interaction coefficients Eqs.(\ref{eq:averaged-J-supat}),(\ref{eq:J-coupl-XXZ}) the ratios of the strengths of 
next-nearest and more distant to the nearest neighbor interaction can be controlled by  
the width of the distribution of the quasimomenta of the initial superposition of Bloch states of mediator atoms.} Assuming the atoms initially prepared in the lowest Bloch band with the Gaussian initial distribution 
$|c_{k_{0\;\nu_{0}=1}}|^{2} \sim e^{-k_{0}^{2}/\kappa_{0}^{2}}$ of quasimomenta, the ratios of interaction coefficients $J^{\bot}_{im}/|J^{\bot}_{m+1,m}|$ can be 
controlled by the distribution width $\kappa_{0}$: $J^{\bot}_{m+2,m}/|J^{\bot}_{m+1,m}|$ (red solid curve), $J^{\bot}_{m+3,m}/|J^{\bot}_{m+1,m}|$ 
(green dashed curve), $J^{\bot}_{m+4,m}/|J^{\bot}_{m+1,m}|$ (blue dotted curve), $J^{\bot}_{m+5,m}/|J^{\bot}_{m+1,m}|$ (pink dashed-dotted curve). 
The interaction coefficients shown in Fig.\ref{fig:J-coupl-Rydb} correspond to $\kappa_{0}=0$.}
}
\end{figure}

The contribution of 
the next-nearest and more distant neighbors can be controlled by preparing initial motional state of the mediator atoms as a superposition of Bloch 
states Eqs.(\ref{eq:Psi-supat}). Assuming the atoms prepared in a Gaussian distribution of quasimomenta of initial Bloch states in the lowest Bloch band with $|c_{k_{0\;\nu_{0}=1}}|^{2} \sim e^{-k_{0}^{2}/\kappa_{0}^{2}}$, 
the ratio of the next-nearest and more distant to the nearest neighbor interaction coefficients can be controlled by the quasimomenta distribution width $\kappa_{0}$, 
as shown in Fig.\ref{fig:JperpimJperp1-XXZ}. In particular, for $\kappa_{0}/(\pi/L_{\rm at})\approx 0.65$ the $J^{\bot,zz}_{im} \ll J^{\bot,zz}_{m+1,m}$ with 
$2 \le |i-m|\le 5$, resulting in the Heisenberg model with nearest neighbor interactions. In the range $0.3 \lesssim \kappa_{0}/(\pi/L_{\rm at}) \lesssim 0.65$ the 
$J^{\bot}_{m+3,m}/J^{\bot}_{m+2,m} \lesssim 0.2$ and $J^{\bot}_{m+2,m}/|J^{\bot}_{m+1,m}| < 1$
and the interaction can be approximated as the $J_{1}-J_{2}$ XXZ model 
$\sum_{l=1,2}J_{l}\left(\hat{S}^{x}_{i}\hat{S}^{x}_{i+l}+\hat{S}^{y}_{i}\hat{S}^{y}_{i+l}+\Delta_{l}\hat{S}^{z}_{i}\hat{S}^{z}_{i+l}\right)$ with 
$J_{1}=J^{\bot}_{m+1,m}<0$,  $J_{2}=J^{\bot}_{m+2,m}>0$, $\Delta_{1}=1$ and $\Delta_{2}\approx 0.27$. The $J_{1}-J_{2}$ XXZ model with both $\Delta_{1}=\Delta_{2}$ and $\Delta_{1} \ne \Delta_{2}$ attracts 
significant interest due to its relevance for description of spin-1/2 frustrated antiferromagnetic copper oxide spin chain compounds such as ${\rm LiCu_{2}O_{2}}$ \cite{LiCu2O2},  
${\rm NaCu_{2}O_{2}}$ \cite{NaCu2O2}, ${\rm PbCuSO_{4}(OH)_{2}}$ \cite{PbCuSO4OH}, ${\rm LiCuSbO_{4}}$ \cite{LiCuSbO4}, etc. 
Numerical analysis of the phase diagram of this model has shown \cite{J1-J2-XXZ-phase-diagram} that 
in the case $\Delta_{1}=\Delta_{2}$ for $J_{1}/J_{2}<-4$ the system is in the ferromagnetic state,  
for $-4<J_{1}/J_{2}<-2.5$ it is in the vector chiral phase, and for $-2.5 <J_{1}/J_{2} < -0.5$ in the Haldane dimer state, i.e. there are two critical points denoting the transitions between the phases. In the 1D copper oxide 
spin chains the ratios $J_{1}/J_{2}$ are fixed at certain values, while in our setup the ratio $J_{1}/J_{2}$ can be set between 
$\approx -1$ to $-\infty$ (keeping the $J_{3}$ and $J_{4}$ small) by 
initially preparing the mediator atoms in the superposition of Bloch states with a certain width $\kappa_{0}$ of the quasimomenta distribution. The $J_{1}-J_{2}$ XXZ model does not have an exact solution yet, 
and the bi-layer setup could potentially allow to study its phases, in particular, near quantum critical points.

The effective spins encoded into Rydberg states have finite lifetimes due to decay by spontaneous emission and interaction with black-body radiation. For the effective 
interactions to be observable their magnitude should exceed the spin states decay rates.  
The $50s_{1/2}$ state of Rb has lifetimes $\tau_{50s_{1/2}}=141.3$ $\mu$s at $T=4.2$ K and $65.2$ $\mu$s at $T=300$ K, and the $50p_{1/2}$ state has lifetimes 
$\tau_{50p_{1/2}}=257.4$ $\mu$s at $T=4.2$ K and $86.5$ $\mu$s at $T=300$ K \cite{Rb-lifetimes}. Since the spin states are equal superpositions 
of the $50s_{1/2}$ and $50p_{1/2}$ states their decay rates will be an average of the $s$ and $p$ decay rates: $(1/\tau_{50s_{1/2}}+1/\tau_{50p_{1/2}})/2$, giving 
the averaged spin states lifetimes  $\approx 182.4$ $\mu$s at $T=4.2$ K and $\approx 74.4$ $\mu$s at room temperature. The effective XXZ interaction strengths 
$\sim 50$ kHz correspond to the interaction times $\sim 2$ $\mu$s, which is more than an order of magnitude smaller than the averaged spin state 
decay times, making the effective interaction observable.

Finally, we come back to the direct dipole-dipole interaction between spin states. Tilting of the spin dipoles allows one 
to cancel the resonant direct dipole-dipole interaction which, for the case of $\theta=0$, when 
the quantization axis is perpendicular to $\vec{R}_{im}$, would be of the order 
$V_{\rm dd}\sim 2d_{{\tilde n}p,{\tilde n}s}^{2}/9R_{im}^{3}\sim 11.5$ MHz, which is three orders of magnitude larger than the effective 
interaction with strengths $\sim$ tens kHz. The non-resonant parts of the direct dipole-dipole interaction $\sim d_{i\pm}d_{mz},d_{iz}d_{m\pm}$ and 
$\sim d_{i\pm}d_{m\pm}$ are, however, not cancelled and are of the order of $d_{{\tilde n}p,{\tilde n}s}^{2}/9\sqrt{2}R_{im}^{3}\sim 4.1$ MHz and 
$d_{{\tilde n}p,{\tilde n}s}^{2}/9R_{im}^{3}\sim 5.7$ MHz, respectively, but they are more than an order of magnitude smaller compared to the 
spin transition frequency $E_{\rm spin}/\hbar=100$ MHz, leading to the probability of the spin changing its state due to the direct 
dipole-dipole interaction being $<10^{-2}$.

\section{Conclusions}

We propose a platform for simulating indirect spin-spin interactions based on polar molecules and/or Rydberg atoms trapped in two parallel 1D optical 
lattices.  The effective spin-1/2 systems are encoded in rotational states of polar molecules or long-lived Rydberg states of ultracold atoms, 
which are tightly 
trapped. The interaction between effective spins is mediated by Rydberg atoms in a parallel shallow lattice to allow for mediator atom motional state to be delocalized and interact simultaneously with several effective spins. The effective interaction is
 therefore realized via direct charge-dipole (dipole-dipole) spin-mediator interactions  
with polar molecule (Rydberg atom) spin encoding. By a particular choice of spin-encoding states XX, Ising and XXZ spin-spin interaction types are realized, with 
$J^{\bot}$, $J^{zz}$ interaction coefficients sign changing with interspin distance analogous to the RKKY interaction. The interactions extend beyond nearest neighbors 
and can reach 
magnitudes $\sim$ 100's kHz, limited by the trapping frequency of the spins optical lattice. 

The bi-layer setup allows to control the relative strengths of the next nearest and more distant 
to nearest neighbor interactions by initially preparing the mediator atoms in a superposition of motional Bloch states with a specific 
distribution (e.g. a Gaussian) of quasimomenta. Additionally, the Rabi frequencies and detuning of spin and mediator MW dressing fields, can be controlled to engineer symmetric Heisenberg interactions for selected pairs of neighbors, e.g. for nearest neighbors.

The bi-layer system can be extended to 2D geometries to simulate not only the Heisenberg/XXZ models, but also the indirect Dzyaloshinskii-Moriya (DM) 
anisotropic spin-spin interaction, which can also be mediated by conduction electrons \cite{DM-indirect,DM-Wiesendanger}, provided spin-orbit interaction 
between internal and motional states of mediator atoms can be incorporated. In this case the DM vector $\vec{D}$ has both the magnitude and the orientation 
oscillating with an interspin distance, which can produce chiral magnetic states with spatially oscillating chirality.

We note an interesting analogy between the XX model, considered in Section III, with the interaction coefficients given by 
Eqs.(\ref{eq:inter-coeff-XY}), 
with the Cook model \cite{Cook-model}, which is the generalized Hopfield model of associative memory \cite{Hopfield-model}, describing a system of $N$ 
interacting neurons, encoding $p$ different patterns. 
The Cook model is predicted to have a phase transition at a certain storage capacity $p/N$ between a self-organized phase, when the stored patterns 
can be reliably retrieved, and a spin glass.

\section{Acknowledgments}

The authors are grateful for fruitful discussions with S. Gopalakrishnan, D. Podolsky and R. Mukherjee. E.K. and I.B. thank the Russian Science 
Foundation (Grant No. 16-12-00028) in the part of derivation of indirect spin-spin interaction in the bi-layer of Rydberg atoms. E.K. is grateful for 
financial support from the Office of Naval Research (Award No. N00014-16-1-3054) and Robert A. Welch Foundation (Grant No. A-1261). S.F.Y. would like to thank the National Science Foundation through the CUA Center grant. S.T.R. acknowledges support from NSF Grant No. PHY-1516421 and funding from the Research Corporation for Science Advancement. H.R.S. was supported by a grant from the NSF to ITAMP.

\begin{appendices}

\section*{Appendix A}
\setcounter{equation}{0} \renewcommand{\theequation}{A.\arabic{equation}} 

The $\hat{V}$ terms can give rise to second order energy 
shifts, having a form of an indirect interaction between the effective spins. This can be shown via the Schrieffer-Wolff transformation
\begin{eqnarray}
e^{\hat{S}}\hat{H}e^{-\hat{S}}=\hat{H}+\left[\hat{S},\hat{H}\right]+\frac{\left[\hat{S},\left[\hat{S},\hat{H}\right]\right]}{2}+O\left(\hat{S}^{3}\right) \nonumber
\end{eqnarray}
in such a way that $\left[\hat{S},\hat{H}_{0}\right]=-\hat{V}$, giving as a result the transformed Hamiltonian
\begin{eqnarray}
\label{eq:Schrieffer-Wolff}
e^{\hat{S}}\hat{H}e^{-\hat{S}}=\hat{H}_{0}+\frac{\left[\hat{S},\hat{V}\right]}{2}+O\left(|\hat{V}|^{3}\right),
\end{eqnarray} 
where the generator $\hat{S}$ has the form:
\begin{widetext}
\begin{align}
\hat{S}= \sum_{i,m=1}^{N}\sum_{q=1}^{N_{a}}\sum_{\substack{ \alpha,\beta,\gamma,\delta=\uparrow,\downarrow \\ k,k',k'' \\ \nu,\nu',\nu''}} 
 \left[-\frac{\ket{{\bf ns},k_{\nu}}_{q}\ket{\alpha_{i}\beta_{m}}V_{{\bf ns},k_{\nu},\alpha;{\bf np'},k'_{\nu'},\gamma}^{iq}\delta_{\beta_{m},\delta_{m}}\bra{\gamma_{i}\delta_{m}}\bra{{\bf np'},k'_{\nu'}}_{q}}{{\cal E}_{{\bf np'}}(k'_{\nu'})-{\cal E}_{{\bf ns}}(k_{\nu})+E_{\gamma}-E_{\alpha}} \right. \nonumber \\ 
 \left. -\frac{\ket{{\bf ns},k_{\nu}}_{q}\ket{\alpha_{i}\beta_{m}}V_{{\bf ns},k_{\nu},\beta;{\bf np'},k'_{\nu'},\delta}^{mq}\delta_{\alpha_{i},\gamma_{i}}\bra{\gamma_{i}\delta_{m}}\bra{{\bf np'},k'_{\nu'}}_{q}}{{\cal E}_{{\bf np'}}(k'_{\nu'})-{\cal E}_{{\bf ns}}(k_{\nu})+E_{\delta}-E_{\beta}}+ \right. \nonumber \\ 
 \left. +\frac{\ket{{\bf ns},k_{\nu}}_{q}\ket{\alpha_{i}\beta_{m}}V_{{\bf ns},k_{\nu},\alpha;{\bf ns},k'_{\nu'},\gamma}^{iq}\delta_{\beta_{m},\delta_{m}}\bra{\gamma_{i}\delta_{m}}\bra{{\bf ns},k'_{\nu'}}_{q}}{{\cal E}_{{\bf ns}}(k'_{\nu'})-{\cal E}_{{\bf ns}}(k_{\nu})+E_{\gamma}-E_{\alpha}}+ \right. \nonumber \\
 \left. +\frac{\ket{{\bf ns},k_{\nu}}_{q}\ket{\alpha_{i}\beta_{m}}V_{{\bf ns},k_{\nu},\beta;{\bf ns},k'_{\nu'},\delta}^{mq}\delta_{\alpha_{i},\gamma_{i}}\bra{\gamma_{i}\delta_{m}}\bra{{\bf ns},k'_{\nu'}}_{q}}{{\cal E}_{{\bf ns}}(k'_{\nu'})-{\cal E}_{{\bf ns}}(k_{\nu})+E_{\delta}-E_{\beta}}+ \right. \nonumber \\ 
 \left. +\frac{\ket{{\bf np'},k'_{\nu'}}_{q}\ket{\alpha_{i}\beta_{m}}V_{{\bf np'},k'_{\nu'},\alpha;{\bf np''},k''_{\nu''},\gamma}^{iq}\delta_{\beta_{m},\delta_{m}}\bra{\gamma_{i}\delta_{m}}\bra{{\bf np''},k''_{\nu''}}_{q}}{{\cal E}_{{\bf np''}}(k''_{\nu''})-{\cal E}_{{\bf np'}}(k'_{\nu'})+E_{\gamma}-E_{\alpha}}+ \right. \nonumber \\ 
\left. +\frac{\ket{{\bf np'},k'_{\nu'}}_{q}\ket{\alpha_{i}\beta_{m}}V_{{\bf np'},k'_{\nu'},\beta;{\bf np''},k''_{\nu''},\delta}^{mq}\delta_{\alpha_{i},\gamma_{i}}\bra{\gamma_{i}\delta_{m}}\bra{{\bf np''},k''_{\nu''}}_{q}}{{\cal E}_{{\bf np''}}(k''_{\nu''})-{\cal E}_{{\bf np'}}(k'_{\nu'})+E_{\delta}-E_{\beta}}\right]-{\rm H.c.} 
\label{eq:generator}
\end{align}
where ${\bf ns}=\{n,l=0,j,m_{j}\}$, ${\bf np'}=\{n,l=1,j',m_{j}'\}$, ${\bf np''}=\{n,l=1,j'',m_{j}''\}$ and the summation is over $j,m_{j}$, $j',m_{j}'$ and 
$j'',m_{j}''$ quantum numbers. 

\end{widetext}

\section*{Appendix B}
\setcounter{equation}{0} \renewcommand{\theequation}{B.\arabic{equation}} 

In this section we give the $K_{\alpha_{i},\beta_{m};\gamma_{i},\delta_{m}}^{q,k_{\nu}}$ coefficients of Eq.(\ref{eq:Veff}): 
\begin{widetext}
\begin{align}
K_{\uparrow \uparrow,\uparrow \uparrow\;im}^{q,k_{\nu}}= \sum_{\substack{ \xi,\eta=i,m}}\sum_{\substack{{\bf nl}={\bf ns},{\bf np}}}\sum_{\substack{ k',\nu'}}(-1)^{l}\left[\frac{V_{{\bf ns},k_{\nu},\uparrow;{\bf nl},k'_{\nu'},\uparrow}^{\xi q}\left(V_{{\bf ns},k_{\nu},\uparrow;{\bf nl},k'_{\nu'},\uparrow}^{\eta q}\right)^{*}}{{\cal E}_{{\bf nl}}(k'_{\nu'})-{\cal E}_{{\bf ns}}(k_{\nu})} 
+\frac{\left|V_{{\bf ns},k_{\nu},\uparrow;{\bf nl},k'_{\nu'},\downarrow}^{\xi q}\right|^{2}}{{\cal E}_{{\bf nl}}(k'_{\nu'})-{\cal E}_{{\bf ns}}(k_{\nu})-E_{\rm spin}} \right], 
\end{align}
\begin{align}
K_{\downarrow \downarrow,\downarrow \downarrow\;im}^{q,k_{\nu}}=\sum_{\substack{ \xi,\eta=i,m}}\sum_{\substack{ {\bf nl}={\bf ns},{\bf np}}}\sum_{\substack{ k',\nu'}}(-1)^{l} \left[\frac{V_{{\bf ns},k_{\nu},\downarrow;{\bf nl},k'_{\nu'},\downarrow}^{\xi q}\left(V_{{\bf ns},k_{\nu},\downarrow;{\bf nl},k'_{\nu'},\downarrow}^{\eta q}\right)^{*}}{{\cal E}_{{\bf nl}}(k'_{\nu'})-{\cal E}_{{\bf ns}}(k_{\nu})}
+\frac{\left|V_{{\bf ns},k_{\nu},\downarrow;{\bf nl},k'_{\nu'},\uparrow}^{\xi q}\right|^{2}}{{\cal E}_{{\bf nl}}(k'_{\nu'})-{\cal E}_{{\bf ns}}(k_{\nu})+E_{\rm spin}} \right], 
\end{align}
\begin{align}
K_{\uparrow \downarrow,\uparrow \downarrow\;im}^{q,k_{\nu}}=\sum_{\substack{ {\bf nl}={\bf ns},{\bf np}}}\sum_{\substack{ k',\nu'}}(-1)^{l} 
\left[\frac{\left|V_{{\bf ns},k_{\nu},\uparrow;{\bf nl},k'_{\nu'},\uparrow}^{i q}+V_{{\bf ns},k_{\nu},\downarrow;{\bf nl},k'_{\nu'},\downarrow}^{m q}\right|^{2}}{{\cal E}_{{\bf nl}}(k'_{\nu'})-{\cal E}_{{\bf ns}}(k_{\nu})}+ \right. \nonumber \\
\left. +\frac{\left|V_{{\bf ns},k_{\nu},\uparrow;{\bf nl},k'_{\nu'},\downarrow}^{i q}\right|^{2}}{{\cal E}_{{\bf nl}}(k'_{\nu'})-{\cal E}_{{\bf ns}}(k_{\nu})-E_{\rm spin}} 
+\frac{\left|V_{{\bf ns},k_{\nu},\downarrow;{\bf nl},k'_{\nu'},\uparrow}^{m q}\right|^{2}}{{\cal E}_{{\bf nl}}(k'_{\nu'})-{\cal E}_{{\bf ns}}(k_{\nu})+E_{\rm spin}}\right], 
\end{align}
\begin{align}
K_{\downarrow \uparrow,\downarrow \uparrow\;im}^{q,k_{\nu}}=\sum_{\substack{ {\bf nl}={\bf ns},{\bf np}}}\sum_{\substack{ k',\nu'}}(-1)^{l} 
\left[\frac{\left|V_{{\bf ns},k_{\nu},\downarrow;{\bf nl},k'_{\nu'},\downarrow}^{i q}+V_{{\bf ns},k_{\nu},\uparrow;{\bf nl},k'_{\nu'},\uparrow}^{m q}\right|^{2}}{{\cal E}_{{\bf nl}}(k'_{\nu'})-{\cal E}_{{\bf ns}}(k_{\nu})}+ \right. \nonumber \\
\left. +\frac{\left|V_{{\bf ns},k_{\nu},\uparrow;{\bf nl},k'_{\nu'},\downarrow}^{m q}\right|^{2}}{{\cal E}_{{\bf nl}}(k'_{\nu'})-{\cal E}_{{\bf ns}}(k_{\nu})-E_{\rm spin}} 
+\frac{\left|V_{{\bf ns},k_{\nu},\downarrow;{\bf nl},k'_{\nu'},\uparrow}^{i q}\right|^{2}}{{\cal E}_{{\bf nl}}(k'_{\nu})-{\cal E}_{{\bf ns}}(k_{\nu})+E_{\rm spin}}\right], 
\end{align}
\begin{align}
K_{\uparrow \downarrow,\downarrow \uparrow\;im}^{q,k_{\nu}}=\sum_{\substack{ {\bf nl}={\bf ns},{\bf np}}}\sum_{\substack{ k',\nu'}}(-1)^{l} 
\left[\frac{V_{{\bf ns},k_{\nu},\uparrow;{\bf nl},k'_{\nu'},\downarrow}^{iq}\left(V_{{\bf ns},k_{\nu},\uparrow;{\bf nl},k'_{\nu'},\downarrow}^{mq}\right)^{*}}{{\cal E}_{{\bf nl}}(k'_{\nu'})-{\cal E}_{{\bf ns}}(k_{\nu})-E_{\rm spin}}
+\frac{V_{{\bf ns},k_{\nu},\downarrow;{\bf nl},k'_{\nu'},\uparrow}^{mq}\left(V_{{\bf ns},k_{\nu},\downarrow;{\bf nl},k'_{\nu'},\uparrow}^{iq}\right)^{*}}{{\cal E}_{{\bf nl}}(k'_{\nu'})-{\cal E}_{{\bf ns}}(k_{\nu})+E_{\rm spin}}\right], 
\end{align}
\begin{align}
K_{\uparrow \uparrow,\uparrow \downarrow\;im}^{q,k_{\nu}}=\sum_{\substack{ {\bf nl}={\bf ns},{\bf np}}}\sum_{\substack{ k',\nu'}}(-1)^{l} 
\left[\frac{\left(V_{{\bf ns},k_{\nu},\uparrow;{\bf nl},k'_{\nu'},\uparrow}^{iq}+V_{{\bf ns},k_{\nu},\uparrow;{\bf nl},k'_{\nu'},\uparrow}^{mq}\right)}{2}\times \right. \nonumber \\
\left. \times \left(V_{{\bf ns},k_{\nu},\downarrow;{\bf nl},k'_{\nu'},\uparrow}^{mq}\right)^{*}\left(\frac{1}{{\cal E}_{{\bf nl}}(k'_{\nu'})-{\cal E}_{{\bf ns}}(k_{\nu})}+\frac{1}{{\cal E}_{{\bf nl}}(k'_{\nu'})-{\cal E}_{{\bf ns}}(k_{\nu})+E_{\rm spin}}\right)+ \right. \nonumber \\
\left. +\frac{V_{{\bf ns},k_{\nu},\uparrow;{\bf nl},k'_{\nu'},\downarrow}^{mq}\left(\left(V_{{\bf ns},k_{\nu},\uparrow;{\bf nl},k'_{\nu'},\uparrow}^{iq}\right)^{*}+\left(V_{{\bf ns},k_{\nu},\downarrow;{\bf nl},k'_{\nu'},\downarrow}^{mq}\right)^{*}\right)}{2}
\left(\frac{1}{{\cal E}_{{\bf nl}}(k'_{\nu'})-{\cal E}_{{\bf ns}}(k_{\nu})}+\frac{1}{{\cal E}_{{\bf nl}}(k'_{\nu'})-{\cal E}_{{\bf ns}}(k_{\nu})-E_{\rm spin}}\right)\right] \nonumber \\
-\frac{V_{{\bf ns},k_{\nu},\uparrow;{\bf ns},k_{\nu},\downarrow}^{mq}\left(\left(V_{{\bf ns},k_{\nu},\uparrow;{\bf ns},k_{\nu},\uparrow}^{iq}\right)^{*}+\left(V_{{\bf ns},k_{\nu},\downarrow;{\bf ns},k_{\nu},\downarrow}^{mq}\right)^{*}\right)}{E_{\rm spin}} 
+\frac{\left(V_{{\bf ns},k_{\nu},\uparrow;{\bf ns},k_{\nu},\uparrow}^{iq}+V_{{\bf ns},k_{\nu},\uparrow;{\bf ns},k_{\nu},\uparrow}^{mq}\right)\left(V_{{\bf ns},k_{\nu},\downarrow;{\bf ns},k_{\nu},\uparrow}^{mq}\right)^{*}}{E_{\rm spin}}, 
\end{align}
\begin{align}
K_{\uparrow \uparrow,\downarrow \uparrow\;im}^{q,k_{\nu}}=\sum_{\substack{ {\bf nl}={\bf ns},{\bf np}}}\sum_{\substack{ k',\nu'}}(-1)^{l} 
\left[\frac{\left(V_{{\bf ns},k_{\nu},\downarrow;{\bf nl},k'_{\nu'},\downarrow}^{iq}+V_{{\bf ns},k_{\nu},\downarrow;{\bf nl},k'_{\nu'},\downarrow}^{mq}\right)}{2}\times \right. \nonumber \\
\left. \times \left(V_{{\bf ns},k_{\nu},\uparrow;{\bf nl},k'_{\nu'},\downarrow}^{iq}\right)^{*}\left(\frac{1}{{\cal E}_{{\bf nl}}(k'_{\nu'})-{\cal E}_{{\bf ns}}(k_{\nu})}+\frac{1}{{\cal E}_{{\bf nl}}(k'_{\nu'})-{\cal E}_{{\bf ns}}(k_{\nu})-E_{\rm spin}}\right)+ \right. \nonumber \\
\left. +\frac{V_{{\bf ns},k_{\nu},\downarrow;{\bf nl},k'_{\nu'},\uparrow}^{iq}\left(\left(V_{{\bf ns},k_{\nu},\uparrow;{\bf nl},k'_{\nu'},\uparrow}^{iq}\right)^{*}+\left(V_{{\bf ns},k_{\nu},\downarrow;{\bf nl},k'_{\nu'},\downarrow}^{mq}\right)^{*}\right)}{2}
\left(\frac{1}{{\cal E}_{{\bf nl}}(k'_{\nu'})-{\cal E}_{{\bf ns}}(k_{\nu})}+\frac{1}{{\cal E}_{{\bf nl}}(k'_{\nu'})-{\cal E}_{{\bf ns}}(k_{\nu})+E_{\rm spin}}\right)\right]+ \nonumber \\
+\frac{V_{{\bf ns},k_{\nu},\downarrow;{\bf ns},k_{\nu},\uparrow}^{iq}\left(\left(V_{{\bf ns},k_{\nu},\uparrow;{\bf ns},k_{\nu},\uparrow}^{iq}\right)^{*}+\left(V_{{\bf ns},k_{\nu},\downarrow;{\bf ns},k_{\nu},\downarrow}^{mq}\right)^{*}\right)}{E_{\rm spin}}
-\frac{\left(V_{{\bf ns},k_{\nu},\downarrow;{\bf ns},k_{\nu},\downarrow}^{iq}+V_{{\bf ns},k_{\nu},\downarrow;{\bf ns},k_{\nu},\downarrow}^{mq}\right)\left(V_{{\bf ns},k_{\nu},\uparrow;{\bf ns},k_{\nu},\downarrow}^{iq}\right)^{*}}{E_{\rm spin}}, 
\end{align}
\begin{align}
K_{\downarrow \downarrow,\uparrow \downarrow\;im}^{q,k_{\nu}}=\sum_{\substack{ {\bf nl}={\bf ns},{\bf np}}}\sum_{\substack{ k',\nu'}}(-1)^{l} 
\left[\frac{\left(V_{{\bf ns},k_{\nu},\downarrow;{\bf nl},k'_{\nu'},\downarrow}^{iq}+V_{{\bf ns},k_{\nu},\downarrow;{\bf nl},k'_{\nu'},\downarrow}^{mq}\right)}{2}\times \right. \nonumber \\
\left. \times \left(V_{{\bf ns},k_{\nu},\uparrow;{\bf nl},k'_{\nu'},\downarrow}^{iq}\right)^{*}\left(\frac{1}{{\cal E}_{{\bf nl}}(k'_{\nu'})-{\cal E}_{{\bf ns}}(k_{\nu})}+\frac{1}{{\cal E}_{{\bf nl}}(k'_{\nu'})-{\cal E}_{{\bf ns}}(k_{\nu})-E_{\rm spin}}\right)+ \right. \nonumber \\
\left. +\frac{V_{{\bf ns},k_{\nu},\downarrow;{\bf nl},k'_{\nu'},\uparrow}^{iq}\left(\left(V_{{\bf ns},k_{\nu},\uparrow;{\bf nl},k'_{\nu'},\uparrow}^{iq}\right)^{*}+\left(V_{{\bf ns},k_{\nu},\downarrow;{\bf nl},k'_{\nu'},\downarrow}^{mq}\right)^{*}\right)}{2}
\left(\frac{1}{{\cal E}_{{\bf nl}}(k'_{\nu'})-{\cal E}_{{\bf ns}}(k_{\nu})}+\frac{1}{{\cal E}_{{\bf nl}}(k'_{\nu'})-{\cal E}_{{\bf ns}}(k_{\nu})+E_{\rm spin}}\right)\right]+ \nonumber \\
+\frac{V_{{\bf ns},k_{\nu},\downarrow;{\bf ns},k_{\nu},\uparrow}^{iq}\left(\left(V_{{\bf ns},k_{\nu},\uparrow;{\bf ns},k_{\nu},\uparrow}^{iq}\right)^{*}+\left(V_{{\bf ns},k_{\nu},\downarrow;{\bf ns},k_{\nu},\downarrow}^{mq}\right)^{*}\right)}{E_{\rm spin}} 
-\frac{\left(V_{{\bf ns},k_{\nu},\downarrow;{\bf ns},k_{\nu},\downarrow}^{iq}+V_{{\bf ns},k_{\nu},\downarrow;{\bf ns},k_{\nu},\downarrow}^{mq}\right)\left(V_{{\bf ns},k_{\nu},\uparrow;{\bf ns},k_{\nu},\downarrow}^{iq}\right)^{*}}{E_{\rm spin}}, 
\end{align}
\begin{align}
K_{\downarrow \downarrow,\downarrow \uparrow\;im}^{q,k_{\nu}}=\sum_{\substack{ {\bf nl}={\bf ns},{\bf np}}}\sum_{\substack{ k',\nu'}}(-1)^{l} 
\left[\frac{\left(V_{{\bf ns},k_{\nu},\downarrow;{\bf nl},k'_{\nu'},\downarrow}^{iq}+V_{{\bf ns},k_{\nu},\downarrow;{\bf nl},k'_{\nu'},\downarrow}^{mq}\right)}{2}\times \right. \nonumber \\ 
\left. \times\left(V_{{\bf ns},k_{\nu},\uparrow;{\bf nl},k'_{\nu'},\downarrow}^{mq}\right)^{*}\left(\frac{1}{{\cal E}_{{\bf nl}}(k'_{\nu'})-{\cal E}_{{\bf ns}}(k_{\nu})}+\frac{1}{{\cal E}_{{\bf nl}}(k'_{\nu'})-{\cal E}_{{\bf ns}}(k_{\nu})-E_{\rm spin}}\right)+ \right. \nonumber \\
\left.+\frac{V_{{\bf ns},k_{\nu},\downarrow;{\bf nl},k'_{\nu'},\uparrow}^{mq}\left(\left(V_{{\bf ns},k_{\nu},\downarrow;{\bf nl},k'_{\nu'},\downarrow}^{iq}\right)^{*}+\left(V_{{\bf ns},k_{\nu},\uparrow;{\bf nl},k'_{\nu'},\uparrow}^{mq}\right)^{*}\right)}{2}
\left(\frac{1}{{\cal E}_{{\bf nl}}(k'_{\nu'})-{\cal E}_{{\bf ns}}(k_{\nu})}+\frac{1}{{\cal E}_{{\bf nl}}(k'_{\nu'})-{\cal E}_{{\bf ns}}(k_{\nu})+E_{\rm spin}}\right) \right]+ \nonumber \\
+\frac{V_{{\bf ns},k_{\nu},\downarrow;{\bf ns},k_{\nu},\uparrow}^{mq}\left(\left(V_{{\bf ns},k_{\nu},\downarrow;{\bf ns},k_{\nu},\downarrow}^{iq}\right)^{*}+\left(V_{{\bf ns},k_{\nu},\uparrow;{\bf ns},k_{\nu},\uparrow}^{mq}\right)^{*}\right)}{E_{\rm spin}} 
-\frac{\left(V_{{\bf ns},k_{\nu},\downarrow;{\bf ns},k_{\nu},\downarrow}^{iq}+V_{{\bf ns},k_{\nu},\downarrow;{\bf ns},k_{\nu},\downarrow}^{mq}\right)\left(V_{{\bf ns},k_{\nu},\uparrow;{\bf ns},k_{\nu},\downarrow}^{mq}\right)^{*}}{E_{\rm spin}}, 
\end{align}
\begin{align}
K_{\uparrow \uparrow,\downarrow \downarrow\;im}^{q,k_{\nu}}=\sum_{\substack{ {\bf nl}={\bf ns},{\bf np}}}\sum_{\substack{ k',\nu'}}(-1)^{l} 
\left[\frac{V_{{\bf ns},k_{\nu},\uparrow;{\bf nl},k'_{\nu'},\downarrow}^{iq}\left(V_{{\bf ns},k_{\nu},\downarrow;{\bf nl},k'_{\nu'},\uparrow}^{mq}\right)^{*}}{2} 
\left(\frac{1}{{\cal E}_{{\bf nl}}(k'_{\nu'})-{\cal E}_{{\bf ns}}(k_{\nu})-E_{\rm spin}}+\frac{1}{{\cal E}_{{\bf nl}}(k'_{\nu'})-{\cal E}_{{\bf ns}}(k_{\nu})+E_{\rm spin}}\right)+ \right. \nonumber \\
\left. +\frac{V_{{\bf ns},k_{\nu},\uparrow;{\bf nl},k'_{\nu'},\downarrow}^{mq}\left(V_{{\bf ns},k_{\nu},\downarrow;{\bf nl},k'_{\nu'},\uparrow}^{iq}\right)^{*}}{2} 
\left(\frac{1}{{\cal E}_{{\bf nl}}(k'_{\nu'})-{\cal E}_{{\bf ns}}(k_{\nu})-E_{\rm spin}}+\frac{1}{{\cal E}_{{\bf nl}}(k'_{\nu'})-{\cal E}_{{\bf ns}}(k_{\nu})+E_{\rm spin}}\right) \right] \nonumber \\
\end{align}
\begin{align}
K_{\alpha \beta,\gamma \delta;im}^{q,k_{\nu}}=\left(K_{\gamma \delta,\alpha \beta;im}^{q,k_{\nu}}\right)^{*}. 
\end{align}
where ${\bf ns}=\{n,l=0,j=1/2,m_{j}\}$, ${\bf np}=\{n,l=1,j',m_{j}'\}$ and the summation is over $j'$, $m_{j}'$ quantum numbers. 

\section*{Appendix C}
\setcounter{equation}{0} \renewcommand{\theequation}{C.\arabic{equation}}

The $\ket{\alpha_{i}\beta_{m}}\bra{\gamma_{i}\delta_{m}}$ can be expressed via the two spin-1/2 variables $\hat{S}_{i}^{\pm,z}\hat{S}_{m}^{\pm,z}$ using relations 
\begin{align}
\ket{\uparrow_{i}\uparrow_{m}}\bra{\uparrow_{i}\uparrow_{m}}=\left(\frac{1}{2}+\hat{S}_{i}^{z}\right)\left(\frac{1}{2}+\hat{S}_{m}^{z}\right)=\frac{1}{4}+\frac{1}{2}\left(\hat{S}_{i}^{z}+\hat{S}_{m}^{z}\right)+\hat{S}_{i}^{z}\hat{S}_{m}^{z}, \nonumber \\
\ket{\downarrow_{i}\downarrow_{m}}\bra{\downarrow_{i}\downarrow_{m}}=\left(\frac{1}{2}-\hat{S}_{i}^{z}\right)\left(\frac{1}{2}-\hat{S}_{m}^{z}\right)=\frac{1}{4}-\frac{1}{2}\left(\hat{S}_{i}^{z}+\hat{S}_{m}^{z}\right)+\hat{S}_{i}^{z}\hat{S}_{m}^{z}, \nonumber \\ 
\ket{\downarrow_{i}\uparrow_{m}}\bra{\downarrow_{i}\uparrow_{m}}=\left(\frac{1}{2}-\hat{S}_{i}^{z}\right)\left(\frac{1}{2}+\hat{S}_{m}^{z}\right)=\frac{1}{4}-\frac{1}{2}\left(\hat{S}_{i}^{z}-\hat{S}_{m}^{z}\right)-\hat{S}_{i}^{z}\hat{S}_{m}^{z}, \nonumber \\
\ket{\uparrow_{i}\downarrow_{m}}\bra{\uparrow_{i}\downarrow_{m}}=\left(\frac{1}{2}+\hat{S}_{i}^{z}\right)\left(\frac{1}{2}-\hat{S}_{m}^{z}\right)=\frac{1}{4}+\frac{1}{2}\left(\hat{S}_{i}^{z}-\hat{S}_{m}^{z}\right)-\hat{S}_{i}^{z}\hat{S}_{m}^{z}, \nonumber \\
\ket{\uparrow_{i}\downarrow_{m}}\bra{\downarrow_{i}\uparrow_{m}}=\hat{S}_{i}^{+}\hat{S}_{m}^{-}, \nonumber \\
\ket{\uparrow_{i}\uparrow_{m}}\bra{\uparrow_{i}\downarrow_{m}}=\left(\frac{1}{2}+\hat{S}_{i}^{z}\right)\hat{S}_{m}^{+}, \nonumber \\
\ket{\uparrow_{i}\uparrow_{m}}\bra{\downarrow_{i}\uparrow_{m}}=\hat{S}_{i}^{+}\left(\frac{1}{2}+\hat{S}_{m}^{z}\right), \nonumber \\
\ket{\downarrow_{i}\downarrow_{m}}\bra{\uparrow_{i}\downarrow_{m}}=\hat{S}_{i}^{-}\left(\frac{1}{2}-\hat{S}_{m}^{z}\right), \nonumber \\
\ket{\downarrow_{i}\downarrow_{m}}\bra{\downarrow_{i}\uparrow_{m}}=\left(\frac{1}{2}-\hat{S}_{i}^{z}\right)\hat{S}_{m}^{-}, \nonumber \\
\ket{\uparrow_{i}\uparrow_{m}}\bra{\downarrow_{i}\downarrow_{m}}=\hat{S}_{i}^{+}\hat{S}_{m}^{+},  
\end{align}
and other states can be obtained using $\ket{\gamma_{i}\delta_{m}}\bra{\alpha_{i}\beta_{m}}=\left(\ket{\alpha_{i}\beta_{m}}\bra{\gamma_{i}\delta_{m}}\right)^{\dagger}$.

The interaction coefficients are given by the following expressions:
\begin{align}
J_{im}^{zz\;q,k_{\nu}}=K_{\uparrow \uparrow,\uparrow \uparrow \; im}^{q,k_{\nu}}+K_{\downarrow \downarrow,\downarrow \downarrow \; im}^{q,k_{\nu}}-K_{\downarrow \uparrow,\downarrow \uparrow \; im}^{q,k_{\nu}}-K_{\uparrow \downarrow,\uparrow \downarrow \; im}^{q,k_{\nu}}= \nonumber \\
=\sum_{\substack{ {\bf nl}={\bf ns},{\bf np}}}\sum_{\substack{ k',\nu'}}(-1)^{l}\left[\frac{\left(V_{{\bf ns},k_{\nu},\uparrow;{\bf nl},k'_{\nu'},\uparrow}^{iq}-V_{{\bf ns},k_{\nu},\downarrow;{\bf nl},k'_{\nu'},\downarrow}^{iq}\right)
\left(\left(V_{{\bf ns},k_{\nu},\uparrow;{\bf nl},k'_{\nu'},\uparrow}^{mq}\right)^{*}-\left(V_{{\bf ns},k_{\nu},\downarrow;{\bf nl},k'_{\nu'},\downarrow}^{mq}\right)^{*}\right)}{{\cal E}_{{\bf nl}}\left(k'_{\nu'}\right)-{\cal E}_{{\bf ns}}\left(k_{\nu}\right)}\right]+{\rm c.c.}, 
\label{eq:Jzz}
\end{align}
\begin{align}
J_{im}^{+-\;q,k_{\nu}}=K_{\uparrow \downarrow,\downarrow \uparrow \; im}^{q,k_{\nu}}
=-\sum_{\substack{ \bf np}} \sum_{\substack{ k',\nu'}}\left( 
\frac{V_{{\bf ns},k_{\nu},\uparrow;{\bf np},k'_{nu'},\downarrow}^{iq}\left(V_{{\bf ns},k_{\nu},\uparrow;{\bf np},k'_{\nu'},\downarrow}^{mq}\right)^{*}}{{\cal E}_{{\bf np}}(k'_{\nu'})-{\cal E}_{{\bf ns}}(k_{\nu})-E_{\rm spin}}+
\frac{V_{{\bf ns},k_{\nu},\downarrow;{\bf np},k'_{\nu'},\uparrow}^{mq}\left(V_{{\bf ns},k_{\nu},\downarrow;{\bf np},k'_{\nu'},\uparrow}^{iq}\right)^{*}}{{\cal E}_{{\bf np}}(k'_{\nu'})-{\cal E}_{{\bf ns}}(k_{\nu})+E_{\rm spin}}\right),
\label{eq:Jpm}
\end{align}
\begin{align}
J_{im}^{z+\; q,k_{\nu}}=K_{\uparrow \uparrow,\uparrow \downarrow \; im}^{q,k_{\nu}}-K_{\downarrow \uparrow,\downarrow \downarrow \; im}^{q,k_{\nu}}=
\sum_{\substack{ {\bf nl}={\bf ns},{\bf np}}}\sum_{\substack{ k',\nu'}}(-1)^{l}\left[ V_{{\bf ns},k_{\nu},\uparrow;{\bf nl},k'_{\nu'},\downarrow}^{mq}\times \right. \nonumber \\
\left. \times \left(\left(V_{{\bf ns},k_{\nu},\uparrow;{\bf nl},k'_{\nu'},\uparrow}^{iq}\right)^{*}-\left(V_{{\bf ns},k_{\nu},\downarrow;{\bf nl},k'_{\nu'},\downarrow}^{iq}\right)\right)\left(\frac{1}{{\cal E}_{{\bf nl}}(k'_{\nu'})-{\cal E}_{{\bf ns}}(k_{\nu})-E_{\rm spin}}+\frac{1}{{\cal E}_{{\bf nl}}(k'_{\nu'})-{\cal E}_{{\bf ns}}(k_{\nu})}\right)+ \right. \nonumber \\
\left. +\left(V_{{\bf ns},k_{\nu},\uparrow;{\bf nl},k'_{\nu'},\uparrow}^{iq}-V_{{\bf ns},k_{\nu},\downarrow;{\bf nl},k'_{\nu'},\downarrow}^{iq}\right)\left(V_{{\bf ns},k_{\nu},\downarrow;{\bf nl},k'_{\nu'},\uparrow}^{mq}\right)^{*}
\left(\frac{1}{{\cal E}_{{\bf nl}}(k'_{\nu'})-{\cal E}_{{\bf ns}}(k_{\nu})+E_{\rm spin}}+\frac{1}{{\cal E}_{{\bf nl}}(k'_{\nu'})-{\cal E}_{{\bf ns}}(k_{\nu})}\right)\right] \nonumber \\
-\frac{V_{{\bf ns},k_{\nu},\uparrow;{\bf ns},k_{\nu},\downarrow}^{mq}\left(\left(V_{{\bf ns},k_{\nu},\uparrow;{\bf ns},k_{\nu},\uparrow}^{iq}\right)^{*}-\left(V_{{\bf ns},k_{\nu},\downarrow;{\bf ns},k_{\nu},\downarrow}^{iq}\right)^{*}\right)}{E_{\rm spin}} 
+\frac{\left(V_{{\bf ns},k_{\nu},\uparrow;{\bf ns},k_{\nu},\uparrow}^{iq}-V_{{\bf ns},k_{\nu},\downarrow;{\bf ns},k_{\nu},\downarrow}^{iq}\right)\left(V_{{\bf ns},k_{\nu},\downarrow;{\bf ns},k_{\nu},\uparrow}^{mq}\right)^{*}}{E_{\rm spin}} \nonumber \\
J_{im}^{++\;q}=K_{\uparrow \uparrow,\downarrow \downarrow \; im}^{q}, \nonumber \\
J_{im}^{z- \; q}=\left(J_{im}^{z+\; q}\right)^{*}, \nonumber \\
J_{im}^{+z \; q}=J_{mi}^{z+ \; q}, \nonumber \\
J_{im}^{-z \; q}=J_{mi}^{z- \; q},  
\end{align}
\begin{align}
\label{eq:biz}
b_{im}^{z\;q,k_{\nu}}=K_{\uparrow \uparrow,\uparrow \uparrow \; im}^{q,k_{\nu}}-K_{\downarrow \downarrow,\downarrow \downarrow \; im}^{q,k_{\nu}}+K_{\uparrow \downarrow,\uparrow \downarrow \; im}^{q,k_{\nu}}-K_{\downarrow \uparrow,\downarrow \uparrow \; im}^{q,k_{\nu}}= \nonumber \\
=\sum_{\substack{ {\bf nl}={\bf ns},{\bf np}}}
\sum_{\substack{ k',\nu'}}(-1)^{l}\left[\frac{2\left|V_{{\bf ns},k_{\nu},\uparrow;{\bf nl},k'_{\nu'},\downarrow}^{iq}\right|^{2}}{{\cal E}_{{\bf nl}}(k'_{\nu'})-{\cal E}_{{\bf ns}}(k_{\nu})-E_{\rm spin}}
-\frac{2\left|V_{{\bf ns},k_{\nu},\downarrow;{\bf nl},k'_{\nu'},\uparrow}^{iq}\right|^{2}}{{\cal E}_{{\bf nl}}(k'_{\nu'})-{\cal E}_{{\bf ns}}(k_{\nu})+E_{\rm spin}}+ 
\frac{2\left(\left|V_{{\bf ns},k_{\nu},\uparrow;{\bf nl},k'_{\nu'},\uparrow}^{iq}\right|^{2}-\left|V_{{\bf ns},k_{\nu},\uparrow;{\bf nl},k'_{\nu'},\uparrow}^{iq}\right|^{2}\right)}{{\cal E}_{{\bf nl}}(k'_{\nu'})-{\cal E}_{{\bf ns}}(k_{\nu})}+ \right. \\
\left.+\frac{\left(V_{{\bf ns},k_{\nu},\uparrow;{\bf nl},k'_{\nu'},\uparrow}^{iq}-V_{{\bf ns},k_{\nu},\downarrow;{\bf nl},k'_{\nu'},\downarrow}^{iq}\right)\left(\left(V_{{\bf ns},k_{\nu},\uparrow;{\bf nl},k'_{\nu'},\uparrow}^{mq}\right)^{*}+\left(V_{{\bf ns},k_{\nu},\downarrow;{\bf nl},k'_{\nu'},\downarrow}^{mq}\right)^{*}\right)}{{\cal E}_{{\bf nl}}(k'_{\nu'})-{\cal E}_{{\bf ns}}(k_{\nu})}+{\rm c.c.} \right], \nonumber 
\label{eq:biz}
\end{align}
\begin{align}
b_{im}^{+\;q,k_{\nu}}=K_{\uparrow \uparrow,\downarrow \uparrow \; im}^{q,k_{\nu}}+K_{\uparrow \downarrow,\downarrow \downarrow \; im}^{q,k_{\nu}}=\sum_{\substack{ {\bf nl}={\bf ns},{\bf np}}}\sum_{\substack{ k',\nu'}}(-1)^{l} 
\left[\frac{\left(V_{{\bf ns},k_{\nu},\downarrow;{\bf nl},k'_{\nu'},\downarrow}^{iq}+V_{{\bf ns},k_{\nu},\downarrow;{\bf nl},k'_{\nu'},\downarrow}^{mq}\right)}{2}\times \right. \nonumber \\
\left. \times \left(V_{{\bf ns},k_{\nu},\uparrow;{\bf nl},k'_{\nu'},\downarrow}^{iq}\right)^{*}\left(\frac{1}{{\cal E}_{{\bf nl}}(k'_{\nu'})-{\cal E}_{{\bf ns}}(k_{\nu})}+\frac{1}{{\cal E}_{{\bf nl}}(k'_{\nu'})-{\cal E}_{{\bf ns}}(k_{\nu})-E_{\rm spin}}\right)+ \right. \nonumber \\ 
\left. +\frac{V_{{\bf ns},k_{\nu},\downarrow;{\bf nl},k'_{\nu'},\uparrow}^{iq}\left(\left(V_{{\bf ns},k_{\nu},\uparrow;{\bf nl},k'_{\nu'},\uparrow}^{iq}\right)^{*}+\left(V_{{\bf ns},k_{\nu},\downarrow;{\bf nl},k'_{\nu'},\downarrow}^{mq}\right)^{*}\right)}{2}
\left(\frac{1}{{\cal E}_{{\bf nl}}(k'_{\nu'})-{\cal E}_{{\bf ns}}(k_{\nu})}+\frac{1}{{\cal E}_{{\bf nl}}(k'_{\nu'})-{\cal E}_{{\bf ns}}(k_{\nu})+E_{\rm spin}}\right)+{\rm c.c.}\right]+ \nonumber \\
+\left[\frac{V_{{\bf ns},k_{\nu},\downarrow;{\bf ns},k_{\nu},\uparrow}^{iq}\left(\left(V_{{\bf ns},k_{\nu},\uparrow;{\bf ns},k_{\nu},\uparrow}^{iq}\right)^{*}+\left(V_{{\bf ns},k_{\nu},\downarrow;{\bf ns},k_{\nu},\downarrow}^{mq}\right)^{*}\right)}{E_{\rm spin}}
-\frac{\left(V_{{\bf ns},k_{\nu},\downarrow;{\bf ns},k_{\nu},\downarrow}^{iq}+V_{{\bf ns},k_{\nu},\downarrow;{\bf ns},k_{\nu},\downarrow}^{mq}\right)\left(V_{{\bf ns},k_{\nu},\uparrow;{\bf ns},k_{\nu},\downarrow}^{iq}\right)^{*}}{E_{\rm spin}}+{\rm c.c.}\right],   
\end{align}
\begin{align}
b_{0\;im}^{q,k_{\nu}}=\frac{1}{4}\left(K_{\uparrow \uparrow,\uparrow \uparrow \; im}^{q,k_{\nu}}+K_{\downarrow \downarrow,\downarrow \downarrow \; im}^{q,k_{\nu}}+K_{\uparrow \downarrow,\uparrow \downarrow \; im}^{q,k_{\nu}}+K_{\downarrow \uparrow,\downarrow \uparrow \; im}^{q,k_{\nu}}\right)= \nonumber \\
=\sum_{\xi,\eta=i,m}\sum_{\substack{ {\bf nl}={\bf ns},{\bf np}}}\sum_{\substack{ k',\nu'}}\frac{(-1)^{l}}{4}
\left[\frac{V_{{\bf ns},k_{\nu},\uparrow;{\bf nl},k'_{\nu'},\uparrow}^{\xi q}\left(V_{{\bf ns},k_{\nu},\uparrow;{\bf nl},k'_{\nu'},\uparrow}^{\eta q}\right)^{*}}{{\cal E}_{{\bf nl}}(k'_{\nu'})-{\cal E}_{{\bf ns}}(k_{\nu})}+ \right. \nonumber \\
+\frac{V_{{\bf ns},k_{\nu},\downarrow;{\bf nl},k'_{\nu'},\downarrow}^{\xi q}\left(V_{{\bf ns},k_{\nu},\downarrow;{\bf nl},k'_{\nu'},\downarrow}^{\eta q}\right)^{*}}{{\cal E}_{{\bf nl}}(k'_{\nu'})-{\cal E}_{{\bf ns}}(k_{\nu})} 
\left. +\frac{\left|V_{{\bf ns},k_{\nu},\uparrow;{\bf nl},k'_{\nu'},\downarrow}^{\xi q}\right|^{2}}{{\cal E}_{{\bf nl}}(k'_{\nu'})-{\cal E}_{{\bf ns}}(k_{\nu})-E_{\rm spin}}+\frac{\left|V_{{\bf ns},k_{\nu},\downarrow;{\bf nl},k'_{\nu'},\uparrow}^{\xi q}\right|^{2}}{{\cal E}_{{\bf nl}}(k'_{\nu'})-{\cal E}_{{\bf ns}}(k_{\nu})+E_{\rm spin}}+\right. \nonumber \\
\left. +\frac{\left|V_{{\bf ns},k_{\nu},\uparrow;{\bf nl},k'_{\nu'},\uparrow}^{i q}+V_{{\bf ns},k_{\nu},\downarrow;{\bf nl},k'_{\nu'},\downarrow}^{m q}\right|^{2}}{{\cal E}_{{\bf nl}}(k'_{\nu'})-{\cal E}_{{\bf ns}}(k_{\nu})}+
\frac{\left|V_{{\bf ns},k_{\nu},\downarrow;{\bf nl},k'_{\nu'},\downarrow}^{iq}+V_{{\bf ns},k_{\nu},\uparrow;{\bf nl},k'_{\nu'},\uparrow}^{m q}\right|^{2}}{{\cal E}_{{\bf nl}}(k'_{\nu'})-{\cal E}_{{\bf ns}}(k_{\nu})}+ \right. \nonumber \\
\left. +\frac{\left(\left|V_{{\bf ns},k_{\nu},\uparrow;{\bf nl},k'_{\nu'},\downarrow}^{i q}\right|^{2}+\left|V_{{\bf ns},k_{\nu},\uparrow;{\bf nl},k'_{\nu'},\downarrow}^{m q}\right|^{2}\right)}{{\cal E}_{{\bf nl}}(k'_{\nu'})-{\cal E}_{{\bf ns}}(k_{\nu})-E_{\rm spin}} 
+\frac{\left(\left|V_{{\bf ns},k_{\nu},\downarrow;{\bf nl},k'_{\nu'},\uparrow}^{i q}\right|^{2}+\left|V_{{\bf ns},k_{\nu},\downarrow;{\bf nl},k'_{\nu'},\uparrow}^{m q}\right|^{2}\right)}{{\cal E}_{{\bf nl}}(k'_{\nu'})-{\cal E}_{{\bf ns}}(k_{\nu})+E_{\rm spin}}\right],  
\end{align}
\end{widetext}
where ${\bf ns}=\{n,l=0,j=1/2,m_{j}\}$, ${\bf np}=\{n,l=1,j',m_{j}'\}$ and the summation is over $j'$, $m_{j}'$ quantum numbers.

\section*{\bf Appendix D}
\setcounter{equation}{0} \renewcommand{\theequation}{D.\arabic{equation}} 

\subsection{Calculation of $V_{{\bf ns},k_{0\;\nu_{0}},\alpha;{\bf np},k_{\nu},\beta}^{mq}$ matrix elements for charge-dipole interaction}

The interaction matrix elements for the charge-dipole interaction between an $m^{\rm th}$ spin, encoded in polar molecule rotational states,  
and a $q^{\rm th}$ mediator Rydberg atom 
have the form
\begin{widetext}
\begin{align}
V_{{\bf ns},k_{0\;\nu_{0}},\alpha;{\bf np},k_{\nu},\beta}^{mq}= \bra{ns_{1/2},m_{j},k_{0\;\nu_{0}}}_{q}\bra{\alpha}_{m}\hat{V}_{e^{-}-M}\ket{\beta}_{m}\ket{np_{j'},m_{j}',k_{\nu}}_{q}= \nonumber \\ 
=-e\bra{\alpha}_{m}\vec{d}_{\rm spin}\ket{\beta}_{m}\int dX_{q}\Phi_{ns\;q}^{*}(X_{q},k_{0\;\nu_{0}\;q})\bra{ns,m_{j}}\frac{\vec{R}_{qm}-\vec{r}}{\left|\vec{R}_{qm}-\vec{r}\right|^{3}}\ket{np_{j}',m_{j}'}\phi(X_{q},k_{\nu\;q})=  \\
=-e\bra{\alpha}_{m}\vec{d}_{\rm spin}\ket{\beta}_{m}e^{-i(k-k_{0})X_{m}}\int dX_{qm}u_{k_{0}}^{(\nu_{0})*}(X_{m}+X_{qm})\bra{ns,m_{j}}\frac{\vec{R}_{qm}-\vec{r}}{\left|\vec{R}_{qm}-\vec{r}\right|^{3}}\ket{np_{j'},m_{j}'}u_{k}^{(\nu)}(X_{m}+X_{qm})e^{-i(k-k_{0})X_{qm}}= \nonumber \\
=c_{{\bf ns},k_{0\;\nu_{0}},\alpha;{\bf np},k_{\nu},\beta}^{mq}e^{-i(k-k_{0})X_{m}}. 
\label{eq:V-ch-dip-matr-elem} 
\end{align}
\end{widetext}
Assuming for concreteness that the mediator atom is initially excited to the $\ket{ns_{1/2},m_{j}=1/2}$ 
state, the matrix elements Eq.(\ref{eq:V-ch-dip-matr-elem}) will be non-zero for virtual excitations only to the $m_{j}'=\pm 1/2$ sublevels of the 
$np_{j'}=65p_{1/2}$ state, which can be expanded in terms of the $l,m_{l}$ states as follows:
\begin{widetext}
\begin{align}
\ket{np_{1/2},m_{j}= \frac{1}{2}}= -\sqrt{\frac{1}{3}}\ket{n,l=1,m_{l}=0;s=\frac{1}{2},m_{s}=\frac{1}{2}}+ \nonumber 
+\sqrt{\frac{2}{3}}\ket{n,l=1,m_{l}=1;s=\frac{1}{2},m_{s}=-\frac{1}{2}}, \nonumber \\ 
\ket{np_{1/2},m_{j} =  -\frac{1}{2}}=-\sqrt{\frac{2}{3}}\ket{n,l=1,m_{l}=-1;s=\frac{1}{2},m_{s}=\frac{1}{2}}+ \nonumber 
 +\sqrt{\frac{1}{3}}\ket{n,l=1,m_{l}=0;s=\frac{1}{2},m_{s}=-\frac{1}{2}},  
\end{align}
\end{widetext}
which can be used to calculate $c_{{\bf ns},k_{0\;\nu_{0}},\uparrow;{\bf np},k_{\nu},\downarrow}^{qm}$ coefficients.

\begin{figure}[h]
\center{
\includegraphics[width=7.cm]{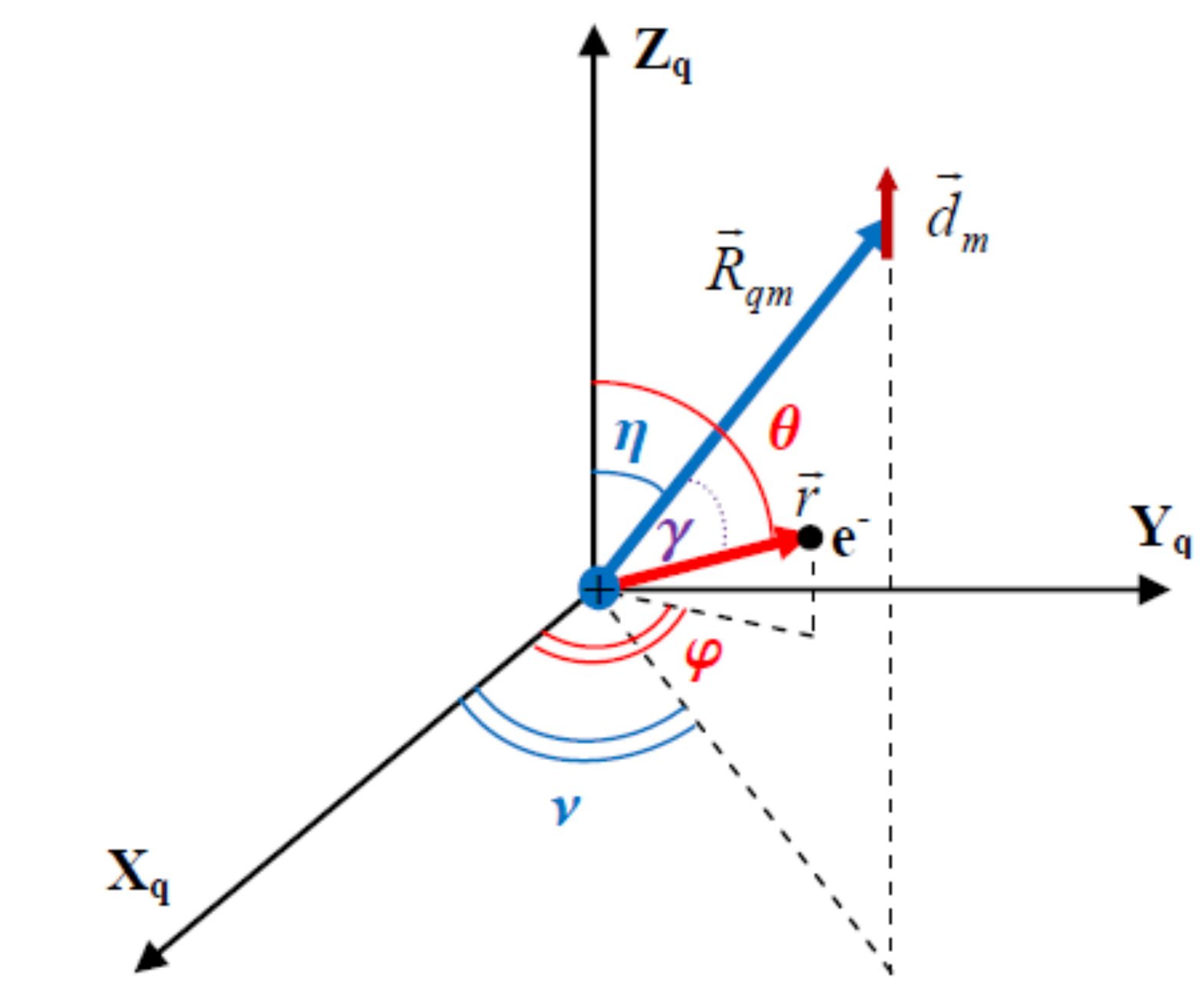}
\caption{\label{fig:spin-spin-setup} Angles of the vectors $\vec{R}_{qm}$ and 
$\vec{r}$ in the case of a general orientation of the Rydberg atom with respect to the molecule.
}
}
\end{figure}

The interaction coefficient corresponding to the resonance between the $\ket{np_{1/2},m_{j}=1/2}-\ket{ns_{1/2},m_{j}=1/2}$ transition of the 
q$^{\rm th}$ atom and the spin transition of the 
m$^{\rm th}$ molecule are given by the expression 
\begin{align} 
c_{{\bf ns},k_{0\;\nu_{0}},\uparrow;{\bf np},k_{\nu},\downarrow}^{mq}= \nonumber \\
=-\frac{e\bra{\uparrow}\vec{d}_{\rm spin}\ket{\downarrow}_{m}}{\sqrt{3}}
\int dX_{qm}\left \langle ns \right| \frac{\vec{R}_{qm}-\vec{r}}{|\vec{R}_{qm}-\vec{r}|^{3}}\left| np,m_{l}=0 \right \rangle \times \nonumber \\
\times u_{k_{0}}^{(\nu_{0})*}\left(X_{m}+X_{qm}\right)u_{k}^{(\nu)}\left(X_{m}+X_{qm}\right)e^{-i(k-k_{0})X_{qm}}, 
\end{align}
where the Bloch functions are normalized as $\int \phi_{k}^{(\nu)*}(X)\phi_{k'}^{(\nu')}(X)dX=\delta_{n,n'}\delta_{k,k'}$. Let us 
first analyze the integral over Rydberg electron's coordinates:
\begin{eqnarray}
I_{ns;np,0}^{mq}=-e\bra{\uparrow}\vec{d}_{\rm spin}\ket{\downarrow}_{m}\bra{ns}\frac{\vec{R}_{qm}-\vec{r}}{\left|\vec{R}_{qm}-\vec{r}\right|^{3}}\ket{np,m_{l}=0}. \nonumber
\end{eqnarray}
For the effective spin states $\ket{\downarrow}=\ket{J=0,m_{J}=0}$ and $\ket{\uparrow}=\ket{J=1,m_{J}=0}$ the spin dipole moment has only the $z$ 
component $\bra{\uparrow}\vec{d}_{\rm spin}\ket{\downarrow}=\vec{e}_{z}\bra{\uparrow}d_{\rm z\; spin}\ket{\downarrow}$. One then can use the following expression 
\cite{PRA-paper}:
\begin{widetext}
\begin{align}
-e\bra{\uparrow}d_{\rm z\;spin}\ket{\downarrow}_{m}\frac{\left(\vec{R}_{qm}-\vec{r}\right)_{z}}{\left|\vec{R}_{qm}-\vec{r}\right|^{3}}=4\pi e\bra{\uparrow}d_{\rm z\;spin}\ket{\downarrow}_{m}\cos \eta \times \nonumber \\
\times \left\{ \begin{array}{ccc} \sum_{l''=0}^{\infty}-\frac{l''+1}{2l''+1}\frac{r^{l''}}{R_{qm}^{l''+2}}\sum_{m''=-l''}^{l''}Y_{l''}^{m''}(\theta,\phi)Y_{l''}^{m''\;*}(\eta,\nu) & {\rm for} & r<R_{qm} \\
\sum_{l''=0}^{\infty}\frac{l''}{2l''+1}\frac{R_{qm}^{l''-1}}{r^{l''+1}}\sum_{m''=-l''}^{l''}Y_{l''}^{m''}(\theta,\phi)Y_{l''}^{m''\;*}(\eta,\nu) & {\rm for} & r>R_{qm}  \end{array} \right. \nonumber \\
-4\pi e \bra{\uparrow}d_{\rm z\; spin}\ket{\downarrow}_{m}\frac{\sin \eta}{R_{qm}} \times \nonumber \\
\times \left\{ \begin{array}{ccc} \sum_{l''=0}^{\infty}\frac{1}{2l''+1}\frac{r^{l''}}{R_{qm}^{l''+1}}\sum_{m''=-l''}^{l''}Y_{l''}^{m''}(\theta,\phi)\frac{\partial Y_{l''}^{m''\;*}(\eta,\nu)}{\partial \eta} & {\rm for} & r<R_{qm} \\
\sum_{l''=0}^{\infty}\frac{1}{2l''+1}\frac{R_{qm}^{l''}}{r^{l''+1}}\sum_{m''=-l''}^{l''}Y_{l''}^{m''}(\theta,\phi)\frac{\partial Y_{l''}^{m''\;*}(\eta,\nu)}{\partial \eta} & {\rm for} & r>R_{qm}  \end{array}, \right.
\label{eq:int-expansion}
\end{align} 
\end{widetext}
where $\theta$ and 
$\phi$ are the Rydberg electron's angular coordinates with respect to the ionic core, $\eta$ and $\nu$ are the angular coordinates of the core-molecule vector $\vec{R}_{qm}$ 
with respect to the quantization axis, chosen to be perpendicular to the spin and mediator lattices and parallel to $\vec{\rho}$ 
(see Fig.\ref{fig:spin-spin-setup}). When calculating the matrix element 
$I_{ns;np,0}^{mq}$ between the $\ket{ns}$ and $\ket{np,m_{l}=0}$ states, there will be integrals over three spherical harmonics, expressed via 3j-symbols 
\begin{eqnarray}
\int_{0}^{2\pi}d\phi \int_{0}^{\pi}\sin \theta d\theta Y_{0}^{0\;*}Y_{l''}^{m''}Y_{1}^{0}= \nonumber \\
=\sqrt{\frac{3(2l''+1)}{4\pi}}\left(\begin{array}{ccc} 0 & l'' & 1 \\
                                                                                                                                                0 & 0 & 0 \end{array} \right) 
\left(\begin{array}{ccc} 0 & l'' & 1 \\
                         0 & m'' & 0 \end{array} \right),
\end{eqnarray}
which are non-zero only for $l''=1,m''=0$, with the corresponding integral given by 
$\int_{0}^{2\pi}d\phi \int_{0}^{\pi}\sin \theta d\theta Y_{0}^{0\;*}Y_{1}^{0}Y_{1}^{0}=1/\sqrt{4\pi}$.

As a result, 
\begin{widetext}
\begin{align}
I_{ns;np,0}^{mq}=4\pi e \bra{\uparrow}d_{\rm z\;spin}\ket{\downarrow}_{m}\cos \eta Y_{1}^{0\;*}(\eta,\nu)\left(-\frac{2}{3\sqrt{4\pi}R_{qm}^{3}}\int_{0}^{R_{qm}}r^{3}R_{ns}R_{np}dr+\frac{1}{3\sqrt{4\pi}}\int_{R_{qm}}^{\infty}R_{ns}R_{np}dr \right) \nonumber \\
-4\pi e \bra{\uparrow}d_{\rm z\;spin}\ket{\downarrow}_{m}\frac{\sin \eta}{R_{qm}}\frac{\partial Y_{1}^{0\;*}}{\partial \eta}\left(\frac{1}{3\sqrt{4\pi}R_{qm}^{2}}\int_{0}^{R_{qm}}r^{3}R_{ns}R_{np}dr+\frac{R_{qm}}{3\sqrt{4\pi}}\int_{R_{qm}}^{\infty}R_{ns}R_{np}dr\right), \nonumber 
\end{align}
\end{widetext}
where $\frac{\partial Y_{1}^{0\;*}}{\partial \eta}=-\frac{1}{2}\left(\sqrt{2}\left(Y_{1}^{-1}\right)^{*}e^{-i\nu}-\sqrt{2}\left(Y_{1}^{1}\right)^{*}e^{i\nu}\right)=-\frac{1}{2}\sqrt{\frac{3}{\pi}}\sin \eta$,
and $Y_{1}^{0}=\frac{1}{2}\sqrt{\frac{3}{\pi}}\cos \eta$, $Y_{1}^{\pm 1}=\mp \frac{1}{2}\sqrt{\frac{3}{2\pi}}\sin \eta e^{\pm i\nu}$. After rearrangement, 
\begin{align}
I_{ns;np,0}^{mq}=\frac{e\bra{\uparrow}d_{\rm z\;spin}\ket{\downarrow}_{m}}{\sqrt{3}}\times \nonumber \\
\times \left(\frac{\sin^{2}\eta-2\cos^{2}\eta}{R_{qm}^{3}}\int_{0}^{R_{qm}}r^3R_{ns}R_{np}dr+\int_{R_{qm}}^{\infty}R_{ns}R_{np}dr\right),
\end{align} 
where $\cos^{2}\eta=\frac{\rho^{2}}{R_{qm}^{2}}$, $\sin^{2}\eta=\frac{(X_{qm})^{2}}{R_{qm}^{2}}$. 

This gives the following expression for the $c^{qm}$ coefficient
\begin{widetext}
\begin{align}
c_{{\bf ns},k_{0\;\nu_{0}},\uparrow;{\bf np},k_{\nu},\downarrow}^{qm}=\frac{e\bra{\uparrow}d_{\rm z\;spin}\ket{\downarrow}_{m}}{3}\int dX_{qm}\left(\frac{\sin^{2}\eta-2\cos^{2}\eta}{R_{qm}^{3}}\int_{0}^{R_{qm}}r^3R_{ns}R_{np}dr+ \right. \nonumber \\
\left. +\int_{R_{qm}}^{\infty}R_{ns}R_{np}dr\right)u_{k_{0}}^{(\nu_{0})*}\left(X_{m}+X_{qm}\right)u_{k}^{(\nu)}\left(X_{m}+X_{qm}\right)e^{-i(k-k_{0})X_{qm}}.
\end{align}
\end{widetext}
For the distances between 
spin and mediator arrays $\rho \sim 500$ nm, which we consider, the $\frac{1}{R_{qm}^{3}}\int_{0}^{R_{qm}}r^{3}R_{ns}R_{np}dr$ term will be much larger than 
the $\int_{R_{qm}}^{\infty}R_{ns}R_{np}dr$ term, such that the latter can be neglected. At these distances one can also approximate $\int_{0}^{R_{qm}}r^{3}R_{ns}R_{np}dr\approx \int_{0}^{\infty}r^{3}R_{ns}R_{np}dr$. As a result, 
the dependence 
on the mediator position of the $c^{qm}$ coefficients will be given by the factor $(\sin^{2}\eta-2\cos^{2}\eta)/R_{qm}^{3}$, which shows that at 
such distances the charge-dipole interaction can be approximated by the dipole-dipole one.

Next, we calculate the coefficients for the spin-mediator interaction involving the $\ket{\downarrow}-\ket{\uparrow}$ and the 
 $\ket{65p_{1/2},m_{j}=-1/2}-\ket{65s_{1/2},m_{j}=1/2}$ transitions
\begin{widetext}
\begin{eqnarray}
c_{{\bf ns},k_{0\;\nu_{0}},\uparrow;{\bf np},k_{\nu},\downarrow}^{mq}=\sqrt{\frac{2}{3}}e\bra{\uparrow}\vec{d}_{\rm spin}\ket{\downarrow}_{m}
\int dX_{qm}\bra{ns}\frac{\vec{R}_{qm}-\vec{r}}{|\vec{R}_{qm}-\vec{r}|^{3}}\ket{np,m_{l}=-1}\times \nonumber \\
\times u_{k_{0}}^{(\nu_{0})*}\left(X_{m}+X_{qm}\right)u_{k}^{(\nu)}\left(X_{m}+X_{qm}\right)e^{-i(k-k_{0})X_{qm}}.
\end{eqnarray}
\end{widetext}
When averaging the $-\frac{4\pi e\bra{\uparrow}d_{\rm z\; spin}\ket{\downarrow}_{m}\left(\vec{R}_{qm}-\vec{r}\right)_{z}}{\left|\vec{R}_{qm}-\vec{r}\right|^{3}}$ function 
 over $\ket{ns}$, $\ket{np,m_{l}=-1}$ states the integral over three spherical harmonics will have the form 
$\int_{0}^{2\pi}d\phi \int_{0}^{\pi}\sin\theta d\theta Y_{0}^{0\;*}Y_{l''}^{m''}Y_{1}^{-1}$, which is non-zero only for $l''=1,m''=1$. 
Following the same steps 
as above one obtains
\begin{align}
I_{ns;np,-1}^{mq}=-e\bra{\uparrow}d_{\rm z\; spin}\ket{\downarrow}_{m}\bra{ns}\frac{\left(\vec{R}_{qm}-\vec{r}\right)_{z}}{\left|\vec{R}_{qm}-\vec{r}\right|^{3}}\ket{np,m_{l}=-1}= \nonumber \\
=-e\bra{\uparrow}d_{\rm z\; spin}\ket{\downarrow}_{m}\cos \eta \sin \eta e^{-i\nu}\frac{1}{R_{qm}^{3}}\sqrt{\frac{3}{2}}\int_{0}^{R_{qm}}r^{3}R_{ns}R_{np}dr, 
\end{align}
resulting in
\begin{widetext}
\begin{align}
c_{{\bf ns},k_{0\;\nu_{0}},\uparrow;{\bf np},k_{\nu},\downarrow}^{mq}=-e\bra{\uparrow}d_{\rm z\;spin}\ket{\downarrow}_{m}\times \nonumber \\
\times \int dX_{qm}\left(\sin\eta\cos\eta e^{-i\nu}\frac{1}{R_{qm}^{3}}\int_{0}^{R_{qm}}r^3R_{ns}R_{np}dr\right) 
u_{k_{0}}^{(\nu_{0})*}\left(X_{m}+X_{qm}\right)u_{k}^{(\nu)}\left(X_{m}+X_{qm}\right)e^{-i(k-k_{9})X_{qm}}. 
\end{align}
\end{widetext}

Finally, we discuss the $V^{mq}_{{\bf ns},k_{0\;\nu_{0}},\uparrow;{\bf ns},k_{\nu},\uparrow}=c^{mq}_{{\bf ns},k_{0\;\nu_{0}},\uparrow;{\bf ns},k_{\nu},\uparrow}e^{-i(k-k_{0})X_{m}}$ 
matrix elements, where 
\begin{align}
c^{mq}_{{\bf ns},k_{0\;\nu_{0}},\uparrow;{\bf ns},k_{\nu},\uparrow}= \nonumber \\
=e\bra{\uparrow}\vec{d}_{\rm spin}\ket{\uparrow}_{m}\int dX_{qm}\bra{ns}\frac{\vec{R}_{qm}}{R_{qm}^{3}}-\frac{\vec{R}_{qm}-\vec{r}}{\left|\vec{R}_{qm}-\vec{r}\right|^{3}}\ket{ns}\times \nonumber \\
\times u_{k_{0}}^{(\nu_{0})*}\left(X_{m}+X_{qm}\right)u_{k}^{(\nu)}\left(X_{m}+X_{qm}\right)e^{-i(k-k_{0})X_{qm}},
\end{align}
The integral over Rydberg electron's coordinates
\begin{align}
I^{mq}_{ns;ns}=e\bra{\uparrow}\vec{d}_{\rm spin}\ket{\uparrow}_{m}\bra{ns}\frac{\vec{R}_{qm}}{R_{qm}^{3}}-\frac{\vec{R}_{qm}-\vec{r}}{\left|\vec{R}_{qm}-\vec{r}\right|^{3}}\ket{ns} 
\end{align}
can be calculated using the expansion (\ref{eq:int-expansion}) since $\bra{\uparrow}\vec{d}_{\rm spin}\ket{\uparrow}=\vec{e}_{z}\bra{\uparrow}d_{\rm z \; spin}\ket{\uparrow}$ also 
has only the $z$ component. As a result,
\begin{align}
I^{mq}_{ns;ns}=\frac{e\bra{\uparrow}d_{\rm z \; spin}\ket{\uparrow}\rho}{R_{qm}^{3}}\left(1-\int_{0}^{R_{qm}}R_{ns}^{2}(r)r^{2}dr\right).
\end{align}
We can estimate the term $\int_{0}^{R_{qm}}R_{ns}^{2}(r)r^{2}dr > \int_{0}^{\rho}R_{ns}^{2}(r)r^{2}dr \approx 0.999999999$ for $65s$ of Rb and $\rho=500$ nm, 
resulting in $I^{mq}_{ns;ns}\approx 10^{-10}e\rho \bra{\uparrow}d_{\rm z\; spin}\ket{\uparrow}/R_{qm}^{3}$. This shows that the effective dipole moment of the 
$ns$ state is $e\rho\left(1-\int_{0}^{\rho}R_{ns}^{2}(r)r^{2}dr\right)\approx 10^{-10}e\rho \approx 10^{-6}$ a.u. If the $\ket{\uparrow}$ state is coupled 
to a $\ket{J=2,m_{J}}$ state by a detuned MW field, the dipole moment $\bra{\uparrow}d_{\rm z\;spin}\ket{\uparrow}\sim \left({\tilde \Omega}/{\tilde \Delta}\right)d_{\rm spin}$. 
The interaction coefficients can then be estimated as 
$c^{mq}_{ns,1/2,k_{0\;\nu_{0}},\uparrow;ns,1/2,k_{\nu},\uparrow} \lesssim 10^{-10}e\rho \bra{\uparrow}d_{\rm z\;spin}\ket{\uparrow}/\rho^{3} \sim 10^{-10}e\rho \left({\tilde \Omega}/{\tilde \Delta}\right)d_{\rm spin}/\rho^{3} \sim 10^{-3}$ Hz 
for ${\tilde \Omega}/{\tilde \Delta}\sim 0.1$. It shows that the $|V^{mq}_{ns,1/2,k_{0},\uparrow;ns,1/2,k_{\nu},\uparrow}| \ll E_{\rm rec}/N_{\rm at\;latt}$ and 
$|V^{mq}_{{\bf ns},k_{0\;\nu_{0}},\uparrow;{\bf ns},k_{\nu},\uparrow}|^{2}/E_{\rm rec}/N_{\rm at\;latt} \sim 3 \cdot 10^{-7}$ Hz, which is much less than the $J_{im}^{\bot \;q}$, $b_{i}^{z\;q}$ 
terms in Eq.(\ref{eq:inter-coeff-XY}).     

\subsection{$c^{qm}_{{\bf ns},k_{0\;\nu_{0}},\alpha;{\rm med},k_{\nu},\beta}$ coefficients for the Rydberg spin encoding}

\begin{widetext}
\begin{align}
c^{mq}_{{\bf ns},\uparrow,k_{0\;\nu_{0}};{\rm med},\downarrow,k_{\nu}}=\frac{a_{\uparrow}b_{\downarrow}+b_{\uparrow}a_{\downarrow}}{3}d_{np,ns}d_{{\tilde n}p,{\tilde n}s}d_{\pm}
\int dX_{qm}u_{k_{0}}^{(\nu_{0})*}\left(X_{m}+X_{qm}\right)\frac{X_{qm}\rho}{R_{qm}^{5}}u_{k}^{(\nu)}\left(X_{m}+X_{qm}\right)e^{ikX_{qm}}, \nonumber \\
c^{mq}_{{\bf ns},\downarrow,k_{0\;\nu_{0}};{\rm med},\uparrow,k_{\nu}}=c^{mq}_{{\bf ns},\uparrow,k_{0\;\nu_{0}};{\rm med},\downarrow,k_{\nu}}, \nonumber \\
c^{mq}_{{\bf ns},\uparrow,k_{0\;\nu_{0}};{\rm med},\uparrow,k_{\nu}}=\frac{2a_{\uparrow}b_{\uparrow}}{9}d_{np,ns}d_{{\tilde n}p,{\tilde n}s}d_{\pm}
\int dX_{qm}u_{k_{0}}^{(\nu_{0})*}\left(X_{m}+X_{qm}\right)\frac{1-3\rho^{2}/R_{qm}^{2}}{R_{qm}^{3}}u_{k}^{(\nu)}\left(X_{m}+X_{qm}\right)e^{ikX_{qm}}, \nonumber \\
c^{mq}_{{\bf ns},\downarrow,k_{0\;\nu_{0}};{\rm med},\downarrow,k_{\nu}}=-\frac{2a_{\downarrow}b_{\downarrow}}{9}d_{np,ns}d_{{\tilde n}p,{\tilde n}s}d_{\pm}
\int dX_{qm}u_{k_{0}}^{(\nu_{0})*}\left(X_{m}+X_{qm}\right)\frac{1-3\rho^{2}/R_{qm}^{2}}{R_{qm}^{3}}u_{k}^{(\nu)}\left(X_{m}+X_{qm}\right)e^{ikX_{qm}}, \nonumber 
\end{align}
\end{widetext}

with the corresponding dipole moments between the $\ket{ns_{1/2},m_{j}=1/2}$ and the $\ket{\pm}_{\rm med}$ states:
\begin{align}
\bra{ns_{1/2},m_{j}=\frac{1}{2}}\vec{d}_{\rm Rydb}\ket{+}_{\rm med}=-\frac{d_{+}d_{np,ns}\vec{e}_{z}}{3}, \nonumber \\
\bra{ns_{1/2},m_{j}=\frac{1}{2}}\vec{d}_{\rm Rydb}\ket{-}_{\rm med}=-\frac{d_{-}d_{np,ns}\vec{e}_{z}}{3},
\end{align}

\subsection{Averaging over atomic motional states}

\begin{figure}[h]
\center{
\includegraphics[width=9.cm]{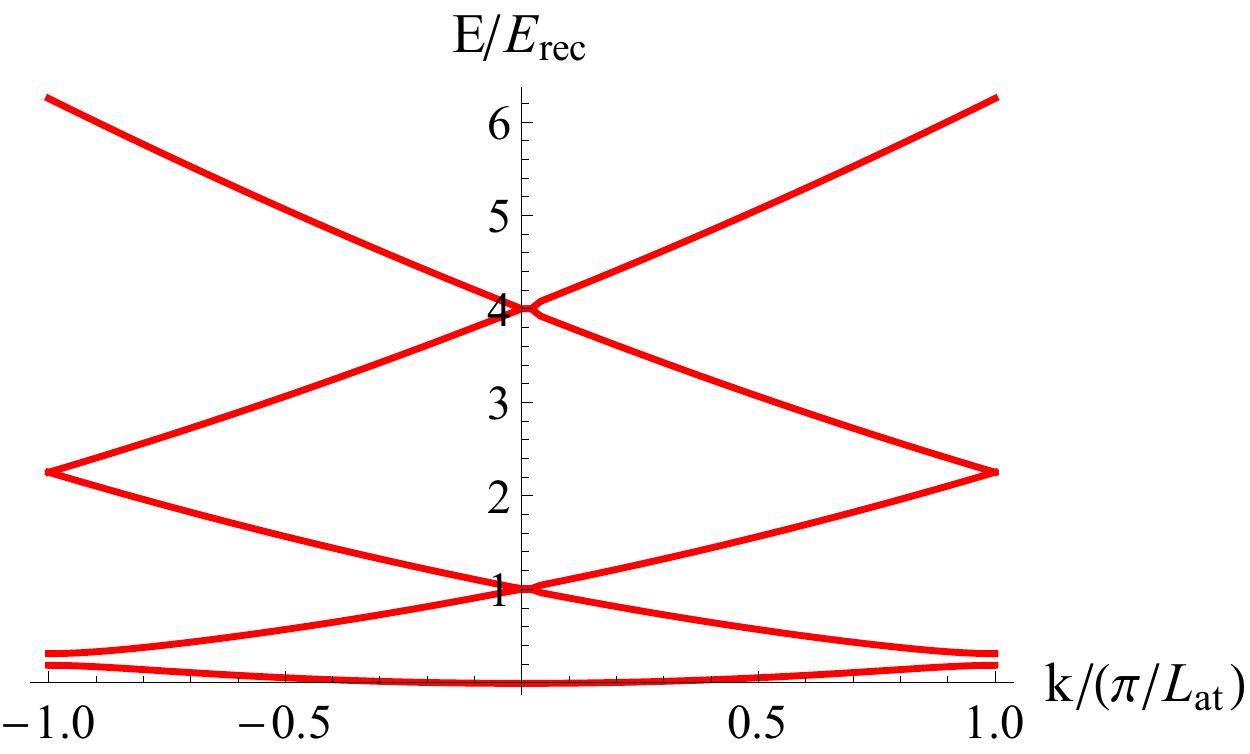}
\caption{\label{fig:bloch-energies} Bloch energies of the lowest six bands of the 1D optical lattice described by the trapping potential $V(X_{q})=V_{0}\cos^{2}(K_{\rm at}X_{q})$ 
with $V_{0}=-E_{\rm rec}$. }
}
\end{figure}

In order to calculate the $c_{{\bf ns},k_{0\;\nu_{0}},\alpha;{\bf np},k_{\nu},\beta}^{mq}$ and $c^{mq}_{{\bf ns},\alpha,k_{0\;\nu_{0}};{\rm med},k_{\nu},\beta}$ 
 coefficients energies and wavefunctions of 
mediator atoms Bloch states are needed.  
In this section we numerically calculate the Bloch functions by analyzing atomic motion in a 1D optical lattice described by the 
trapping potential $V(X_{q})=V_{0}\cos^{2}K_{\rm at}X_{q}$ with the lattice momentum $K_{\rm at}=\pi/L_{\rm at}$ 
and $V_{0}=-E_{\rm rec}$, where $E_{\rm rec}=\hbar^{2}(K_{\rm at})^{2}/2M_{\rm at}$ is the atomic recoil energy in the lattice, 
$M_{\rm at}$ is the atomic mass, $L_{\rm at}$ is the mediator atom lattice period. The Bloch states and energies 
can be found by solving the Schrodinger equation for atomic motion in the 1D lattice:
\begin{align}
-\frac{\hbar^{2}}{2M_{\rm at}}\frac{d^{2}\phi_{k}^{(\nu)}}{dX_{q}^{2}}+V_{0}\cos^{2}(K_{\rm at}X_{q})\phi_{k}^{(\nu)}(X_{q})=E^{(\nu)}(k)\phi_{k}^{(\nu)}(X_{q}), 
\label{eq:Schrodinger-eq}
\end{align} 
where the wavefunction corresponding to the $\nu^{\rm th}$ Bloch band and the quasimomentum $k$ is $\phi_{k}^{(\nu)}(X_{q})=u_{k}^{(\nu)}(X_{q})e^{ikX_{q}}$.  
The periodic function $u_{k}$ can be expanded in terms of the harmonics of the lattice momentum: $u_{k}^{(\nu)}(X_{q})=\sum_{s=-S_{\rm max}}^{S_{\rm max}}c_{s}^{(\nu)}(k)e^{2isK_{\rm at}X_{q}}$ 
with the expansion truncated at some $S_{\rm max}$. The periodic boundary conditions $\phi_{k}^{(\nu)}(X_{q})=\phi_{k}^{(\nu)}(X_{q}+N_{\rm latt \; at}L_{\rm at})$ 
require $k=2\pi \kappa/L_{\rm at}N_{\rm latt \; at}$, $\kappa=-N_{\rm latt\; at}/2,...N_{\rm latt \; at}/2$, where $N_{\rm latt\; at}$ is 
the number of unitary cells in the mediator atoms lattice. 
The coefficients $c_{s}^{(\nu)}(k)$ and Bloch energies $E^{(\nu)}$ can be calculated by numerically solving Eq.(\ref{eq:Schrodinger-eq}):
\begin{align}
\left(\frac{\kappa}{N_{\rm at}}+s\right)^{2}c_{s}^{(\nu)}+\frac{V_{0}}{4E_{\rm rec}}\left(c_{s-1}^{(\nu)}+c_{s+1}^{(\nu)}\right)=\frac{E^{(\nu)}-V_{0}/2}{E_{\rm rec}}c_{s}^{(\nu)}.
\end{align}
Fig.\ref{fig:bloch-energies} shows energies of the five lowest Bloch bands in the case $V_{0}=-E_{\rm rec}$ assuming that the lattice has $N_{\rm latt \;at}=100$ 
unitary cells. The expansion of $u_{k}$ was truncated at $S_{\rm max}=10$. 

We numerically calculated the $u_{k}^{(\nu)}(X_{q})$ functions for five lowest Bloch bands and used them for obtaining the $c^{mq}$ 
coefficients discussed in the subsections A, B. 

\end{appendices}

\end{document}